\documentclass[10pt,a4paper,oneside,twocolumn]{book}
\usepackage{multicol} 

\usepackage[T1]{fontenc}

\usepackage{hyperref}

\usepackage{graphicx}

\usepackage{widetext} 

\usepackage{fancyhdr} 
\usepackage{xcolor}
\usepackage[normalem]{ulem} 
\usepackage{bm}
\usepackage{amsmath}
\usepackage{amssymb}
\usepackage{yhmath}  
\usepackage{enumitem}
\usepackage{cite} 

\usepackage[all]{xy}  
\usepackage[customcolors]{hf-tikz} 
   \hfsetfillcolor{blue!8}
   \hfsetbordercolor{blue!50}

\usepackage{textcomp}

\usepackage{titlesec} 
\titleformat{\chapter}[display]
{\normalfont\Large\filcenter}
{
\vspace{-3.5pc}%
\large\bfseries\MakeUppercase{\chaptertitlename} \thechapter}
{1pc}
{\LARGE\bfseries\MakeUppercase}
[\vspace{1pc}%
]

\titleformat{\section}{\Large\bfseries\sffamily}{\thesection}{0.3em}{}
\titleformat{\subsection}{\large\bfseries\sffamily}{\thesubsection}{0.3em}{}

\usepackage[english]{babel}
\usepackage{ragged2e}

\definecolor{DarkGreen}{rgb}{0,0.4,0}
\definecolor{DarkBlue}{rgb}{0,0,0.8}

\newcommand{\revisit}[1]{\textcolor{black}{#1}}

\newcommand{\ignore}[1]{\relax}

\newcommand{\DS}{{\Delta\hskip -0.2mm S}}

\newcommand{\e}{{\rm e}}
\newcommand{\rmd}{{\rm d}}
\newcommand{\rmi}{{\rm i}}
\newcommand{\kB}{k_{\rm B}}

\newcommand{\dtraj}{\gamma}

\newcommand{\steady}{{\rm steady \hskip -7.2mm \raisebox{-1.1ex}{$\scriptstyle \rm state$}\hskip 1mm}}

\newcommand{\myBoxed}{\boxed{\phantom{j\hskip -1.5mm A^*}}\phantom{\Big|} \hskip -6mm} 

\newcommand{\mytitle}{
ILLUSORY CRACKS
\vskip 2mm 
IN THE
\vskip 2mm 
SECOND LAW 
OF THERMODYNAMICS\vskip 2mm
IN \vskip 3mm
QUANTUM
NANOELECTRONICS
}

\newcommand{\myshorttitle}{Illusory cracks
in the second law of thermodynamics 
in quantum nanoelectronics}

\fancypagestyle{plain}{

}

\begin{document}


\title{\vskip -32mm 
\hrule\vskip 1mm\hrule
\vskip 10mm
{\textbf {\textsf {\huge \mytitle}}}\\
\vskip 14mm
\hrule\vskip 1mm\hrule
\vskip 10mm
{\large \textbf{Robert S. Whitney}}\\
\vskip 3mm
{\small Laboratoire de Physique et Mod\'elisation des Milieux Condens\'es,} 
\vskip -4mm
{\small Universit\'e Grenoble Alpes and CNRS, 25 avenue des Martyrs BP166, 
38042 Grenoble, France.}\\
\vskip -3mm
{\small \texttt{robert.whitney@grenoble.cnrs.fr}}\\
\vskip 5mm
{\small April 6. 2023}
\vskip 8mm
\begin{minipage}{0.7\textwidth}
\centerline{{\textbf {\normalsize ABSTRACT}}}
\vskip 3mm
{\normalsize This is a review of the theory of quantum thermodynamic demons; these are quantum systems that look like they violate the laws of thermodynamics, in analogy with Maxwell's demon. It concentrates on autonomous demons that can be made using nanoelectronics.  
Here ``autonomous'' means that the demon operates without any external measurement or driving, making it possible to model their entire thermodynamic behaviour using
Schr\"odinger's equation. My main aim is to review why cracks in the laws of thermodynamics, sometimes glimpsed in such systems, have turned out to be illusory. 
\vskip 3mm
For this, I work by example, introducing the methods of quantum thermodynamics via an old thought experiment that appears to break the second law. This thought experiment is usually known as Smoluchowski's trapdoor, although it was first proposed by Maxwell.  Smoluchowski showed that the trapdoor's thermal motion complicates its dynamics, but I argue that he did not show that it must obey the second law. This question is now of practical relevance, because 
a nanoelectronic version of the trapdoor can be made with quantum dots. 
The methods presented here show that such a quantum trapdoor obeys the laws of thermodynamics.
\vskip 3mm
This reviews then addresses other types of autonomous demon that superficially appear to break the laws of thermodynamics, but which do not.  These include
an experimental demonstration of a Maxwell demon, and a kind of demon that exploit non-equilibrium resources (the N-demon). It discusses a way of classifying different kinds of autonomous demon. It concludes by briefly reviewing how fluctuations affect nanoelectronics, and how their role in stochastic thermodynamics changes our view of entropy.
\vskip 3mm}
\end{minipage}
\vskip -30mm
}
\date{}
%
\maketitle

\setcounter{page}{2}
\tableofcontents

\mainmatter


\fancyhead[LO]{\textsf{R.S. Whitney --- \myshorttitle}}
\fancyhead[RE]{\textsf{Chapter~\thechapter : \leftmark}}

\setcounter{page}{5}

\chapter[Introduction. From 19th century thermodynamics to 21st century nanoelectronic]{Introduction. \mbox{\ \ \ From 19th century thermodynamics\ \ \ } \mbox{to 21st century nanoelectronics}}
\label{Chap:intro}

\begin{quotation}
\noindent
\textsf{Oh ye seekers after perpetual motion,}

\vskip -1mm
\textsf{how many vain chimeras have you pursued?} 

\hfill \textsf{--- Leonardo da Vinci, 1494} \cite{daVinci-quote}
\end{quotation}

This review is about what we learn from pursuing chimeras that we {\it known} do not exist.  In this case, the chimeras are perpetual-motion machines that generate useful work by violating the second law of thermodynamics -- known as perpetual motion of the second kind \footnote{Perpetual-motion machines of the first kind are ones that violate energy conservation, to make useful work out of nothing. They are immediately ruled out here by the fact I will only consider energy-conserving theories.}. 
The typical example would be a machine capable of generating work from a single heat source, turning heat in that source into useful work, and thereby cooling that heat source, reducing the overall entropy.

I will specifically look at the thermodynamics of electronic flows in nanostructures (quantum dots, etc). The electronic flows exhibit quantum effects, such as quantization, tunnelling, and coherent superpositions, which means we are in the domain of {\it quantum thermodynamics}.
I will review designs of electronic systems that look like they should behave like Maxwell's demon, enabling them to be perpetual-motion machines.  These electronic systems fall into the class of machines that we will call {\it autonomous demons}; they are ``autonomous'' in the sense that they look like they would produce useful work autonomously without any external equipment (no external detection-feedback circuits, and no external driving). 
I will then review correct analyses of such systems, showing that they do not violate the second-law, and thus cannot be perpetual-motion machines. 
Hence, the cracks in the second law that are sometimes glimpsed, turn out on closer inspection to be illusory.

You may well ask why we should pursue perpetual-motion machines, when we know from the start that they are doomed to fail?
My answer is that a good way to understand a physical theory is to test its limits.  In logic and mathematics,
{\it reductio ad absurbum} is a powerful analytical method.
Here I propose something similar, to try to construct quantum machines that violate the second law of thermodynamics, to better understand why the second law is {\it never} violated.

\section{The second law}
\label{sec:2ndlaw}
Before going further, it is important to establish which version of the second law we believe is never violated.
The version that I use is formulated as follows.
\begin{quotation}
\noindent\textsf{\textbf{Second law of thermodynamics:}\\
The \textit{average} entropy cannot decrease, when the average is taken over all possible thermal fluctuations. } 
\end{quotation}
This formulation is much less strong than the traditional version of the second law,
which forbids entropy reduction completely.
We must take this less strong formulation of the second law for nanoscale systems, because it is understood that such systems can exhibit large thermal fluctuations on short timescales, which can occasionally decrease entropy by a few $\kB$ for a short time, even though the entropy grows on average.   

This formulation is based on that of Smoluchowski \cite{Smoluchowski1912}, who drew attention to the fact that thermal fluctuations in simple machines violated the version of the second law that forbids entropy reduction completely.
However, he then said \cite{Smoluchowski1912}
\begin{quotation}
\textsf{
Molecular fluctuation phenomena today give us no reason to overturn completely the second law of thermodynamics, as we have so many other dogmas
of physics. They compel us only to a weakened formulation, if we demand
universal validity for the laws of thermodynamics. Perhaps an apparently quite minor extension of the wording suffices, in so far as one says: ``There can be no automatic device that would produce {\em continuously} usable work at the expense of the lowest temperature''. The brief version [of the Second Law] ``impossibility of a perpetual-motion machine of the second kind'' is even sufficient, for one has transferred the difficulty into the explication of the latter concept.\footnote{Translation into English taken from Ref.~\cite{Earman1998Dec}}}
\end{quotation}
Here Smoluchowski is also clear about what one requires a machine to do, if one wants to argue that it is violating his formulation of the second law.  It is not enough that it reduces the overall entropy from time to time; any small machine exhibiting significant thermal fluctuations will do this for short periods of time in a random and unpredictable manner. Instead, he is defining a violation of the second law as the possibility to create a perpetual-motion machine; a machine that reduces the overall entropy in a {\it continuous} manner on long timescales, specifically timescales much larger than the timescale of thermal fluctuations.

It is easier to understand this formulation of the second law using the fluctuation theorems discovered in the late 20th century.  Take the fluctuation theorem that is easiest to understand; {\it the steady-state fluctuation theorem} of Evans and Searles \cite{Evans1994Aug,Crooks1999Sep},
which only holds under very precise conditions
(see Section~\ref{Sect:Evans-Searles}) but gives a useful intuitive view.
Taking $P(\DS)$ as the probability of the entropy changing by $\DS$,
it reads
\begin{eqnarray}
P(-\DS) = P(\DS) \e^{-\DS}.
\label{Eq:evans-searles-intro}
\end{eqnarray}
where entropy changes are measured in units of the Boltzmann constant, $\kB$.
In other words, there is a finite probability of the total entropy decreasing, but it is exponentially lower than the probability that it increases by the same amount.
This means the entropy is always more likely to  
increase than decrease, so the second law of thermodynamics as formulated above is satisfied.  

In the systems studied here, each thermal fluctuation lasts only a relatively short time, so its properties on long timescale are given by the average over thermal fluctuations.  This means that the above formulation of the second law in terms of the {\it average over thermal fluctuations} is completely equivalent to Smoluchowski's formulation in terms of {\it continuous operation}. A violation of this would imply it is possible to build a perpetual-motion machine.  
It is this formulation of the second law that most experts believe is never violated,
and that tells us that perpetual-motion machines are impossible.

\section{Maxwell's demon}
\label{Sect:intro-Maxwell}

Efforts to build perpetual-motion machines have been around for centuries, 
as has the consensus that such machines are impossible. 
Intriguingly, this consensus significantly predates any clear 
formulations of the laws of thermodynamics, 
which are the physical laws that we now use to rule out perpetual motion.
However, I would argue that the modern debate around perpetual-motion machines,
starts with Maxwell's demon in 1867, which was arguably the first serious proposal to come from someone who
deeply understood the laws of thermodynamics \cite{Maxwelldemon,Maxwell1871}.

To set the scene, we  recall that in the early 1850s, the first formulations of the second law of thermodynamics
(by Clausius and Thomson/Kelvin \footnote{
William Thomson's earlier works carry the name Thomson, while his later works carry the name Kelvin. because he followed British aristocratic traditions, and took a new name (Kelvin) when he was made a lord in 1892. Intriguingly he was made a lord in recognition for his achievements in thermodynamics and for his opposition to 
``Home rule for Ireland''. While his contributions to thermodynamics have survived wonderfully, his opposition to political autonomy for Ireland (which was then part of the UK) is hard to understand now.}) quantified the fact that natural processes are usually {\it irreversible}.
Many of those who had already accepted Carnot's theorem (1824), rapidly concluded that these laws of thermodynamics were fundamental laws of nature. 

However, it was soon noted that these laws were hard to reconcile
with the developing knowledge of the kinetic theory of gases (with significant contributions in the late 1850s from Clausius and Maxwell, amongst others),
and attempts to explain Brownian motion.  The critical point there was that gas particles were assumed to undergo elastic collisions (as in Newton's laws) at the microscopic scale, implying the gas dynamics were reversible at that microscopic scale. Yet gas dynamics appeared irreversibility at the macroscopic scale---with work being easy to convert into heat, but heat being difficult to convert into work (as exemplified by Carnot's theories). 
So how did irreversibility emerge from reversibility?
Was the second law of thermodynamics a fundamental law  of nature (like energy conservation was believed to be),
or was it merely an approximation that was well-justified for macroscopic systems?
In the late 1860s, Maxwell proposed a thought experiment that 
very clearly showed that the response to such questions were unanswered.  
For this, he proposed a way of violating the second law of thermodynamics, rapidly named {\it Maxwell's demon} (by Thomson/Kelvin).

This thought experiment appeared in Maxwell's correspondences with various other experts in thermodynamics in the late 1860s, where Maxwell's own explanation of this thought experiment was extremely clear. 
In a letter to Tait \footnote{Peter Guthrie Tait was a well-known expert in thermodynamics, who is now better remembered for his contributions to knot theory. Maxwell and Tait had been friends since childhood, having gone to the same school in Edinburgh.} in 1867 he considered
a diaphragm (i.e. a wall) between gas A and gas B, where gas A is hotter than gas B,
he then wrote \cite{Maxwelldemon}
\begin{quotation}
\textsf{
Now conceive a finite being who knows the paths and velocities of all the
molecules by simple inspection but who can do no work except open and close a
hole in the diaphragm by means of a slide without mass.
Let him first observe the molecules in A and when he sees one coming the
square of whose velocity is less than the mean sq. vel. of the molecules in
B let him open the hole and let it go into B. Next let him watch for a molecule of B,
the square of whose velocity is greater than the mean sq. vel. in A, and when it
comes to the hole let him draw the slide and let it go into A, keeping the slide
shut for all other molecules.}

\textsf{
Then the number of molecules in A and B are the same as at first, but the
energy in A is increased and that in B is diminished, that is, the hot system has got
hotter and the cold colder and yet no work has been done, only the intelligence of
a very observant and neat-fingered\footnote{Here ``neat-fingered'' is a less common expression for ``nibble-fingered'' meaning ``agile''.} being has been employed.
}
\end{quotation}
A similar (but slightly less explicit) text was later published in Maxwell's book \cite{Maxwell1871}.  
The {\it neat-fingered being} is what we would now call {\it Maxwell's demon}
(a term invented by Thomson/Kelvin). This demon would clearly violate the second law of thermodynamics (including the modern formulation given in section~\ref{sec:2ndlaw} above).
This demon could easily be used to make a perpetual-motion machine, one simply needed to connect gas A and B
to a conventional heat engine to use the temperature difference to generate useful work.
As the work is generated through a violation of the second law, it is a perpetual-motion machine of the second kind.

\section{Trapdoor as a demon}
\label{sect:intro-trapdoor}
A second undated letter from Maxwell to Tait \cite{Maxwelldemon}, 
noted that Thomson/Kelvin had given it the nickname ``demon'', 
and pointed out an even simpler version of 
the demon that would contradict the second law:
\begin{quotation}
\textsf{
... less intelligent demons can produce a difference in pressure as well as
temperature by merely allowing all particles going in one direction while stopping
all those going the other way. This reduces the demon to a valve. ...  Call him no more a demon but a valve like that of the hydraulic ram,
suppose.}
\end{quotation}
The valve that Maxwell is referring to is often called a non-return valve or a check valve, which is a critical component in a hydraulic ram.

This is exactly the sort of valve later proposed by Smoluchowski  as a trapdoor, shown in Fig.~\ref{fig:trapdoor}.
If it worked as Maxwell suggested, it would clearly produce useful work, 
the pressure difference it generates between the two gases could be used to provide mechanical motion to a turbine, making a perpetual-motion machine.  It is a little harder to see which law of thermodynamics is being violated.  One has to go back to the ideal gas equations to show that this demon conserves the total energy of the gases but reduces their total entropy. So it is violating the second law (without violating the first law), 
again making it a perpetual-motion machine of the second kind.

At this point, we already see that Maxwell was speculating that the demon did not need to be particularly observant,
or intelligent to violate the second law, maybe it could simply be an autonomous mechanical machine such as a valve or trapdoor.  It is also clear that he was not arguing in favour of perpetual motion. Instead he was 
drawing attention to an apparent contradiction between the second law of thermodynamics and the kinetic theory of gases, as a way of saying there was more to understand about both of them.

\section{Autonomous demons}

In the years after Maxwell proposed his demon, many people proposed autonomous devices that appeared to obey the laws of mechanics, and yet play the role of a demon.
These designs appeared to be able to extract useful work from the thermal motion in a fluid of particles at a single temperature (which would make them perpetual-motion machines of the second kind). 
Some extremely serious and brilliant scientists considered such possibilities; including\footnote{It gives us an idea of the calibre of these scientists, that three of them later won Nobel prizes; 
Lippmann -- Nobel Prize for Physics in 1908; 
Ostwald -- Nobel Prize for Chemistry in 1909, and Svedberg -- Nobel Prize for Chemistry in 1926.}
Louis Georges Gouy\footnote{Louis Georges Gouy usually signed his papers ``G. Gouy'', however citations to him often accidentally replace the initial by ``L''.  His works (at the Sorbonne 1880-1883, and in Lyon 1883-1924) on Brownian motion, optics and magnetism were numerous. They included introducing the distinction between group velocity and wave velocity one year before Rayleigh, and inventing the Gouy balance.} \cite{Gouy1888},
Gabriel Lippmann \cite{Lippmann1900}, 
Wilhelm Ostwald \cite{Ostwald1907}, 
Theodor Svedberg \cite{Svedberg1907}, 
and Franz Richarz \cite{Richarz1907}.
Just like Maxwell, they were not arguing in favour of perpetual motion. Instead they were proposing their machines as thought-experiments that drew attention to an apparent contradiction between the second law of thermodynamics 
and the idea that the
heat in a fluid is nothing but the random motion of its constituent 
molecules.\footnote{Ref.~\cite{Earman1998Dec} gives an interesting interpretation of part of the history of this apparent contradiction.}
However, there were also proposals for perpetual-motion machines,
such as John Gamgee's {\it zeromotor} \cite{Pollard2017Apr}, 
which received support from the US Navy and US President Garfield in the 1880s.

\begin{figure}
\centerline{\includegraphics[width=0.85\columnwidth]{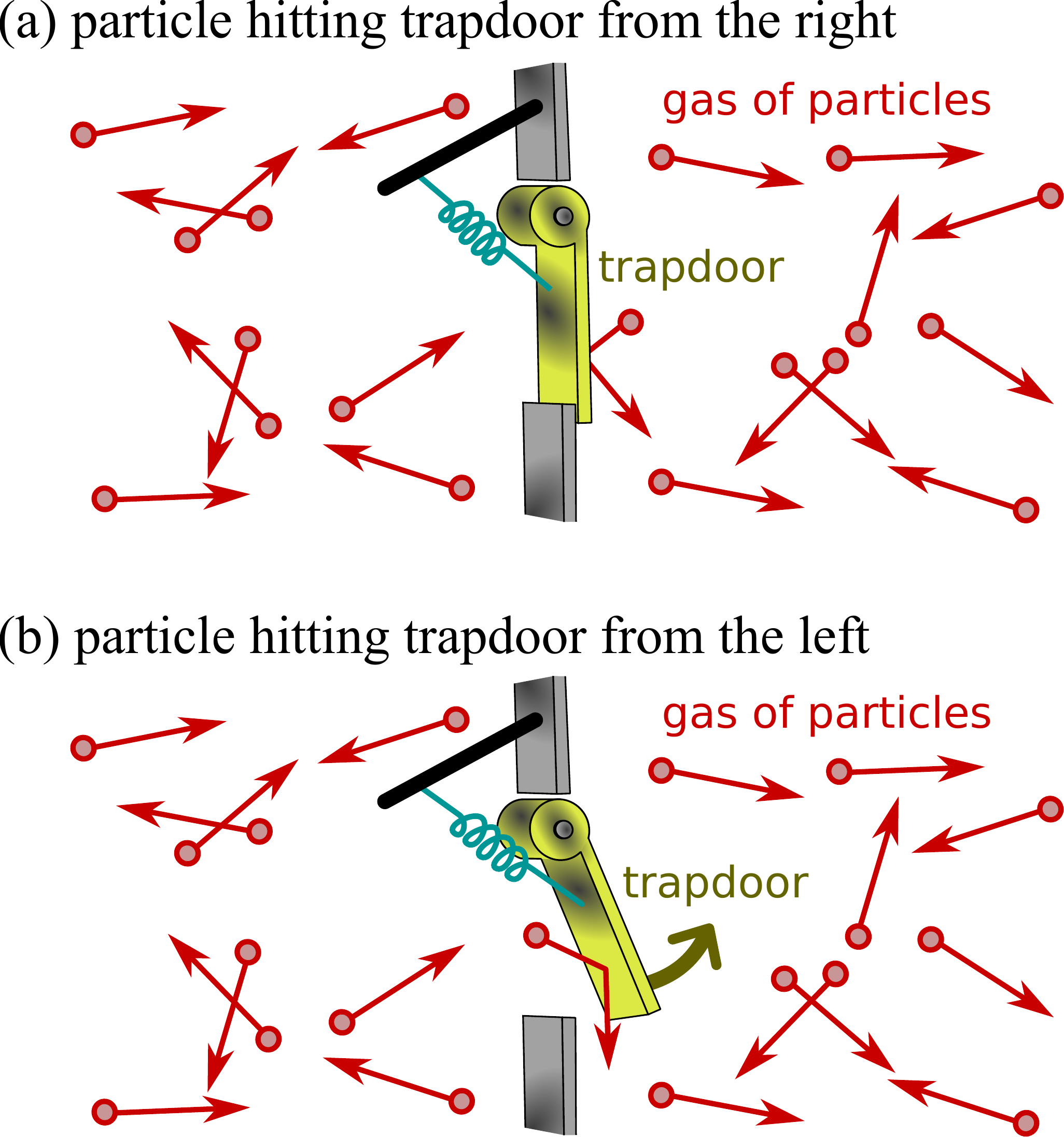}}
\caption{\label{fig:trapdoor} 
The thought experiment known as Smoluchowski's trapdoor; the lightweight trapdoor only opens to the right, and is pulled closed by the spring.
Thus a particle arriving from the right, as in (a), cannot pass to the left.
In contrast a particle arriving from the left will push the trapdoor open upon hitting it, and may thereby go to the right.} 
\end{figure}

\begin{figure}
\centerline{\includegraphics[width=0.75\columnwidth]{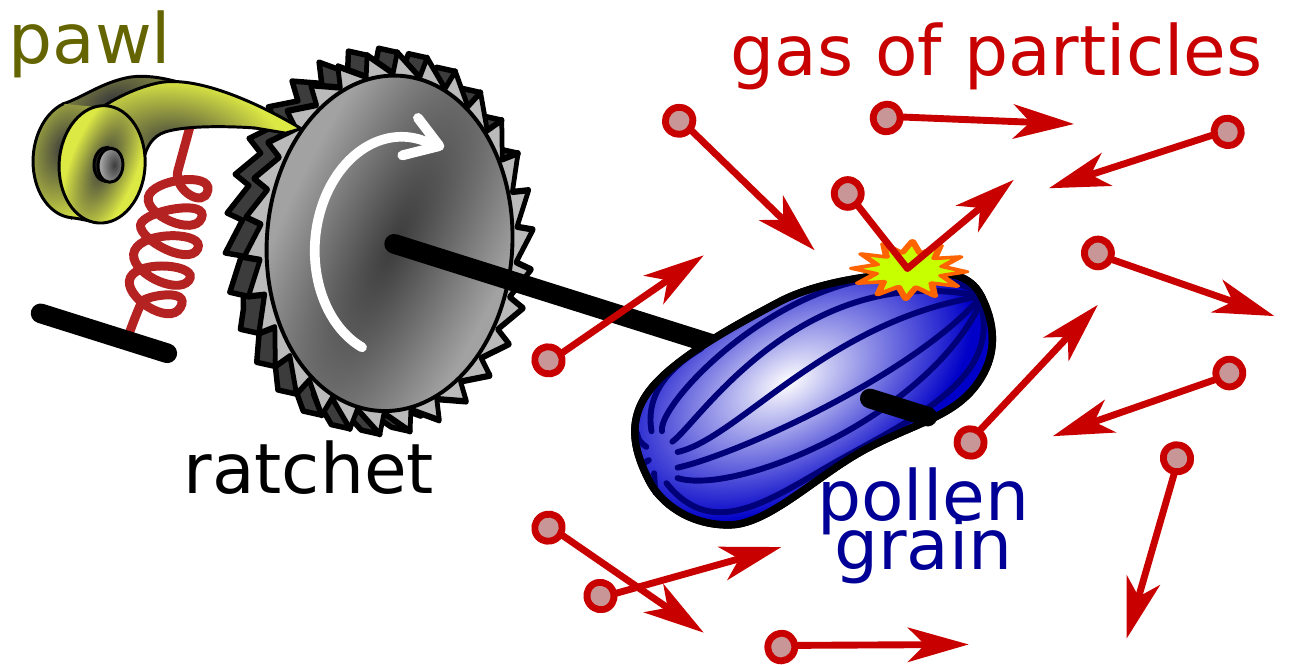}}
\caption{\label{fig:ratchet} 
The thought-experiment that has become known as the {\it Feynman ratchet}.
Here I draw Gouy's version of this thought experiment\cite{Gouy1888}, which dates from 1888, making it the oldest version that I know of. 
A grain of pollen undergoes a spinning Brownian motion due to collisions with gas particles in thermal motion. In principle it could be 
connected by  a narrow rigid wire to a very light ratchet (\textsf{``un fil d'un diam\`etre tr\`es petit par rapport au sien, \`a une roue \`a rochet tr\`es l\'eg\`ere''} \cite{Gouy1888}).  The ratchet's mechanism (the pawl) ensures the spinning can only go in one direction.  Gouy pointed out that this would produce useful work from a single heat source in contradiction of Carnot's principle.
} 
\end{figure}

In 1912 Smoluchowski \cite{Smoluchowski1912,Smoluchowski1914,Smoluchowski-collected-works} made a huge step in resolving the
apparent contradiction between the second law and the idea that heat was the random motion of molecules. By this time it was universally accepted that heat was indeed such motion, and was now well-described by Boltzmann's statistical mechanics theory.  Firstly, Smoluchowski gave an improved definition of the second law, that was not violated by fluctuations, as explained in section~\ref{sec:2ndlaw} above. 
Secondly, he realised that the previous proposals for autonomous machines designed to violate the second law had overlooked the thermal fluctuations of the machine itself, and that these thermal fluctuations greatly hinder their functioning.

In this context, he considered various simple thought-experiments, to show how the thermal fluctuations of the machine hindered their functioning.
The two most famous thought-experiments were: 
\begin{itemize}
\item[(1)] The machine in Fig.~\ref{fig:trapdoor}, often called the {\it Smoluchowski trapdoor}, a version of the valve mentioned by Maxwell in his second letter to Tait \cite{Maxwelldemon}, and later mentioned by Lippmann\cite{Lippmann1900}. 
It is a trapdoor at a partition in the gas which can open to the right but not the left.  It is lightly spring-loaded, so it is opened by a high momentum gas particle hitting it from the right, allowing that particle to go through, however particles hitting it from the left cannot open it to go through. A quantum nanoelectronic version of this (in which the trapdoor is a two-level system) is the subject of chapter~\ref{Chap:rate-eqns}.
\item[(2)] The machine in Fig.~\ref{fig:ratchet}, proposed by Gouy in 1888 \cite{Gouy1888}, which Smoluchowski \cite{Smoluchowski1912} described in much the same manner\footnote{Smoluchowski \cite{Smoluchowski1912} cited Lippmann \cite{Lippmann1900} but not Gouy \cite{Gouy1888}, and some have claimed that the idea originated with Lippmann. However Lippman's work only mentioned the word ``rochet" (ratchet), when Gouy had given a detailed proposal for the device,
which Smoluchowshi later described in much the same manner. As Lippmann's work contained no citations at all (except to Maxwell's book), I speculate that Lippmann was implicitly referring to Gouy when he used the word ``rochet'', and that Smoluchowski learnt of the details of Gouy's proposal from someone (perhaps Lippmann) without becoming aware of Gouy's paper. While this is pure speculation on my part, one should not forget that all three had worked at the Sorbonne, although not at the same time.} in 1912. It later became widely known as 
the {\it Feynman ratchet}, and was analysed in detail in Refs.~\cite{FeynmanLecturesVolOne,Parrondo1998Nov}.  This is much harder to implement in a nanoelectronic system, and I will not consider it further here.
\end{itemize}
In both cases\footnote{To quote Smoluchowski in a rather approximative translation from the German of 
Ref.~\cite{Smoluchowski1912} ``Make a hole of microscopically small dimensions and provide it with a small unidirectional valve, or with a ring of fine elastic hairs (eyelashes), which would allow the particles of an emulsion to pass through in one direction only. Such a device would automatically create a permanent pressure difference and would thus be a source of usable energy at the expense of the heat of the environment.'' This quote is then followed by the proposal for the ratchet in our Fig.~\ref{fig:ratchet} in similar terms to Gouy's proposal. One should not take this quote out of context,
Smoluchowski's goal in Ref.~\cite{Smoluchowski1912} was to show why it was flawed to think the flow would be in one direction only, and to argue that it would not provide usable energy at the expense of the heat of the environment.}, Smoluchowski pointed out that certain components of the machine were required to have a similar mass to a single particle in the gas that they were trying to extract energy from.
In the trapdoor case, the trapdoor must have a similar mass to a single gas particles, if it is to be opened by the force of a particle hitting it from the left.  In the ratchet case, the pawl must have a similar mass to a single gas particle, if it is to be lifted by the force of a particle hitting the pollen grain. 
As a result, those components of the machine would exhibit similarly large thermal fluctuations as the particles in the gas. Smoluchowski performed calculations of the magnitude of these fluctuations in a number of case.  He then argue that it was a grave error to neglect these fluctuations in the context autonomous machines intended to function like Maxwell's demons, because they  
greatly impede the functioning of the machine. 
Smoluchowski argued that this meant that such machines could not violate the second law in a continuous manner.

However, I believe that Smoluchowski only provided a plausibility argument against perpetual-motion machines; his argument \cite{Smoluchowski1912,Smoluchowski1914} 
appears too qualitative to {\it prove} that no violation of the second law is possible in such machines.\footnote{However, I may have missed some subtleties, because I required online tools to translate the German to English.}  His calculations made it clear that 
thermal fluctuations in the machines
would greatly reduce their capacity to extract work from a single heat bath.  However, he gave no analysis to show that the work extracted is reduced to zero, and the only way to rule out perpetual motion is to show that the work output is strictly zero.   
Feynman supplied such a quantitative theory for the second machine above\cite{FeynmanLecturesVolOne}, showing that it neither produces continuous work nor violates the second law\footnote{This is a justification for calling it the 
{\it Feynman ratchet}.  There are those who call it the {\it Feynman-Smoluchowski ratchet}, however then I would argue that  Gouy's name should be added as well.}  \cite{FeynmanLecturesVolOne}.
Feynman's analysis has since been criticised and improved upon, but the conclusion remains the same \cite{Parrondo1998Nov}.
In contrast, little progress has been made on quantitative analysis of 
the Smoluchowski trapdoor.  The only work that I am aware of is a numerical simulation using molecular dynamics \cite{Skordos1998Jun}, which is very nice, but does not constitute a proof that it can never violate the second law.

\section{\mbox{Demons as microscopic} \mbox{measurement-feedback loops}}

Another way to look at a Maxwell demon is as a measurement-feedback system.
However the distinction from other such measurement-feedback systems is that the measurement is at the {\it microscopic} level; that is to say the measurement must capture information about an {\it individual} particle in the fluid
and then use that information to perform feedback.
In Maxwell's original proposal, the measurement is the demon observing the momentum of an individual particle arriving at its door, and the feedback is it opening the door if that particle has the desired momentum. 

The microscopic nature of the measurement is critical, because 
the general idea is to detect if a given microscopic thermal fluctuation 
will increase or decrease the entropy.  If a fluctuation will increase the entropy, then feedback is used to stop it from occurring (for example closing a door), but if it will decrease the entropy, then the feedback is used to allow it to occur (for example opening the door).
One can thereby think of it as rectifying random thermal fluctuations into a directed motion that can produce useful work.

In many cases, a measurement-feedback loop requires some sort of
circuitry (often the measurement results are processed by an electronic circuit which then supplies the feedback). This circuitry usually requires energy
and generates entropy; these must be evaluated to be included in the balance sheet of energy and entropy, before one can see if one has a perpetual-motion machine or not.
What is nice about autonomous demons is that they include a complete mechanical model of the measurement-feedback loop, which makes it possible to evaluate its energy and entropy for inclusion in the balance sheet.

For Smoluchowski's trapdoor, the trapdoor both measures the individual particles and does the feedback on them. It does this by responding in one manner if the particle hits it from the left (it opens), and responding another way if the particle hits if from the right (it remains closed).
The energy necessary to make this trapdoor work can only come from the particle hitting it, i.e. from the heat source itself.
For Feynman's ratchet, the situation is the same.  The pawl acts on the ratchet wheel in exactly the same way that the trapdoor acts on the gas; it detects if the wheel is trying to turn one way or the other way, in one case it lets it turn, and in the other case it blocks its motion.  Again the energy source for the motion of the ratchet and pawl comes from the heat source itself (gas particles hitting the pollen grain).

In the autonomous demons discussed here, the feedback mechanism is as microscopic as the measurement.  By ``microscopic'', I mean acting on a single particle (or a small number of particles) rather than acting on a macroscopic parameter like volume or pressure.
However this raises the question of whether a Maxwell demon requires the feedback mechanism to be microscopic or not, given that the measurement is microscopic. The Szilard engine \cite{Szilard1929Nov} is an example in which the measurement is clearly microscopic; i.e. measuring if a single particle is in the left half of a container or not. However, the feedback appears to be a macroscopic action, i.e. inserting a partition if the particle is in the left half of the container (and doing nothing otherwise) and then let the particle apply a pressure on the partition that cause the partition to move, thereby doing work.  Superficially, letting pressure on the partition do useful work, also appears to be a macroscopic action.
However, the subtlety is that the Szilard engine works best when there is only one particle in the container.  A macroscopic version would thus involve $N$ such systems in parallel (i.e. $N$ particles in $N$ containers) for large $N$.  Then one sees that one needs $N$ independent partitions that can be independently inserted (or not) into each container based on the measurement of the particle position in that container.  This hints that a Maxwell demon requires the feedback mechanism to be about as microscopic as the measurement.  However, I believe that this is still an open question.

\section{\mbox{Quantum systems and} \mbox{nanoelectronics}}

In the previous section, I argued that a Maxwell demon required 
microscopic measurements of individual particles.
This was not really conceivable until the late 20th century,
which is why all works on the Maxwell demon from Maxwell to Smoluchowski and Szilard were posed as thought-experiments to better understand the limits of the laws of thermodynamics.

However in the late 20th century, the situation changed drastically.
Suddenly we had access to control of individual particles, initially in the field of quantum optics, and more recently in the field of nanoelectronics.
However, those individual particles are very different from those that Maxwell, Smoluchowski and the others, had imagined.
They were thinking of classical particles, when we now know that such individual particles are quantum, and exhibit quantum effects such as quantization, coherent superpositions, and entanglement.

Nanoelectronics is perhaps the most direct method for the
creation of autonomous Maxwell demons. It consists of engineering the nanostructure to generate a time-independent potential in which there are single electron states; 
a nanostructure engineered to have a single electron state is known as a {\it quantum dot}.  
The engineering of this potential allows us to tune 
(i) the rate of tunnelling between these states, 
(ii) the energy of these states, and 
(iii) the magnitude of the interactions between electrons in these states.
Once the potential has been engineered, no further energy is supplied to 
maintain it.
These states are then coupled to thermal reservoirs of free-electrons, which are the only source of heat or work.
If those reservoirs are all at the same temperature, then any steady-state work production would imply a violation of the second-law.\footnote{In contrast, if one wants to construct an autonomous demon using cold atoms or ion traps, the potential felt by the particles
is usually supplied by lasers, and these lasers consume a substantial amount of power; more than the autonomous demon could possibly produce. At a conceptual level a laser-generated time-independent potential (one that does not change any particle's total energy) can be treated as any other time-independent potential, and should not enter the calculations of work and heat for the particles.  However more practically, when building an experimental device, it can be hard to be sure the laser is not also supplying some energy to the particles.  Thus if the machine produces a small steady-state power from a single thermal bath, it is hard to be sure it is a violation of the second-law.  It could simply be that the atoms are absorbing a small amount of energy from the laser, which is thereby acting as a hidden work source.}
An example of such an experiment is that of \cite{Koski2015Dec},  discussed in chapter \ref{Chap:experiments} of this review.

\section{Bringing Smoluchowski's trapdoor into the 21st century}
The primary objective of this review is to introduce and discuss a set of questions and techniques that have emerged from the domain of quantum thermodynamics in recent years.  I believe a pleasing way to do this is to take the example of Smoluchowski's trapdoor, update it for the quantum era,
and then discuss its quantum thermodynamics.

In this context,the principle change I will make the Smoluchowski's trapdoor is to give it only two discrete energy-levels, one corresponding to the trapdoor being open, and the second to the trapdoor being closed. This two-level trapdoor is simpler than Smoluchowski's trapdoor, which was described by a continuous variable; the angle of the opening of the trapdoor.\footnote{This is similar in spirit to a discrete version of the Feynman ratchet proposed in Ref.~\cite{Jarzynski1999Jun}.} However, our main motivation for considering it is that it will be straightforward to build it using quantum dots.  Unlike Smoluchowski's trapdoor, the two-level trapdoor will be quantum, so it can be in a superposition of these two levels. Thus the trapdoor can be in a coherent quantum superposition of open and closed.

This two-level trapdoor system is much like other electronic nanostructures constructed using quantum dots. Thus its quantum thermodynamics will be very similar to systems already discussed in the scientific literature. This makes is a nice pedagogical example to guide the reader through some of the ideas and methods within quantum thermodynamics, while following directly in the footsteps of the greatest thermodynamicists of the 19th century.

Chapter~\ref{Chap:rate-eqns} analyses it using the rate equation method, showing that it cannot work as a Maxwell demon, unless there is some dissipative process damping the thermal motion of the trapdoor.
This dissipative process requires the presence of an additional cold reservoir.
Such processes obey the laws of thermodynamics, and the entropy reduction 
achieved by the demon is compensated by a entropy increase in this cold reservoir.

The two-level trapdoor is a close cousin of quantum dot systems that have already been built to resemble Maxwell demons.  
Chapter~\ref{Chap:experiments} discusses these experiments, and the physics that they reveal.

\subsection{Beyond the trapdoor: N-demons}

Chapter~\ref{Chap:N-demon} discusses novel devices that would use non-equilibrium distributions in ways that looks like they are violating the laws of thermodynamics, but are actually exploiting the fact that the laws of thermodynamics can be non-local \cite{Whitney2016Jan,Sanchez2019Nov,Hajiloo2020Oct,Ciliberto2020Nov,Deghi2020Jul,Freitas2021Mar}
We call these {\it N-demons} (``N'' for non-equilibrium). 
I will discuss such N-demons, and make the connection with other potential ways to exploit non-equilibrium resources.  I then explain how N-demons differ from other types of Maxwell demon, using a classification inspired by Ref.~\cite{Freitas2021Mar}.

\subsection{The quantum trapdoor}
Chapter~\ref{Chap:quantum-trapdoor} reviews a fully quantum analysis of the trapdoor system using the real-time diagrammatic method. This method can go beyond 
the rate equations, and be applied is situations where quantum interference and entanglement in the trapdoor system are significant.  However here I restrict myself to using it to prove that there is a natural parameter regime in which the rate equations used for the trapdoor in chapter~\ref{Chap:rate-eqns}
(which are classical rate equations acting on quantum states)
are a correct description. 

\subsection{A new view of entropy}
Finally, chapters~\ref{Chap:when-fluctuations-are-important} and \ref{Chap:fluctuations} discuss fluctuations about the average quantities,
when the rest of this review discussed on average quantities.
While these fluctuations cannot be used to
to turn heat into work in a continuous fashion, they are extremely interesting.
Indeed, I will argue that stochastic thermodynamic theories are revising (for the first time in about 100 years) our view of what entropy is. Entropy is no longer a number. It is a distribution!

\chapter{Non-autonomous demons and information}
\label{Chap:non-autonomous}

This review concentrates on autonomous \cite{autonomous} demons, but non-autonomous demons should also be mentioned.
They are demons like 
the intelligent being that was imagined by Maxwell.  
A non-autonomous demon could equally be some sort of living but fairly unintelligent organism, such as an ant or an amoeba. A more modern alternative, would be for the demon to be a computer. The crucial point is that there is a macroscopic being or system with the means to observe
the microscopic state of degrees-of-freedom in the fluid, and analyze this information to decide when to perform feedback on that system (such as opening and closing the trapdoor).
There many ways to implement such non-autonomous demons in nano-electronics, for example Refs.~\cite{Koski2014Jul,Koski2016Dec,Elouard2017Jun,Rossello2017Aug,Cottet2019Apr,Ryu2022Mar,Naghiloo2018Jul}. 
However, the problem is that the macroscopic being or system that records the microscopic information and applies the feedback is too complex to be modelled using classical or quantum mechanics. 
As such, the entropy produced by the demon remains outside the analysis, which makes it hard to say if the second law is obeyed or not.


Various people, including Smoluchowski\cite{Smoluchowski1914}, 
raised this problem.\footnote{Ref.~\cite{Earman1999Mar} reviews this aspect of Smoluchowski's work.}
They pointed out that biological organisms, be they human or amoeba, obey thermodynamic processes which produce relatively large amounts of entropy.  Following Schr\"odinger~\cite{WhatisLife} and many others, I summarize the situation as follows. Plants extract useful energy from the entropy-producing process of heat flow from hot sunlight (black body radiation at about 6000K) to a colder environment (ambient temperature at about 300K). In doing this they obey the same thermodynamic laws as any heat engine, and cannot exceed Carnot efficiency. Part of that useful energy is dissipated as heat during the plant's lifetime, part goes into growth and is dissipated when the plant dies and decomposes, and part may be trapped in the environment (sometimes for hundreds of millions of years in the form of coal or oil). Animals get useful energy by eating plants (or eating other animals that eat plants), and dissipate most of that energy as heat.  For example,
as an adult human, I produce $10^{24} \kB$ of entropy per second (assuming all the calories I eat per day are dissipated as heat at 300K).
If I was to do the job of a demon once per second, then
I could reduce the fluid's entropy by only about $1 \kB$ per second, while my existence is increasing the entropy of my surroundings by about $10^{24} \kB$ per second. Clearly this is not violating the second law.

You may object and say we should not consider all the entropy I produce, but only the extra entropy that I produce when doing the demon's job, compared to when I am at rest. If that was the case, one would have to analyse my entropy production to about 1 part in $10^{24}$ to see if my demonic action was violating the second law; such precision is beyond any conceivable experiment. 

If we could train an ant to act as the demon; we could reduce this entropy production to $10^{16}\kB$ per second (based on an ant's daily calorie consumption \cite{Fire-ants}). If we could train an amoeba to act as the demon, we could reduce this entropy production to about $10^{10}\kB$ per second.\footnote{
I did not find an amoeba's calorie consumption, but 
I guess it is $10^{-6}$ times an ant's, since its mass is $10^{-6}$ times an ant's.}
However, even then to observe a violation of the second law one would need to measure a change of order $1\kB$ in $10^{10}\kB$  in the amoeba's entropy production (between its entropy production when it is acting as a demon and it entropy production when it is at rest), and this also seems a hopeless task.
We could replace the amoeba by a computer chip capable of processing a signal from a measurement, and determining whether to apply feedback or not.  However even the smallest computer chip produces much more entropy than the above estimate for an amoeba; each 1\,$\mu$W of power consumption is about $10^{14}\kB$ of entropy produced per second. So we face the same hopeless task with a computer chip as with an ant or an amoeba.

The main problem with this argument of considering smaller and smaller organisms (or computer chips) is that one needs a way to
predict what is the minimal entropy production of a organism or computer chip that would be capable of acting as a Maxwell demon. Without this lower bound on the demon's entropy production, we cannot see if the demon could violate the second law or not.
Remarkably, Bennett \cite{Bennett1982Dec} came up with an elegant way of establishing such a lower bound.

\section{\mbox{Bennett's information} \mbox{argument}}
\label{Sect:Bennett}
Bennett \cite{Bennett1982Dec} had the idea to attack the problem from a completely different angle; an angle that developed into the idea of entropy of computation and information. 
Detailed reviews of such demons and the connection between thermodynamics and information
include Refs.~\cite{Parrondo2015Feb,Maruyama2009Jan}.

In short, Bennett argued that we do not know all sources of entropy inside a biological organism (or computer) that acts as the demon, but we do know that the demon must {\it at least} expend the entropy necessary to reset its memory.  To be specific, the Maxwell demon must measure the momentum of a particle arriving at the door, and record at least one bit of information (for example a ``1'' if the particle has the desired momentum, and ``0'' if it does not).
It uses this bit of information to decide if the door should be opened or not.
The demon must then re-set its memory (for example to the state ``0'') to be ready to record whether the next particle that arrives has the desired momentum or not. 
For this it must erase one bit of information from its memory. Erasing information is a process with a thermodynamic cost; either the erasure generates entropy, or work must be performed to mitigate this entropy production.
Bennett used Landauer's erasure principle \cite{Landauer1961Jul} to place a lower bound on this thermodynamic cost, the erasure requires increasing the entropy of a reservoir by $\kB \ln[2]$, which may require turning some work into heat.  He argued that adding it to the balance sheet of total work and entropy production was enough to restore the second law in simple cases. 

Much more recently, Sagawa and Ueda \cite{Sagawa2009Jun} generalised Bennett's information argument. They showed there is not necessarily a bound on the work done to erase the information in the memory. Instead, there is a bound on the sum of the work done during the measurement and work done during the erasure. This sum must be larger than  or equal to $k_{\rm B}T I$, where $I$ is the mutual information content between the measured system and the memory, and $T$ is the temperature of the thermal bath.

\subsection{\mbox{A comment on Earman \& Norton's} \mbox{criticism of this argument}}
Earman and Norton \cite{Earman1999Mar} criticised Bennett's argument for its circular nature. They pointed out that Landauer's erasure principle is derived from the second-law, and thus Bennett's proof that a Maxwell demon cannot violate the second law, implicitly assumes that the second-law is not be violated! While Bennett admitted that this is true, I feel that it misses the main point of his work. Bennett's argument implies that a Maxwell demon cannot violate the second law, unless the erasure of a one bit of information from the demon's memory violates the second law (irrespective of whether that memory is made of a silicon chip, neurons, a qubit, or something else).
The thermodynamics of erasing one bit of information from a memory is much simpler than the thermodynamics of a whole demon.  The erasure of a one bit of information is a dissipative process much like any other, and we can use classical or quantum mechanics to decide if there is a process by which the erasure of one bit of information can violate the second law. 

I think most physicists would argue that if such a simple process violates the second law, then we would see violations of the second law all over the place.
This gives a strong argument that erasing a memory does not violate the second law. Bennett's work then means that this becomes a reasonable argument to say that a non-autonomous Maxwell demon would not violate the second law of thermodynamics.

To be more precise,
the use of classical or quantum mechanics to evaluate the thermodynamic cost of erasing the information in a one-bit memory, requires identifying the degrees of freedom that store the information.  We must then model the irreversible process of
those degrees of freedom going to a given final state (say the state representing ``0'') irrespective of whether their initial state was the state representing ``0'' or ``1''. This dissipative process can be studied using the methods used in this review to study autonomous demons. However, we warn the reader that erasing information in a one-bit memory is often harder to model than in these autonomous demons.
The reason is that most protocols for erasing the information in a memory involves work sources that provide time-dependent potentials acting on the degrees of freedom that act as that memory; these work sources and time-dependences must be accounted for in the  thermodynamic analysis.  
Autonomous \cite{autonomous} demons avoid this complication, because they have neither work-sources nor time-dependent potentials.
None the less, I know of no serious proposal to erase a one-bit memory that is believed to violate the second law.

\subsection{Difficulty with autonomous demons}

These information arguments are much harder to apply to autonomous nanoelectronic machines (like the two-level trapdoor  in chapter~\ref{Chap:rate-eqns}) than other methods of calculating entropy production.  Thus we do not consider these information arguments further in this review. 

In contrast, each time someone finds a way to make a non-autonomous demon into an autonomous one, we learn a great deal about how such demons function; the best known example being those discussed in chapter~\ref{Chap:experiments}.  Some recently proposed non-autonomous demons that we could learn a lot about if we could invent autonomous versions include a gambling demon \cite{Manzano2021Feb},  and a quantum consensus demon for qubits\cite{Ryu2022Mar}.

\chapter{A mini-overview of nanoelectronics} 


To build an autonomous \cite{autonomous} Maxwell demon, we need degrees of freedom that interact with each other.  For example, to make a system that acts like the Smoluchowski trap-door in Fig.~\ref{fig:trapdoor}, we need degrees of freedom for the gas of particles, and others for the trapdoor.  This review concentrates on using electrons
as these degrees of freedom, using the engineering of electronic nanostructures
to control the electrons. Reviews of this subject include~\cite{
Imry2008Oct,Datta1995Sep,Akkermans2007May,Nazarov2009May,Moskalets-scattering-book,Benenti2017Jun,Whitney2019Apr}.

A critical ingredient in such autonomous Maxwell demons is that the degrees-of-freedom must interact in a very controlled manner.  For example, the trapdoor degree-of-freedom must interact with the gas degrees-of-freedom in a very precise manner.  This chapter briefly reviews how electrons interact, and how that interaction can be controlled with various nanostructures.

\section{\mbox{How do electrons interact} \mbox{in nanostructures}}

Electrons always repel each other, however this repulsive interaction is counter-intuitive; it is \textit{strong}
when the electron density is \textit{low}, and it is \textit{weak} when the electron density is \textit{high!}  
This surprising fact is due to the fact that electrons are fermions, and are described 
by Landau-Fermi liquid theory.  However, it not necessary to know this theory to understand why high-density electrons only interact weakly, it is sufficient to compare  their kinetic energy to the energy of their repulsive interaction.
If we consider $N$ free electrons in a $d$-dimensional box of size $L$, they form a Fermi surface whose Fermi momentum $p_F$ is given by the equation $N \sim (p_FL/h)^d$ (we drop all factors of order one).
Thus the kinetic energy per electron at the Fermi surface goes like  $N^{2/d}$.
In contrast, the energy of the Coulomb repulsion goes like $1/r$, and the typical distance between particles $r$ goes like $N^{-1/d}$,  hence the Coulomb energy per electron goes like $N^{1/d}$.
Thus for large enough electron number $N$ in any fixed box size, the kinetic energy will dominate over the Coulomb interaction energy, and the repulsive interaction will have almost no effect on the electrons' trajectories.  Landau-Fermi liquid theory provides a quantitative theory of this, and it tells us that screening effects further reduce the effects of the repulsive interaction at high-densities.

\begin{figure}[t]
\centerline{\includegraphics[width = 0.8\columnwidth]{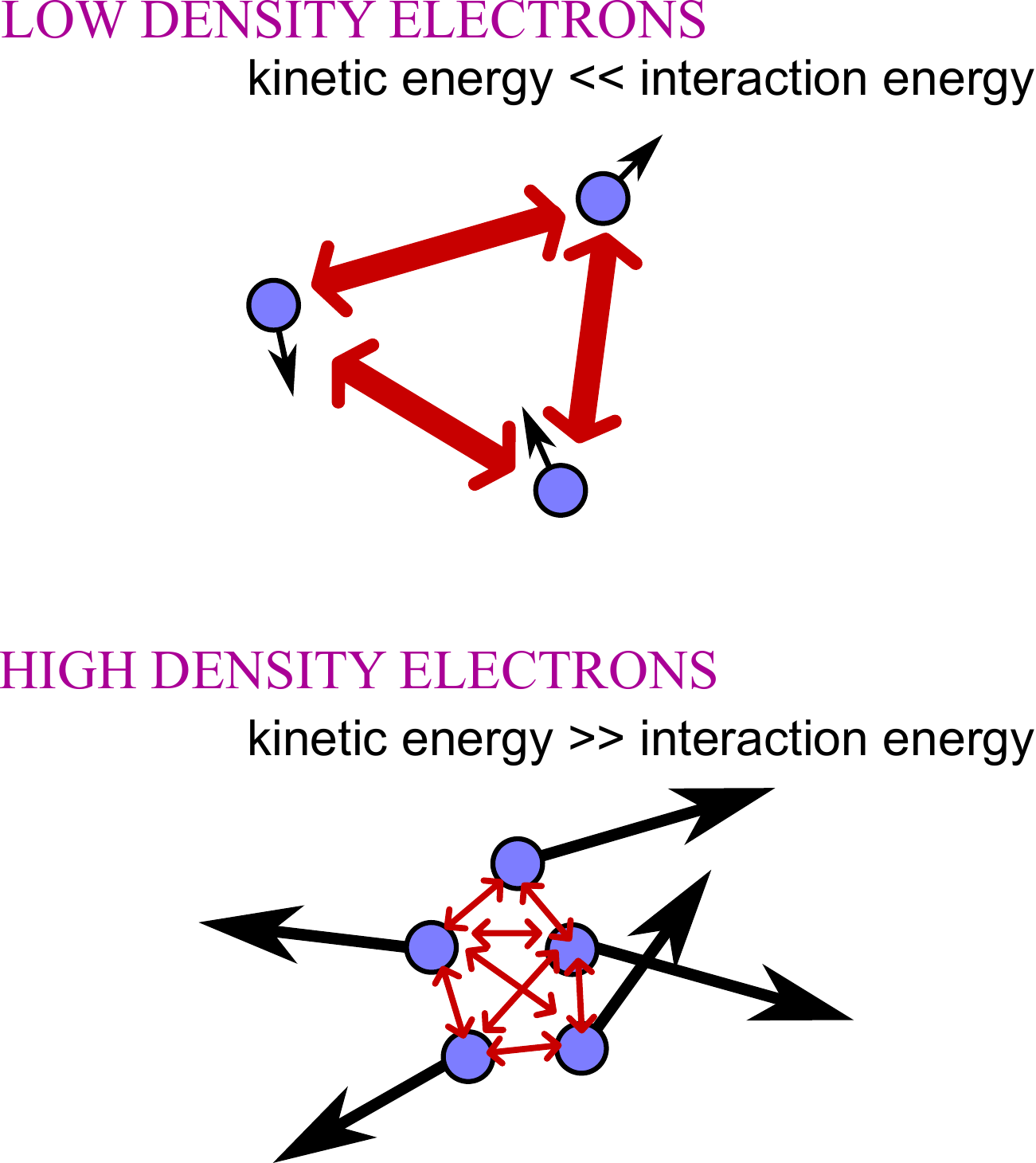}}
\caption{\label{fig:when-interaction}
A sketch showing the relative weights of kinetic energy (black arrows) and interaction energy (red arrows) in an electron gas. Surprisingly,  the interaction energy dominates when the electrons are far apart in a dilute gas. and the kinetic energy dominates when they are close together in a dense gas.
} 
\end{figure}

In the context of the systems considered in this review, we have reservoirs with a high-enough density of electrons that Coulomb repulsion can be neglected, and we can treat the electrons in the reservoir as non-interacting.  However, the nanostructure itself is a different story.  It typically has a much lower density of electrons than the reservoirs, and those electrons are often confined in a manner that reduces their screening.  
The exact strength of the Coulomb repulsion in a given nanostructure can often be hard to predict, because it depends a lot on the screening due to the electron gases in the vicinity of the nanostructures (for example the electron gases in the reservoirs).  It is thus typically quantified in terms of phenomenological capacitances, which must be measured in the nanostructure in question. 

If the capacitance is low, then the Coulomb repulsion energy for $N+1$ 
electrons in a given region of the nanostructure may be significantly more than for 
$N$ electrons in that region.  If the energy difference between $N$ and $N+1$ 
is bigger than temperature or bias, then
it will have a strong effect on the electronic transport, and will induce single-electron interaction effects such as \textit{Coulomb blockade}.
We will take advantage of such Coulomb blockade effects in the two-level trapdoor model in chapter~\ref{Chap:rate-eqns}.


\section{Molecular nanostructures}

\begin{figure}
\centerline{\includegraphics[width = 0.95 \columnwidth]{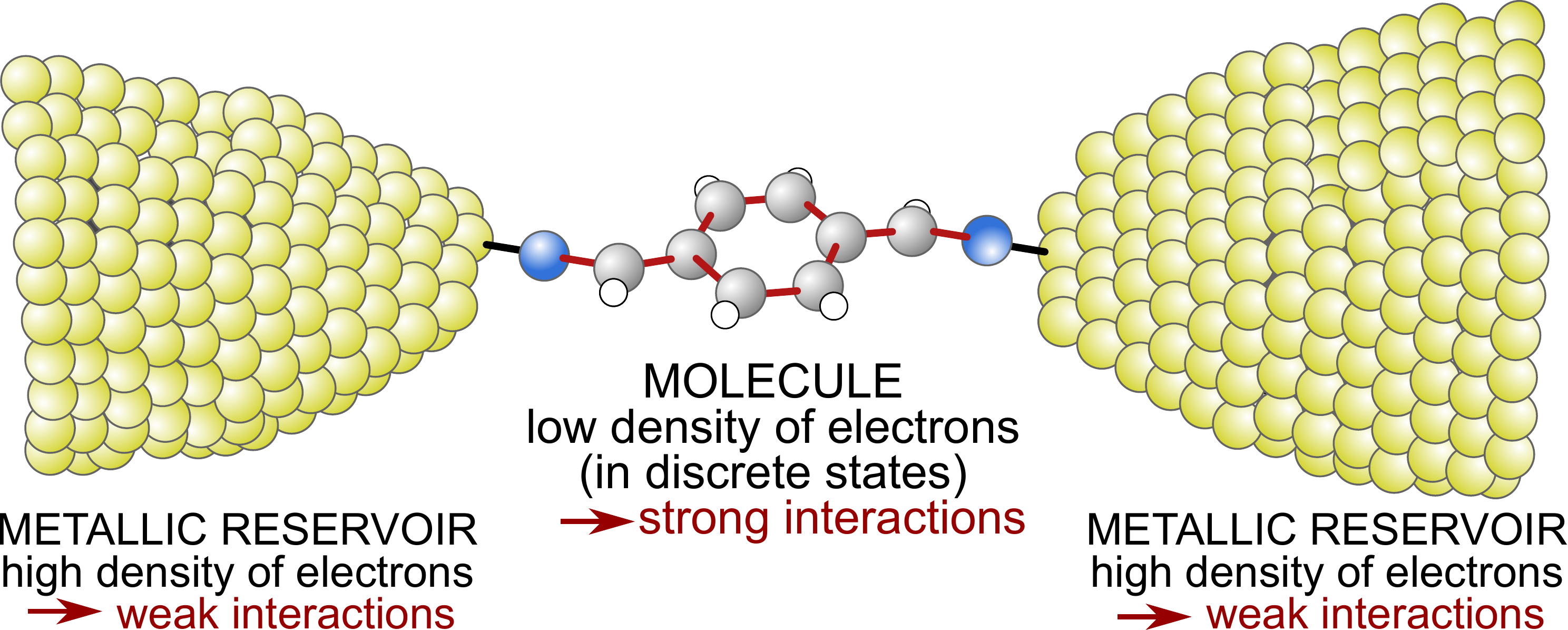}}
\caption{\label{fig:breakjunction}
A sketch of an electronic nanostructure made of a molecule between two metallic tips. The tips are often made by stretching a metal wire slowly until it just breaks, leaving two tips almost touching each other; this is called a break-junction.  The molecule is then place between the two tips.
} 
\end{figure}

The easiest electronic nanostructures to explain are molecular nanostructures in a break-junction.
A break junction is created when you take a metallic wire and slowly pull it until it breaks.
At the moment when it breaks the two pieces of metal wire are separated by a tiny nanometre-sized gap.  If one stops pulling at this moment, this tiny gap will stay there.  One can then take a suitable molecule (for example a benzene ring, a polymer, or a bucky-ball) and  spray a very low density of such molecules on the wire. With luck one will lodge in the nanometre-sized gap.
This then makes an electronic nanostructure where the molecule is sandwiched between
two large metallic reservoirs of electrons.

Another method is often used with carbon nanotubes to achieve the same effect.
A single nanotube is placed on a insulating surface, and then metallic contacts are deposed onto it.  This allows one to connect the molecule (the nanotube) to multiple metallic reservoirs of electrons.  It is commonly used to connect the nanotube to four or more metallic reservoirs.

The physics that will be observed in such set-ups will depend on the energy level structure of the molecule, so it is important to choose the right molecule.  Indeed the weakness of this approach is that if one wants to study a particular energy-level structure,
then one will have to find a molecule which has the desired structure.

The density of electrons is high in the metallic reservoir, but low in the molecule.  Hence interactions between electrons can be neglected in the reservoirs, but may be significant in the molecule.  With the right molecule, one can have very non-trivial dynamics in which one electron entering the molecule pushes another electron in the molecule into a different state, which  will in-turn affect the dynamics of the first electron.  Thus it has the basic ingredient to construct an autonomous Maxwell demon.  However, it would take an expert in chemistry to choose the right molecule.

\section{\mbox{Two-dimensional nanostructures} and quantum dots }

\begin{figure}
\centerline{\includegraphics[width = 0.95\columnwidth]{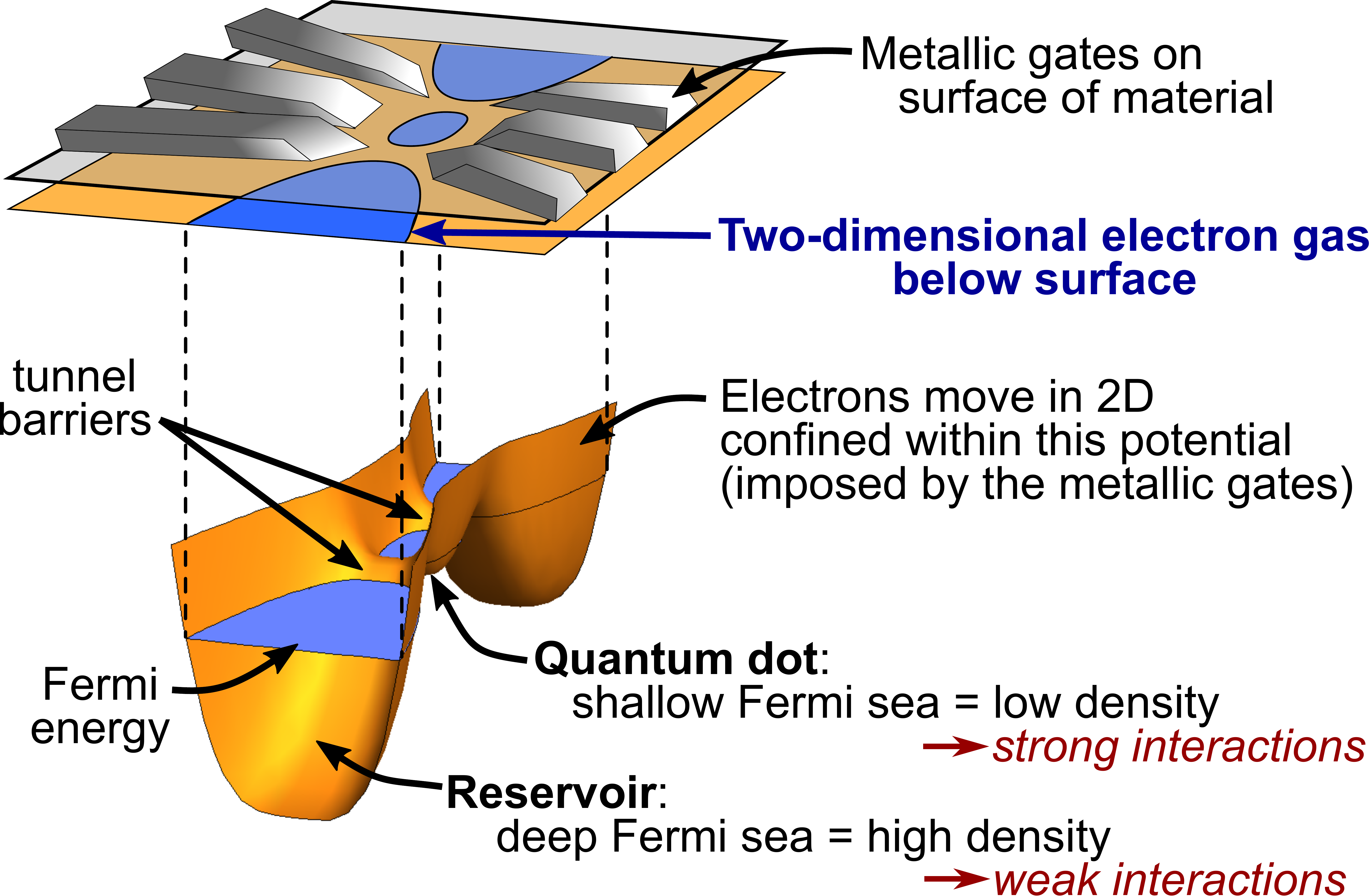}}
\caption{\label{fig:q-dots-2D}
A sketch showing how a nanostructure (in this case a single quantum dot) is made 
from a two-dimensional electron gas.  The depth of the Fermi sea is large in the reservoirs, so the electron density is high there, and the interactions between electrons are tiny.  In contrast, the depth of the Fermi sea is small in the quantum dot,
so the electron density is low there, and interactions between electrons can be strong.  
} 
\end{figure}

Two-dimensional electron gases (``2DEGs'' for short) 
allow the experimentalist a superior level of control
of the nanostructure.  They are widely used, and new ones are regularly discovered. They are mostly based on engineering a sandwich of materials such that there is a two-dimensional gas of electrons at the interface between two materials (when the rest of the sandwich of materials is electrically insulating).  The gas is two-dimensional because it is so strongly confined in the third direction that all the electrons have wavefunctions in the lowest mode in that third direction.  

Since the two-dimensional electron gas is at an interface 
just below the surface of the material, one can change the electrostatic 
potential felt by the electrons by placing metallic gates on the surface, and giving them a negative bias to expel the electrons from under those metallic gates.
Fig.~\ref{fig:q-dots-2D} shows metallic gates used to define an electrostatic potential for the electrons corresponding to a quantum dot between two large reservoirs of electrons.  Some gates are used to determine the shape and size of the quantum dot, while others determine the height of the tunnel-barriers, and thereby control the rate at which electrons tunnel into and out of the quantum dot. As a theoretical physicist, I am always amazed by what experimentalists can do with such techniques, they regularly make nanostructures consisting of dozen or more quantum dots, while controlling the size of each one, and the rate of tunnelling between them.
The size is a critical parameter; if the quantum dot is big then it will have a small level-spacing and will contain many electrons; if it is small, then the level spacing will be so large that there may only be one energy level to occupy.  Magnetic fields are often used to push these nanostructures into the quantum Hall regime, because experimentalists often have even better control of the nanostructure's properties in the quantum Hall regime (for reasons I will not address here).

\section{Metallic nanostructures}
\label{sect:metal}

Typically, metallic nanostructures are made of nanometre-sized grains of metal.
Most metals have about one free electron per atom, so the
Fermi wavelength in a grain is of order an angstrom, while the grain is tens of angstroms across.  The grain thus contains thousands of free-electrons with a level spacing that is often small compared to the temperature and bias of the reservoirs it is coupled to.

\begin{figure}
\centerline{\includegraphics[width = 0.6\columnwidth]{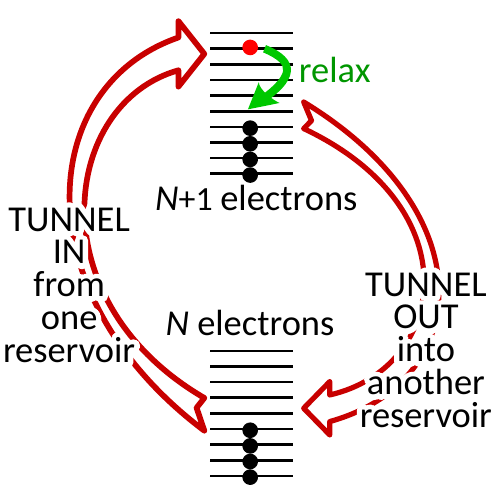}}
\caption{\label{Fig:blockade sketch}
A sketch of one possible tunnelling process into and out of a metallic grain containing thousands of electrons in the Coulomb blockade regime. 
A hot-electron tunnels in, and scatters from other electrons
and or phonons, so it relaxes before tunnelling out into another reservoir, leaving some of its initial energy in the grain.
} 
\end{figure}

In contrast, charging effects can be of order or larger than the bias or temperature (assuming the experiment is performed at sub-Kelvin temperatures).
So the metallic grain can be in the Coulomb blockade regime sketched in Fig.~\ref{Fig:blockade sketch}.  The process in Fig.~\ref{Fig:blockade sketch} is an inelastic scattering process (the electron leaving the grain has a different energy from that entering the grain), induced by relaxation processes within the grain.
However, even when electrons do not interact with anything while in the grain, one can have inelastic effects! An electron can tunnel into one level in the grain, but then a different electron in a different level of the grain can tunnel out,
leaving the grain with a different electron distribution amongst its levels, and hence a different energy.  As electrons are indistinguishable, these effects look inelastic; it is like an electron has travelled through the grain, but has exchanged energy with the grain as it does so.
These sorts of effects make quantitative analysis of metallic grains harder than 
molecules of quantum dots with only a small number of levels.

\section{Theories of electronic transport in nanostructures}

There are multiple theoretical methods applicable to small quantum systems that are coupled to macroscopic reservoirs.  In the context of electronic systems, there are 
two main classes of model; those that treat the electrons as independent of each other, and those which account for interactions between the electrons (or between electrons and other particles like phonons and photons).

\begin{figure}
\centerline{\includegraphics[width = 0.9\columnwidth]{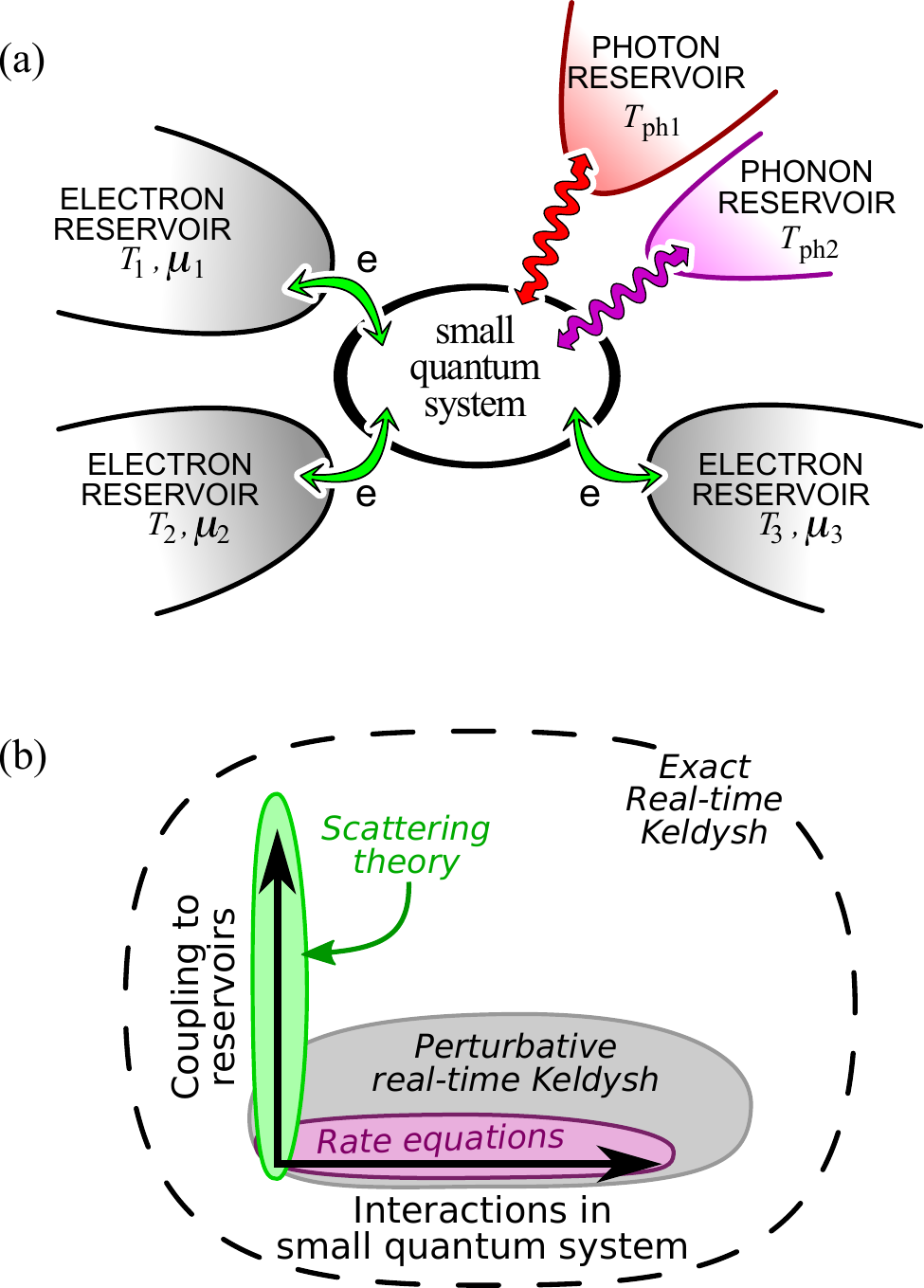}}
\caption{\label{fig:regimes}
(a) The theoretical view of an electronic nanostructure: a small quantum system is coupled to multiple reservoirs of electrons, photons or phonons. In general, each reservoir might have a different temperature, $T_i$, and each electron reservoir might have a different electrochemical potential $\mu_i$.  Then the small quantum system is driven out of equilibrium by its coupling to these reservoirs. 
(b) Sketch of regimes of applicability of the three theoretical methods discussed in this review.
} 
\end{figure}

This review discusses works that use three of these methods, outlined below, 
each with its domain of validity.  These domains of validity are sketched in 
Fig.~\ref{fig:regimes}.
One sees that the domain of validity of these theories depends not only on whether
electrons interact with each other (or with other particles), but also on how strongly coupled the quantum system is to the reservoirs.
Typically to have a \textit{simple} theoretical description, we need a small parameter to perform an expansion that we treat to lowest order; typically zeroth order or first order. 
The two parameters one can take to be small are (i) the strength of the interactions or (ii) the strength of the coupling of the system to the reservoirs.
If the strength of the interactions are sufficiently small, 
then one can get a simple theory; Landauer-B\"uttiker scattering theory.
If the strength of the coupling to the reservoirs is sufficiently small, 
then one can get another simple theory: rate equations.

If neither of these parameters are small (as is often the case in experimental systems), then there is no 
simple theory, one requires a complicated theory of quantum non-equilibrium systems \cite{Reimer2019Jul}, such as formulated in terms of Keldysh theory.  There are different ways to formulate Keldysh theory, and we present 
one of them below, which is called  {\it real-time diagrammatics} or {\it real-time transport theory}.\footnote{While this name is the one generally used, its real-time nature is not necessarily at the heart of what differentiates it from other Keldysh theories, like non-equilibrium Green functions (NEGF).}
However, formulating the problem in terms of a Keldysh theory is usually not sufficient to solve the problem in question, further approximations are usually required, and 
progress is difficult.
Thus even in experimental situations where we believe that neither parameter is small,
it is useful to apply one (or both) of the simple theories discussed here, as a phenomenological manner to get a basic understanding of the physics at play, before thinking about developing a complicated Keldysh theory.

\subsection{\mbox{Landauer scattering theory for} \mbox{independent electrons}}
Independent electrons are relatively easy to treat theoretically, giving rise to a 
theory called Landauer scattering theory. The earliest  literature on this theory includes Refs.~\cite{Landauer1957Jul,Landauer1970Apr,Landauer1975Sep}, with applications to heat flows is \cite{Engquist1981Jul,Pendry1983Jul,Sivan1986Jan}, but readers should be warned that papers written before the late 1980s 
were confused about profound issues, such as whether a perfect conductor had a finite resistance or not.\footnote{This was resolved by a theory of four-terminal measurements in addition to two-terminal ones \cite{Buttiker1986Oct}. A brief view of the confusion in the earlier papers was given in section~4.2 of Ref.~\cite{Benenti2017Jun}.}   For general reviews see Refs~\cite{Imry2008Oct,Datta1995Sep,Akkermans2007May,Nazarov2009May,Moskalets-scattering-book}, while
the thermodynamics of scattering theory is reviewed in chapters 4 to 6  of Ref.~\cite{Benenti2017Jun}.  This theory can be used to model many interesting quantum effects (from interference effects to quantum Hall edge-states) using relatively simple algebra. 
This treats the small quantum system as a scatterer, described in terms of the probability that an electron entering the small quantum system at energy $E$ from reservoir $\alpha$ will end up transmitted into another reservoir $\beta$ (or reflecting back into the original reservoir $\alpha$).
For simple geometries of the scatterer this can be calculated with pen and paper.
However for more complicated systems, one can use the open-source software kwant (\href{https://kwant-project.org/}{kwant-project.org}) developed at the CEA Grenoble \cite{Groth2014Jun}, which efficiently implements the Laudauer scattering theory for almost any imaginable nanostructure.

The Landauer scattering theory is based on the idea that the electrons transverse
the scatterer rapidly, without interacting with any other electrons.  However this does not mean it assumes the electrons never interact with each other; it is assumed  that electrons spend so long in the reservoir that they eventually interact enough to relax to a thermal distribution.
We require that the relaxation processes in reservoirs is weak enough that we can treat electrons in the reservoir as non-interacting when they arrive at and when they leave the scatterer.
However, we also require that these relaxation processes are strong enough that electrons injected into a reservoir relax completely to a local
thermal distribution (a Fermi distribution determined by that reservoir's temperature and electrochemical potential)
before they can return to the scatterer. In this manner, we can assume that all electrons arriving at the scatterer from a given reservoir come from the Fermi
distribution with the temperature and electrochemical potential of that reservoir.
Beyond these two requirements, we do not need to make any specific assumption  about the nature of the relaxation in the reservoirs.
A crucial point is that scattering theory can never violate the laws of thermodynamics, so long as the scatterer is described by a physical set of transmission probabilities (no negative probabilities, probabilities sum up to one, etc) \cite{Nenciu2007Mar}, see also Refs.~\cite{Whitney2013Mar,Benenti2017Jun}

There are tricks to add inelastic scattering effects in a phenomenological manner, 
known as B\"uttiker voltage probes, dephasing probes or temperature probes \cite{Buttiker1986Oct,Meair2014Jul}.  
These involve adding fictitious reservoirs (in which electrons relax to a thermal state) to mimic the
role of relaxation or decoherence that may occur during the scattering.
However, the proof that such systems never violate the laws of thermodynamics  still holds \cite{Whitney2013Mar}.
For more details, see section 5.5 of Ref.~\cite{Benenti2017Jun}.
Superconducting leads can equally be added to the scattering theory \cite{Beenakker1997Jul,Benenti2017Jun},
with heat flows and thermodynamics treated in various works, 
including Refs.~\cite{Jacquod2010Oct,Jacquod2012Oct,Whitney2013Mar,Mazza2015Jun}.
However, I will not discuss these further here. 

\subsection{\mbox{Rate equations for interacting} \mbox{electrons}}

The rate equation approach requires assuming that the small quantum system contains
relatively few levels, and is very weakly coupled to the reservoirs.  Then one can solve the dynamics of the small quantum system (including all interactions effects) exactly by diagonalizing its many-body Hamiltonian. After this, the coupling to the reservoirs
is added as perturbation.
Remarkably, when this is done one finds that all of the quantum physics is restricted to determining the many-body eigenstates, and the transitions between these states is governed by classical rate equations.
These rate equations are reviewed in detail in Chapters 8 and 9 of \cite{Benenti2017Jun}, or \cite{VandenBroeck2013,VandenBroeck2015Jan,Seifert2012Nov}.  
They have been shown to always obey the laws of thermodynamics,
and to satisfy fluctuation theorems like Eq.~\ref{Eq:evans-searles-intro}.
These rate equations will be used for the two-level trapdoor in Chapter~\ref{Chap:rate-eqns}.

\subsection{\mbox{Real-time diagrammatics for} \mbox{interacting electrons}}

\revisit{
The general theory of a small quantum system driven out of equilibrium by its coupling to reservoirs is always complicated; there are a number of different formal approaches \cite{Reimer2019Jul}. These approaches are often exact, but are usually too complicated to exactly calculate observables for concrete systems.  
So perhaps the most interesting differences between the different approaches is that they invite different approximations or simplifications. 
All these theories work with density matrices, and the time-evolution of density matrices is an evolution on the Keldysh contour.
To see this, consider the propagation of the density matrix from time $t_0$ to time $t$.
In general, the value of $ij$th element of the density matrix at time $t$ can depend depends on the value of the $kl$th element of the density matrix at time $t_0$.
Hence one sees one has a double-index structure evolving in time.
In the Keldysh way of seeing this, this is a double-line evolving in time (the Keldysh contour), in which the upper-line is the first index on the density matrix and lower line    
is the second. In other contexts, this double-index structure is called a super-operator, 
however we will always think of it in terms of the Keldysh contour.
} 

\revisit{
In this context there are two kinds of diagrammatic Keldysh theories for such systems.
The best known is called the  {\it non-equilibrium Green functions} (NEGF) technique, but it is \textit{not} the one used in this review, so I simply mention a few of its characteristics here. It is a diagrammatic expansion built upon the limit of independent electrons, as reviewed in Ref.~\cite{Haug-Jauho-book} or Chapter 10 of \cite{Kamenev2011Sep}. It is analogous to taking the 
Landauer scattering theory as a foundation and then adding more and more diagrams
on the Keldysh contour to account for interactions; electron-electron, electron-photon and electron-phonon interactions. If one keeps all diagrams to all orders, then this theory is exact. However, in practice one often needs to make approximation that keep only certain classes of diagrams (Hartree approximation, Hartree-Fock approximation, GW, etc), then the theory is expected to work well if the interactions are not too strong, but is still likely to fail to capture the physics for very strong-interactions. For a very incomplete selection of works that do this, see Refs.~\cite{Haug-Jauho-book,Lopez2002Jan,Darancet2007Feb,vonDelft2007Nov,Starke2012Feb,Michelini2017Mar}
and references therein. 
What is nice about this theory is that one can typically treat arbitrary size nano-structures with relative ease;
it is built upon the dynamics of independent electrons, which are no more complicated in large structures than small structures (electrons are always in eigenmodes, which become close to plane waves in the limit of large nanostructures). 
One important ingredient is currently missing for this non-equilibrium Green functions approach, that is a proof that systems described by it do not violate the second law,  although some strides have been made in this direction \cite{Ludovico2014Apr,Esposito2015Feb,Bruch2016Mar,Ludovico2016Jul}.
}

Now let us turn to the method used in this review, known as {\it real-time diagrammatics} or {\it real-time transport theory} \cite{Schoeller1994Dec,Konig1996Dec,Schoeller1997,Konig2000,Governale2008Apr,Saptsov2012Dec,Saptsov2014Jul,Schulenborg2016Feb,Whitney2018Aug,Lindner2019May}.
Its starting point is the exact many-body eigenbasis of the small quantum system
in the {\it absence} of the reservoirs.
This typically limits it to systems small enough (or simple enough) that their many-body eigenbasis can be calculated analytically or numerically using exact diagonalization.
This typically means treating a system with a small number of levels, or that Coulomb interactions are strong enough that it is reasonable to truncate the system in certain charge states.\footnote{An example of this would be a system that is restricted to the subspace with $N$ or $N+1$ electrons in the system (for some $N$), because Coulomb charging effects push all other states to a much higher energy.  Then it is a reasonable approximation to truncate the state-space to keep only the sub-space of $N$ or $N+1$ electrons, giving a simple enough eigenbasis to work with.}
One then introduces the reservoirs via a series of diagrams on the Keldysh contour,
with each diagram representing the exchange of particles or energy with one of the reservoirs.  If one keeps all diagrams to all orders, then the theory is exact.
However, in practice one often assumes the tunnelling is weak enough to only keep diagrams at the lowest orders in the tunnelling (the first order being sequential tunnelling, the second order being cotunnelling, etc). To capture the physics of strong coupling (such as Kondo physics) within this real-time diagrammatic approach, one can use real-time renormalization group, for example see Refs.~\cite{Saptsov2012Dec,Lindner2019May}.

This real-time diagrammatic method will be discussed in chapter~\ref{Chap:quantum-trapdoor} of this review.
However it is worth mentioning here that Ref.~\cite{Whitney2018Aug} showed that all diagrams on the Keldysh contour have a simple relationship to their symmetric partner rotated by 180$^{\circ}$ on the Keldysh contour.
This relationship ensures that the dynamics produced by this theory will always respect the laws of thermodynamics (something not know for the Keldysh NEGF technique mentioned above).
I believe that this symmetry applies no matter how many levels there are in the ``small quantum system'' (so it does not need to be small), and no matter how strongly it is coupled to the reservoirs (although the proof only allows certain types of coupling
to the reservoirs).  This means that no system exactly modelled in this way should violate the laws of thermodynamics. Furthermore if one makes approximations that respect this symmetry between diagrams, then that approximate theory will also never violate the laws of thermodynamics.  This is discussed more in chapter~\ref{Chap:quantum-trapdoor}.

\chapter{The two-level trapdoor}
\label{Chap:rate-eqns}

Here I introduce a simplified version of the Smoluchowski trapdoor, 
that could be constructed with nanoscale electronics.
While Smoluchowski had a classical trapdoor described by its angle of opening (a continuous degree of freedom), here I consider a quantum trapdoor described by a two-level system with only two possible states, corresponding to open or closed. Thus I call it a {\it two-level trapdoor}. 

This chapter gives a demonstration of the rate equation approach to quantum thermodynamics for this two-level trapdoor, showing that 
it works as a rectifier but not a Maxwell demon.  It then shows that 
damping the trapdoor's oscillations by coupling the trapdoor to a cold bath can make it act in the desired way (driving a current against a potential gradient),
but it does not violate the laws of thermodynamics because it produces entropy in the cold bath.

The {\it rate-equation method} used here was extensively reviewed in sections 8 and 9 of Ref.~\cite{Benenti2017Jun}; its section 8 shows that this rate-equation method obeys fluctuation theorems, and thus cannot violate the laws of thermodynamics, and its section 9 gives various examples of applications of the method.
The rate-equation method is a quantum approach that does not look very quantum.
It starts from the quantum many-body eigenstates of the small quantum system
(which can include almost any imaginable quantum effects), but it then treats 
transitions between these eigenstates with a rate equation that is very classical, 
neglecting the possibility of the system being in a superposition of these eigenstates. 
Thus it is sometimes called a {\it classical rate equation method}, to distinguish it from more quantum methods.
In chapter~\ref{Chap:quantum-trapdoor} of this review, we will show why this is permissible; showing that these classical-looking rate-equations emerges naturally when one considers the full quantum problem in the limit of weak coupling between the system and reservoir (where ``system'' and ``reservoir'' are defined precisely below).
However, the rate-equation method can also be thought of as about the simplest phenomenological model that one could construct for the physics in question.
As such, it makes sense to use it to get a first idea of the physics of a given situation before performing more complicated modelling.  This is why we present the use of the rate-equation method to analyze a trapdoor and a demon {\it before} presenting the fully quantum approach which allows us to both derive the rate-equation method (see chapter~\ref{Chap:quantum-trapdoor}), and calculate quantum corrections to it.

\begin{figure*}[t]
\centerline{\includegraphics[width=\textwidth]{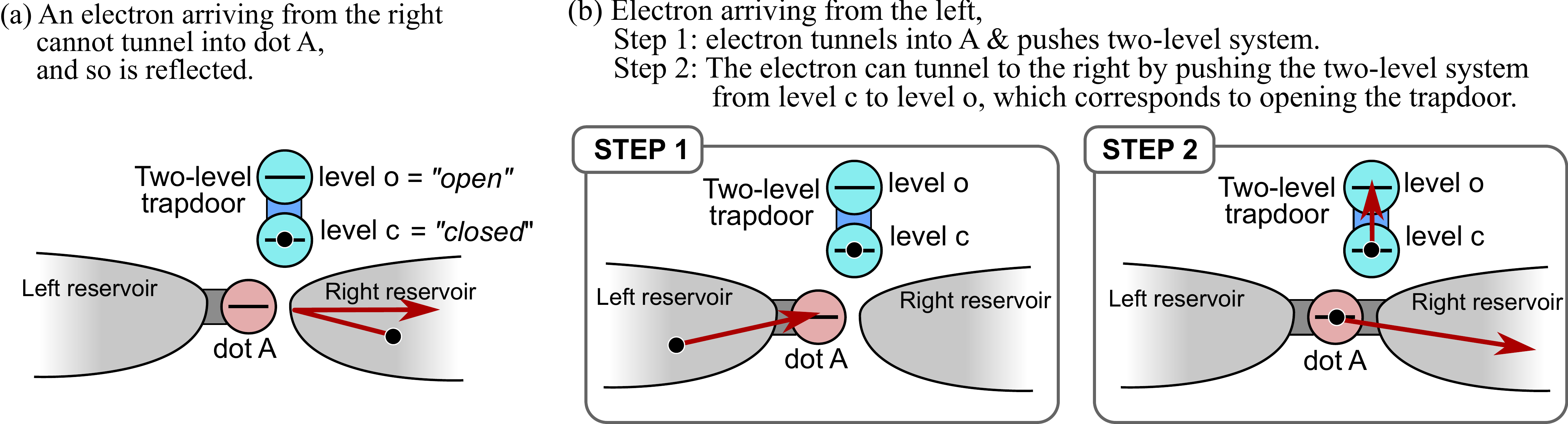}}
\caption{\label{fig:two-level} 
Sketches showing the functioning of the two-level trapdoor;  
the two-levels of the trapdoor are labelled "c" for closed and ``o'' for open. 
Level c has lower energy than o, so it is energetically favourable for the 
two-level system to be in level c.
When the trapdoor is closed (in the level c), it blocks any tunnelling 
between dot A and the right reservoir, but when it is open (in level o) it allows this tunnelling. 
This occurs because the two-level system is a negatively charged electron in either level c or level o. When it is in level c it is close enough to the tunnel-barrier between
the right reservoir and dot A, that its electric field raises the tunnel-barrier,
and make tunnelling between the right reservoir and dot A impossible.
In contrast, when the two-level system is in level o, it is far enough from
the tunnel-barrier that it does not prevent tunnelling between  right reservoir and dot A.
In (a) an electron arrives from the right reservoir but cannot tunnel into dot A, because the trapdoor is closed, so it is certain that this electron is reflected.  
In contrast in (b) an electron arrives from the left reservoir tunnels into dot A, where its repulsive interaction with the two-level-system can push the two-level system from level c to level o (i.e. from closed to open). This then allows tunnelling between dot A and the right reservoir, so there is a finite change the electron in dot A will go into the right reservoir. Comparing (a) and (b) it looks like there will be a net particle from from left to right, even if the right reservoir has a higher electro-chemical potential than the left reservoir, which would violate the second law of thermodynamics. 
} 
\end{figure*}

\begin{table*}[h]
\noindent\fbox{\hskip 3mm\parbox{0.9\textwidth}{\vskip 2mm
{\bf Hamiltonian of the trapdoor system (Fig.~\ref{fig:two-level}) without the reservoirs:}
\begin{eqnarray}
\hat{H}_{\rm sys} = 
E_{\rm A}\hat{d}_{\rm A}^\dagger \hat{d}_{\rm A}
+ E_{\rm c}\hat{d}_{\rm c}^\dagger \hat{d}_{\rm c} 
+ E_{\rm o}\hat{d}_{\rm o}^\dagger \hat{d}_{\rm o} + \gamma \Big(\hat{d}_{\rm c}^\dagger \hat{d}_{\rm o}+\hat{d}_{\rm o}^\dagger \hat{d}_{\rm c} \Big)
+ U \hat{d}_{\rm A}^\dagger \hat{d}_{\rm A} \hat{d}_{\rm c}^\dagger \hat{d}_{\rm c} 
\label{Eq:Hsys}
\end{eqnarray}
where $\hat{d}_{\rm A}^\dagger$, $\hat{d}_{\rm c}^\dagger$ and $\hat{d}_{\rm o}^\dagger$ are creation operators for an electron in levels A, c and o respectively. 
The first three terms are the energies of levels A, c and o,
the fourth term is tunnelling between c and o,
and the last term is the repulsive Coulomb interaction, $U$, between an electron in level A and an electron in level c. 
I define $|s_{\rm A},s_{\rm c},s_{\rm o}\rangle$ as the state where $s_{\rm A}$ labels whether dot A is empty ($s_{\rm A}=0$) or contains one electron ($s_{\rm A}=1$), with $s_{\rm c}$ and $s_{\rm o}$ doing the same for levels c and o. 
Then $\hat{H}_{\rm sys}$ has eight many-body eigenstates.
Two of them have dot A empty, with one electron shared between
levels c and o; one has the electron predominantly in  level c
(labelled $\widetilde{{\rm c}}$), and the other has the electron predominantly 
in level o (labelled $\widetilde{{\rm o}}$);
\begin{subequations}
\label{Eqs:manybody-eigenstates}
\begin{eqnarray}
\big|0\widetilde{{\rm c}}\,\big\rangle &\equiv& 
\phantom{-}\!\cos \left(\tfrac{1}{2}\theta_0\right) \big|0,1,0\big\rangle  
+\sin\left(\tfrac{1}{2}\theta_0\right)\big|0,0,1\big\rangle  \, ,
\label{Eqs:manybody-eigenstates-1+}
\\
\big|0\widetilde{{\rm o}}\,\big\rangle &\equiv& 
-\sin \left(\tfrac{1}{2}\theta_0\right) \big|0,1,0\big\rangle  
+\cos\left(\tfrac{1}{2}\theta_0\right)\big|0,0,1\big\rangle  \, ,
\\
& &\hskip -3mm \mbox{with energies }\ \  
E_{\rm 0\tilde{c}}\,=\, \tfrac{1}{2}(E_{\rm o}+E_{\rm c}) - \sqrt{\tfrac{1}{4}\left(E_{\rm o}-E_{\rm c}\right)^2 + \gamma^2\,},
\\
& &\hskip -3mm \phantom{\mbox{with energies }}\ \  
E_{0\tilde{o}}\,=\, \tfrac{1}{2}(E_{\rm o}+E_{\rm c}) + \sqrt{\tfrac{1}{4}\left(E_{\rm o}-E_{\rm c}\right)^2 + \gamma^2\,},
\end{eqnarray}
where $\theta_0= \arctan\big[2\gamma\big/\big(E_{\rm o}-E_{\rm c}\big)\big]$, see Fig.~\ref{fig:energy-level}b.
Another two many-body eigenstates have one electron in dot A, with one electron shared between
levels c and a;
\begin{eqnarray}
\big|1\widetilde{{\rm c}}\,\big\rangle &\equiv& 
\cos \left(\tfrac{1}{2}\theta_1\right) \big|1,1,0\big\rangle  
-\sin\left(\tfrac{1}{2}\theta_1\right)\big|1,0,1\big\rangle\, ,  
\\
\big|1\widetilde{{\rm o}}\,\big\rangle &\equiv& 
\sin \left(\tfrac{1}{2}\theta_1\right) \big|1,1,0\big\rangle  
+\cos\left(\tfrac{1}{2}\theta_1\right)\big|1,0,1\big\rangle \, ,  
\\
& &
\hskip -3mm
\mbox{with energies }\ \  
E_{\rm 1\tilde{c}}\,=\, \tfrac{1}{2}(2E_{\rm A}+E_{\rm o}+E_{\rm c}+U) 
+ \sqrt{\tfrac{1}{4}\left(U-E_{\rm o}+E_{\rm c}\right)^2 + \gamma^2\,},
\\
& &
\hskip -3mm
\phantom{\mbox{with energies }}\ \  
E_{\rm 1\tilde{o}}\,=\, \tfrac{1}{2}(2E_{\rm A}+E_{\rm o}+E_{\rm c}+U) 
- \sqrt{\tfrac{1}{4}\left(U-E_{\rm o}+E_{\rm c}\right)^2 + \gamma^2\,}.
\end{eqnarray}
where $\theta_1= \arctan\big[2\gamma\big/\big(U-E_{\rm o}+E_{\rm c}\big)\big]$, see Fig.~\ref{fig:energy-level}b.
\end{subequations}
The four other many-body eigenstates have a total of either zero or two electrons in levels c and o; they will play no role in our analysis.

\vskip 2mm
{\bf Hamiltonian of the reservoirs, and their tunnel couplings to dot A:}
\begin{eqnarray}
\hat{H}_{res} =  \sum_p 
\epsilon^{\rm (L)}_p \hat{c}_{{\rm L};p}^\dagger \hat{c}_{{\rm L};p}\,+\, 
\epsilon^{\rm (R)}_p \hat{c}_{{\rm R};p}^\dagger \hat{c}_{{\rm R};p}\,+\, 
\gamma^{\rm (L)}_p \Big(\hat{c}_{{\rm L};p}^\dagger \hat{d}_A + \hat{d}_A^\dagger \hat{c}_{{\rm L};p} \Big) 
\,+\, 
\gamma^{\rm (R)}_p 
\hat{d}_{\rm o}^\dagger \hat{d}_{\rm o}\Big(\hat{c}_{{\rm R};p}^\dagger \hat{d}_A + \hat{d}_A^\dagger \hat{c}_{{\rm R};p} \Big) \, .
\label{Eq:Hres}
\end{eqnarray}
where mode $p$ of reservoir $j$ has creation operator
$\hat{c}_{j;p}^\dagger$, energy $\epsilon^{\rm (j)}_p$, and tunnel coupling  $\gamma^{\rm (j)}_p$ to dot A.
The crucial term that makes the trapdoor is $d_{\rm o}^\dagger d_{\rm o}$; it stops tunnelling to R when level c is occupied. 
\vskip 2mm
}\hskip 3mm}
\end{table*}

\noindent\fbox{\hskip 3mm\parbox{0.9\columnwidth}{\vskip 2mm
\noindent
\textsf{
A comment on the terminology used here:
\begin{itemize}
\item The \textit{two-level trapdoor} refers the two-level system consisting of level c (trapdoor closed) and level o (trapdoor open), see  Fig.~\ref{fig:two-level}.
\item The \textit{trapdoor system} refers the two-level trapdoor and dot A together (including the coupling between them). Dot A is a critical part of the trapdoor system, because the trapdoor is {\it triggered} to open by electron entering dot A; it is also the electron entering dot A that provides the energy to open the trapdoor.
\end{itemize}
}
\vskip 1mm
}\hskip 3mm}
\vskip 3mm

Here, I will use the rate-equation method to do the following:
\begin{itemize}
\item[(i)] 
I show that if one neglects the fluctuations of the two-level trapdoor, then it does appear to act as a autonomous mechanical version of a Maxwell demon.
It appears to violate the second law by continuously producing work from a single heat source.

\item[(ii)]
I use the rate-equation method to treat the dynamics of the trapdoor (including its thermal fluctuations).  We will take the long-time limit (steady-state limit) of this system, and show that it does not produce any work. Thermal fluctuations stop it acting as a demon, and stop it violating the second law.  So the illusion of this system opening a crack in the second law is dispersed by thermal fluctuations.

\item[(iii)]
This does not mean the two-level trapdoor is useless. 
I will show that It is one of the simplest systems that acts as a rectifier (diode) when placed between two electron gases at different biases.
This confirms Smoluchowski's observation that a trapdoor can act as a rectifier, when a density distribution is imposed by external means.

\item[(iv)]
Having shown that this system fails to work as a demon because of the thermal motion of the trapdoor, one can imagine making it work as a demon by damping out this thermal motion.  Microscopically, this can be done by coupling the trapdoor to a cold phonon reservoir.  Then the trapdoor does work as a demon, making a current flow against a potential, and thereby reducing its entropy and producing electrical power. However, damping the thermal excitation of the trapdoor requires a heat flow into the cold phonon reservoir,
which increases its entropy. This restores the second law, and makes the demon look like a heat engine, generating work from a heat flow from hot to cold.
\end{itemize}

\section{\mbox{Apparent violation of} \mbox{the second law}}

Fig.~\ref{fig:two-level} sketches the mechanism by which it appears that the two-level trapdoor could violate the second law.  Here I give more detail on this.

The two-level system is a double-dot containing a single electron. When that 
electron is in the level c, it is close enough to the tunnel-barrier between 
the right reservoir and dot A that its electric field slightly increases the tunnel-barrier height\footnote{Let us assume the two-level system has no effect on the tunnel-barrier between the left reservoir and dot A}.
As the tunnelling probability is exponentially sensitive to the barrier height, this can greatly reduce the tunnelling probability.  The simplest case to picture is when this makes tunnelling between the right reservoir and dot A impossible when the two-level system is in level c, like a closed trapdoor between dot A and the right reservoir.

I assume that when dot A is empty, then it is energetically favourable for the two-level system to be in level c. In contrast, when dot A is occupied by an electron, it is energetically favourable for the two-level system to be is level o (because an electron in dot A repels an electron in level c, pushing up the energy of level c). 
Defining $\epsilon_{\alpha \sigma}$ as the energy of level $\sigma$ in the two-level system, when dot A is in level $\alpha$, we have
\begin{eqnarray}
\epsilon_{0o} > \epsilon_{0c}
\\
\epsilon_{1o} < \epsilon_{1c} 
\end{eqnarray}
where subscript $1$ indicates that dot A contains an electron, and $0$ indicates that contains no electrons.
Then an electron arriving in dot A from the left will push the trapdoor open, giving up some of its energy to do so.

Now if I look only at the sketch in Fig.~\ref{fig:two-level}, it seems clear that this trapdoor will act as a demon. When a particle arrives from the right, the trapdoor stays closed, but when a particle arrives from the left, the trapdoor is pushed open, allowing the article to go to the right.
Thus it seems to allow a particle flow from left to right, even when the electrochemical potential is higher on the right.  It would thus be creating a flow against its natural direction, thereby generating electrical power. 
As energy is conserved, this must be converting heat in the reservoir into electrical work, so the reservoirs would cool down, spontaneously lowering their entropy.
This would be a clear violation of the second law of thermodynamics.

\subsection{\mbox{Smoluchowski's argument is not} \mbox{enough to restore the second law}}

Given the works of Smoluchowski discussed in chapter~\ref{Chap:intro}, we can already guess why the above argument is wrong.
For a thermal particle to have enough energy to push the trapdoor open, then energy gap between the two levels of the trapdoor (open and close) must be of order temperature $T$. So even if the trapdoor starts closed it will soon acquire thermal excitations
from the particles flowing through it, and equilibrate with them at temperature $T$, meaning it will be open much of the time. At these moments, particles are free to pass in both directions, so the net flow will be from the reservoir with the higher electrochemical potential to the one with the lower electrochemical potential
(in our case from reservoir R to reservoir L).  Thus he argued that the trapdoor could not work as a Maxwell demon.

Yet, let us look at his argument critically.  Let us imagine that dot A is empty, and the trapdoor is in a thermal state at temperature $T$. The trapdoor still has a higher probability to be closed than open (as the closed state has lower energy).  Thus an electron arriving at this trapdoor has a higher probability to find it closed than open.  This implies that the demon-like action of the trapdoor
(allowing electrons to flow from left to right but not vice-versa)
is not completely suppressed by the fact the trapdoor is thermally excited. 
Yet, even the smallest asymmetry between the probability to go from left to right and that to go from right to left, 
then there can be a small violation of the second law.  

There is no doubt that Smoluchowski's argument makes it clear that 
thermal fluctuations will greatly degrade the operation of the trapdoor as a demon, compared to simplistic picture in Fig.~\ref{fig:two-level}.
However, I do not see how his argument is enough to rule out a small
asymmetry between flow from left to right and flow from right to left, and even a small asymmetry allows violations of the second law.
For me, the only way to prove the second law is not violated is to write down a full
(if simplified) model of the trapdoor, to analyse it, and show that the asymmetry is strictly zero.

\begin{figure}
\centerline{\includegraphics[width=0.75\columnwidth]{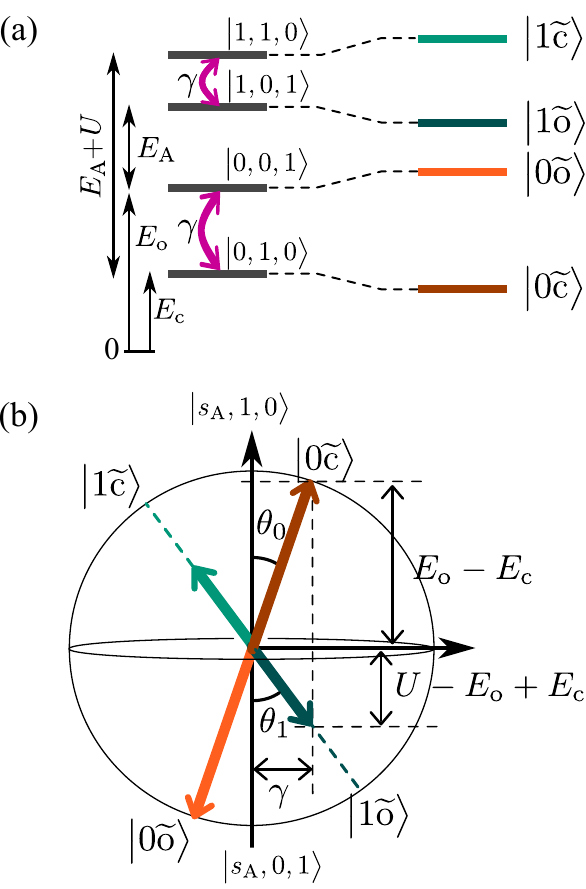}}
\caption{\label{fig:energy-level} 
The many-body eigenstates of the Hamiltonian of the trapdoor system in the absence of the reservoirs, given in Eq.~(\ref{Eq:Hsys}).  (a) The basis is constructed of states of the form $|s_{\rm A},s_{\rm c},s_{\rm o}\rangle$, where $s_{\rm A}=0$ if dot A is empty, and $s_{\rm A}=1$ if dot A is occupied by one electron. Similarly $s_{\rm c}$ and $s_{\rm o}$ are 0 or 1 depending on whether levels c and o are empty or occupied.    The tunnelling coupling, $\gamma$, between levels o and c, causes transitions between $|s_{\rm A},0,1\rangle$ and  $|s_{\rm A},1,0\rangle$.
Then the many-body {\it eigenstates} are the bonding and anti-bonding combination of these two states. I label these states as follows; $|s_{\rm A}\widetilde{\rm c}\rangle$ is the state that is more closed than open, and $|s_{\rm A}\widetilde{\rm o}\rangle$ is the state that is open than closed (where $s_{\rm A}=0,1$).
Their exact forms are given in Eqs.~(\ref{Eqs:manybody-eigenstates}). (b) This  construction can be seen graphically by taking $|s_{\rm A},1,0\rangle$ and  $|s_{\rm A},0,1\rangle$ as the north and south pole of a Bloch sphere. Then the four many-body eigenstates given in Eqs.~(\ref{Eqs:manybody-eigenstates}) are the four vectors shown on the Bloch sphere; $|0\widetilde{\rm c}\rangle$,  $|0\widetilde{\rm o}\rangle$, $|1\widetilde{\rm c}\rangle$ and   $|1\widetilde{\rm o}\rangle$.
} 
\end{figure}

\subsection{Comment on the assumptions}
Here I will always assume that when the trapdoor is closed, it is completely closed.
In other words, tunnelling between the right reservoir and dot A is assumed to be impossible when the two-level system is in level c. However, the same logic would apply
whenever the tunnelling is less likely when the trapdoor is in level c than when it is it in level o.
Defining $\Gamma_{j}^{(\sigma)}$ as the rate of tunnelling between reservoir $j$ and dot A, while the two-level system is in level $\sigma$, it is enough to have 
\begin{eqnarray}
\Gamma_{\rm R}^{\rm (c)} < \Gamma_{\rm R}^{\rm (o)}\,    
\qquad \Gamma_{\rm L}^{\rm (c)} = \Gamma_{\rm L}^{\rm (o)}.   
\end{eqnarray}
and the physics will be qualitatively similar.  The asymmetry between the flow for left to right and the flow from right to left (if and when such an asymmetry exists) will be smaller than in the case I take.  But even a very small asymmetry would be enough to violate the second law.

\section{Setting up the rate equation}
Both the rate equation method used here, and the fully quantum approach (used in the chapter~\ref{Chap:quantum-trapdoor}) start by dividing the set-up into ``system'' and ``reservoirs''.
The system is small enough to have a discrete set of energy levels,
and is written in terms of the many-body eigenstates of its Hamiltonian.
Each reservoir is taken to be large and contains a continuum of states corresponding to free particles. The free particle states in reservoir $i$ are assumed to be occupied with a thermal distribution (Fermi distribution for electrons and Bose distribution for phonons and photons) with a temperature $T_i$ and electro-chemical potential $\mu_i$.

Imagine the trapdoor-system is in the many-body eigenstate $|0\widetilde{{\rm c}}\,\rangle$
given by Eq.~(\ref{Eqs:manybody-eigenstates-1+}), when an electron tunnels into dot A. This changes dot A 's state from 0 to 1, but does not modify the state of levels o and c, Hence,the trapdoor-system's state after the tunnelling event is 
\begin{eqnarray}
\cos \left(\tfrac{1}{2}\theta_0\right) \big|1,1,0\big\rangle  
+\sin\left(\tfrac{1}{2}\theta_0\right)\big|1,0,1\big\rangle  \, ,
\end{eqnarray}
With the aid of the sketch in Fig.~\ref{fig:energy-level}, we can see that this is a linear combination of many-body eigenstates,
\begin{eqnarray}
\sin \left(\tfrac{1}{2}\theta_0+\tfrac{1}{2}\theta_1\right) \big|1\widetilde{{\rm c}}\big\rangle  
+\cos\left(\tfrac{1}{2}\theta_0+\tfrac{1}{2}\theta_1)\right)\big|1\widetilde{{\rm o}}\big\rangle  \, .
\end{eqnarray}
Indeed, every time that there is a tunnelling event involving reservoir L, there is a transition of this type between the many-body eigenstates of the trapdoor system, These tunnelling events have the following amplitudes
\begin{eqnarray}
\big\langle  1\widetilde{{\rm c}} \big| 
\,\hat{d}^\dagger_{\rm A} \big| 0\widetilde{{\rm c}}\big\rangle 
\,=\, 
\big\langle  0\widetilde{{\rm c}} \big| 
\,\hat{d}_{\rm A} \big| 1\widetilde{{\rm c}}\big\rangle 
\!\!&=&\! \cos \left(\tfrac{1}{2}\theta_0+\tfrac{1}{2}\theta_1\right)\! ,
\\
\big\langle  1\widetilde{{\rm o}} \big| 
\,\hat{d}^\dagger_{\rm A} \big| 0\widetilde{{\rm c}}\big\rangle 
\, =\,  
\big\langle  0\widetilde{{\rm c}} \big| 
\,\hat{d}_{\rm A} \big| 1\widetilde{{\rm o}}\big\rangle
\!\!&=&\!  \sin \left(\tfrac{1}{2}\theta_0+\tfrac{1}{2}\theta_1\right)\! ,
\\
\big\langle  1\widetilde{{\rm c}} \big| 
\,\hat{d}^\dagger_{\rm A} \big| 0\widetilde{{\rm o}}\big\rangle 
\, =\,  
\big\langle  0\widetilde{{\rm o}} \big| 
\,\hat{d}_{\rm A} \big| 1\widetilde{{\rm c}}\big\rangle 
\!\!&=&  \!\!\!\!-\sin \left(\tfrac{1}{2}\theta_0+\tfrac{1}{2}\theta_1\right)\! , \qquad\ 
\\
\big\langle  1\widetilde{{\rm o}} \big| 
\,\hat{d}^\dagger_{\rm A} \big| 0\widetilde{{\rm o}}\big\rangle 
\,=\,  
\big\langle  0\widetilde{{\rm o}} \big| 
\,\hat{d}_{\rm A} \big| 1\widetilde{{\rm o}}\big\rangle 
\!\!&=&\!   \cos \left(\tfrac{1}{2}\theta_0+\tfrac{1}{2}\theta_1\right)\! .
\end{eqnarray}
Similarly, tunnelling events involving reservoir R, have the following amplitudes
\begin{eqnarray}
\big\langle  1\widetilde{{\rm c}} \big| 
\,\hat{d}^\dagger_{\rm o} \hat{d}_{\rm o} \hat{d}^\dagger_{\rm A} \big| 0\widetilde{{\rm c}}\big\rangle 
&=&  
\big\langle  0\widetilde{{\rm c}} \big| 
\,\hat{d}^\dagger_{\rm o} \hat{d}_{\rm o} \hat{d}_{\rm A} \big| 1\widetilde{{\rm c}}\big\rangle 
\nonumber \\
&=& \!-\sin \left(\tfrac{1}{2}\theta_0\right)\sin \left(\tfrac{1}{2}\theta_1\right), \qquad
\\
\big\langle  1\widetilde{{\rm o}} \big| 
\,\hat{d}^\dagger_{\rm o} \hat{d}_{\rm o} \hat{d}^\dagger_{\rm A}
 \big| 0\widetilde{{\rm c}}\big\rangle 
&=&  
\big\langle  0\widetilde{{\rm c}} \big| 
\,\hat{d}^\dagger_{\rm o} \hat{d}_{\rm o} \hat{d}_{\rm A}
 \big| 1\widetilde{{\rm o}}\big\rangle 
\nonumber \\
&=& \sin \left(\tfrac{1}{2}\theta_0\right)
\cos \left(\tfrac{1}{2}\theta_1\right),
\\
\big\langle  1\widetilde{{\rm c}} \big| 
\,\hat{d}^\dagger_{\rm o} \hat{d}_{\rm o} \hat{d}^\dagger_{\rm A} 
\big| 0\widetilde{{\rm o}}\big\rangle 
&=&  
\big\langle  0\widetilde{{\rm o}} \big| 
\,\hat{d}^\dagger_{\rm o} \hat{d}_{\rm o} \hat{d}_{\rm A} 
\big| 1\widetilde{{\rm c}}\big\rangle 
\nonumber \\
&=&  - \cos \left(\tfrac{1}{2}\theta_0\right)
\sin \left(\tfrac{1}{2}\theta_1\right),
\\
\big\langle  1\widetilde{{\rm o}} \big| 
\,\hat{d}^\dagger_{\rm o} \hat{d}_{\rm o} \hat{d}^\dagger_{\rm A} 
\big| 0\widetilde{{\rm o}}\big\rangle 
&=&  
\big\langle  0\widetilde{{\rm o}} \big| 
\,\hat{d}^\dagger_{\rm o} \hat{d}_{\rm o} \hat{d}_{\rm A} 
\big| 1\widetilde{{\rm o}}\big\rangle 
\nonumber \\
&=&  \cos \left(\tfrac{1}{2}\theta_0\right)
\cos \left(\tfrac{1}{2}\theta_1\right),
\end{eqnarray}
In what follows, 
I need to give formulas containing these overlaps in a compact manner, 
thus I define 
\begin{eqnarray}
L_\beta^\alpha &=& 
\left\{\begin{array}{ccl}
\cos \left(\tfrac{1}{2}\theta_0+\tfrac{1}{2}\theta_1\right) & \phantom{\Big|}\hbox{for} & \alpha=\beta,
\\
\sin\left(\tfrac{1}{2}\theta_0+\tfrac{1}{2}\theta_1\right) & \phantom{\Big|}\hbox{for} & \alpha\neq \beta,
\end{array}\right.
\label{Eq:overlap_L}
\end{eqnarray}
where $\alpha$ and $\beta$ are $\widetilde{{\rm c}}$ or $\widetilde{{\rm o}}$, and
\begin{eqnarray}
R_\beta^\alpha \!\!&=&\!\!\!\!
\left\{\begin{array}{ccl}
\sin \left(\tfrac{1}{2}\theta_0\right) \sin\left(\tfrac{1}{2}\theta_1\right) & \phantom{\Big|}\hbox{\!for\!\!} & \alpha=\beta=\widetilde{{\rm c}},
\\
\sin \left(\tfrac{1}{2}\theta_0\right) \cos\left(\tfrac{1}{2}\theta_1\right) & \phantom{\Big|}\hbox{\!for\!\!} & \alpha=\widetilde{{\rm c}}, \beta=\widetilde{{\rm o}},
\\
\cos \left(\tfrac{1}{2}\theta_0\right) \sin\left(\tfrac{1}{2}\theta_1\right) & \phantom{\Big|}\hbox{\!for\!\!} & \alpha=\widetilde{{\rm o}}, \beta=\widetilde{{\rm c}},
\\
\cos \left(\tfrac{1}{2}\theta_0\right)\cos \left(\tfrac{1}{2}\theta_1\right) & \phantom{\Big|}\hbox{\!for\!\!} & \alpha=\beta=\widetilde{{\rm o}}.
\end{array}\right.\qquad
\label{Eq:overlap_R}
\end{eqnarray}

\begin{figure}[t]
\centerline{\includegraphics[width=0.85\columnwidth]{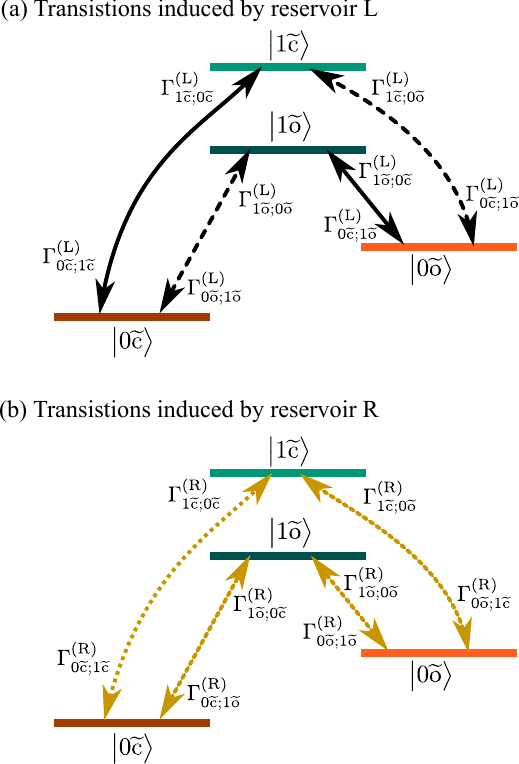}}
\caption{\label{fig:trajectories} 
The many-body eigenstates in Fig.~\ref{fig:energy-level}a shown 
with the transitions induced by the coupling to reservoirs L and R.
Each transition occurs at different rates in each direction; the rate in a given direction is noted next to the relevant arrowhead.
If one is in the limit of small $\gamma$ (small coupling between trapdoor levels c and o), 
then the solid lines have a transition rate of order $\gamma^0$, 
the dashed lines have a transition rates of order $\gamma^2$, 
and the dotted line have transition rates of order $\gamma^4$.
The rates also depend on the Fermi functions of the reservoir modes at the energy of the transition, and so obey local detailed balance, 
Eq.~(\ref{Eq:local-detailed-balance}). 
} 
\end{figure}

Fermi's golden rule tells us that a transition from many-body eigenstate $|0\alpha\rangle$ to many-body eigenstate $|1\beta\rangle$ occurs when an electron with energy $E_{1\beta}-E_{0\alpha}$ tunnels from reservoir $j$ to dot A. 
I define the tunnelling amplitude for this electron as $\gamma_j(E_{1\beta}-E_{0\alpha})$
as that given by $\gamma_p^{(j)}$ in Eq.~(\ref{Eq:Hres}), for the $p$ that corresponds to this electron energy. 
For reservoirs L and R respectively, these Golden rule transition rates are
\begin{subequations}
\label{Eq:golden-rule-rates}
\begin{eqnarray}
\Gamma^{\rm (L)}_{1\beta;0\alpha} \! &=& \!\pi \,\big|L_\beta^\alpha \  \gamma_{\rm L}(E_{1\beta}-E_{0\alpha})\big|^2 \,\nu_{\rm L}(E_{1\beta}-E_{0\alpha})\qquad
\nonumber \\
& &\times  f_{\rm L}(E_{1\beta}-E_{0\alpha}), 
\\
\Gamma^{\rm (R)}_{1\beta;0\alpha} \! &=& \!\pi \,\big|R_\beta^\alpha \  \gamma_{\rm R}(E_{1\beta}-E_{0\alpha})\big|^2  \,\nu_{\rm R}(E_{1\beta}-E_{0\alpha}) \,
\nonumber \\
& &\times f_{\rm R}(E_{1\beta}-E_{0\alpha}), \qquad
\end{eqnarray}
where $f_j(E)$ is the reservoir's occupation (Fermi function) at energy $E$, and $\nu_j(E)$ is the reservoir's density of states at energy $E$. 
The reverse transition between many-body states occurs when an electron tunnelling from dot A into a state in reservoir $j$ with energy $E_{1\beta}-E_{0\alpha}$.
For reservoirs L and R respectively, this occurs at a rate
\begin{eqnarray}
\Gamma^{\rm (L)}_{0\alpha;1\beta} \!&=&\!\!\pi \,\big|L_\beta^\alpha \  \gamma_{\rm L}(E_{1\beta}-E_{0\alpha})\big|^2\, \nu_{\rm L}(E_{1\beta}-E_{0\alpha})\qquad
\nonumber \\
& &\times \Big(1-f_{\rm L}(E_{1\beta}-E_{0\alpha})  \Big),\qquad\quad
\\
\Gamma^{\rm (R)}_{0\alpha;1\beta} \!&=&\!\!\pi \,\big|R_\beta^\alpha \  \gamma_{\rm R}(E_{1\beta}-E_{0\alpha})\big|^2\, \nu_{\rm R}(E_{1\beta}-E_{0\alpha}) 
\nonumber \\
& &\times \Big(1-f_{\rm R}(E_{1\beta}-E_{0\alpha})  \Big).\qquad\quad
\end{eqnarray}
\end{subequations}
A fully quantum calculation, such as that in chapter~\ref{Chap:quantum-trapdoor} shows that Fermi's golden rule only applies when the rates in Eqs.~(\ref{Eq:golden-rule-rates}) are much less than all other energy-scales in the problem, specifically much smaller than the energy gaps between the many-body eigenstates.

The crucial ingredient of Eqs.~(\ref{Eq:golden-rule-rates}) is a simple relationship between the rate at which reservoir $j$ induces
transition $|0\alpha\rangle \to |1\beta\rangle$,
and the rate of its time reverse $|1\beta\rangle \to |0\alpha\rangle$.
It comes from the fact that the two rates only differ in the terms containing Fermi functions, which obey $f_j(E)\big/\big(1-f_j(E)\big) = \exp [-(E-\mu_j)/T_j]$, 
where reservoir $j$ has electro-chemical potential $\mu_j$ 
and temperature $T_j$. 
The relationship is called {\it local detailed balance}, and it is so important for thermodynamics, that I put it in a box:
\begin{eqnarray}
\boxed{\  
\begin{array}{l}
\!\!\mbox{Local detailed balance: \phantom{\Big|} } \\
\phantom{\Bigg|}\Gamma^{\rm (j)}_{1\beta;0\alpha}  
\, =\, \Gamma^{\rm (j)}_{0\alpha;1\beta} 
\ \exp\!\left[-\dfrac{E_{1\beta}-E_{0\alpha}-\mu_j}{T_j} \right]\!. 
\end{array}\!\! }
\label{Eq:local-detailed-balance}
\end{eqnarray}
This relation is the crucial ingredient in proving that such a system cannot violate the second law of thermodynamics; see section~\ref{Sect:rate-eqn-2nd law} below.

\begin{figure}
\centerline{\includegraphics[width=0.95\columnwidth]{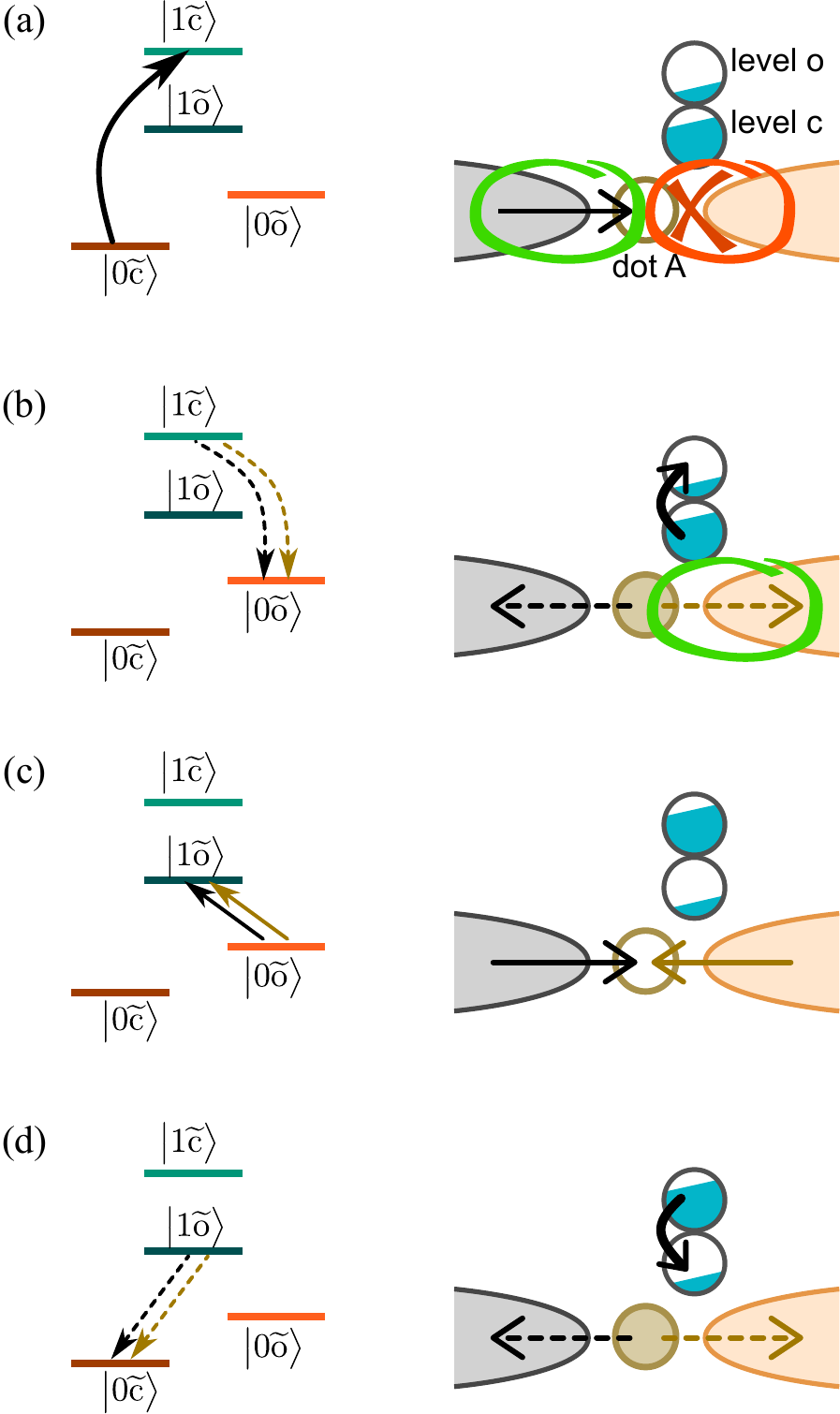}}
\caption{\label{fig:step-by-step} 
On the left is one possible evolution of the trapdoor system, in the 
small $\gamma$ limit (see section~\ref{sect:small-gamma-limit}), 
with the equivalent physical process sketched on the right. 
In the sketches on the right, the state $\widetilde{{\rm c}}$ is shown as being partially in both level c and o, but more in level c, 
while state $\widetilde{{\rm o}}$ is more in level o.
Solid arrows are transitions of order $\gamma^0$, dashed arrows are 
transitions of order $\gamma^2$, and I assume transitions of order $\gamma^4$ can be neglected.
The absence of a transition highlighted by the red cross in Fig.~\ref{fig:two-level}a, corresponds to the reflection of all electrons arriving at dot A from reservoir R.
The transitions highlighted in green correspond to flow from reservoir L 
to reservoir R, as sketched in Fig.~\ref{fig:two-level}b.
} 
\end{figure}

\section{\mbox{The trapdoor limit:} \mbox{weak-coupling between} \mbox{trapdoor levels}}
\label{sect:small-gamma-limit}

Here I take a limit which makes the trapdoor simpler to analyse, while allowing it to work as intended.  I take the coupling, $\gamma$, between the two trapdoor levels 
(fourth term in Eq.~\ref{Eq:Hres}) to be small compared to the energy gaps between the trapdoor levels.
In other words, I take $\gamma \ll (E_{\rm o}-E_{\rm c})$ and $\gamma \ll (U-E_{\rm o}+E_{\rm c})$.
I will justify taking this limit after the fact, by noting that it gives the desired effect of allowing particle flow from left to right, while hindering particle flow from right to left, see Fig.~\ref{fig:step-by-step}.
However, one can already see that the opposite limit (large $\gamma$) is  {\it unlikely} to work as a demon, since then both the trapdoor system's eigenstates would have the trapdoor open about half the time. 

For small $\gamma$, we have $\theta_0,\theta_1 \ll 1$ so
the overlaps in Eqs.~(\ref{Eq:overlap_L},\ref{Eq:overlap_R}) 
reduce to
\begin{eqnarray}
\big|L_\beta^\alpha\big|^2 &\simeq& 
\left\{\begin{array}{ccl}
1-\tfrac{1}{4}\left(\theta_0+\theta_1\right)^2 & \phantom{\Big|}\hbox{for} & \alpha=\beta,
\\
\tfrac{1}{4}\left(\theta_0+\theta_1\right)^2 & \phantom{\Big|}\hbox{for} & \alpha\neq \beta,
\end{array}\right.
\\
\big|R_\beta^\alpha\big|^2 \!\!&\simeq&\!\!\!\!
\left\{\begin{array}{ccl}
0 & \phantom{\Big|}\hbox{\!for\!\!} & \alpha=\beta=\widetilde{{\rm c}},
\\
\tfrac{1}{4}\theta_0^2 & \phantom{\Big|}\hbox{\!for\!\!} & \alpha=\widetilde{{\rm c}}, \beta=\widetilde{{\rm o}},
\\
\tfrac{1}{4}\theta_1^2 & \phantom{\Big|}\hbox{\!for\!\!} & \alpha=\widetilde{{\rm o}}, \beta=\widetilde{{\rm c}},
\\
1-\tfrac{1}{4}\theta_0^2-\tfrac{1}{4}\theta_1^2 & \phantom{\Big|}\hbox{\!for\!\!} & \alpha=\beta=\widetilde{{\rm o}}.
\end{array}\right.\qquad
\end{eqnarray}
where I drop terms of order $\gamma^4$ or smaller.
After substituting these into the transition rates, 
one can make the sketch of transitions in Fig.~\ref{fig:step-by-step};
the fastest transitions (those that are ${\cal O}[\gamma^0]$) 
are marked by solid arrows, 
while slower ones (those that are ${\cal O}[\gamma^2]$)
are marked by dashed arrows.
In this small $\gamma$ limit, the transitions in 
Fig.~\ref{fig:step-by-step} have the asymmetry in the flows,
meaning the electron flow from left to right appears faster than 
the flow from right to left.  This makes it look like
it could act as a Maxwell demon in the manner shown in Fig.~\ref{fig:two-level}.

\section{Steady-state of trapdoor system}

Now  that we have constructed the rates of transitions between the different 
many-body eigenstates of the trapdoor system, it is rather easy to construct its dynamics.  The rate of change of the occupation of a given eigenstate
is given by the rate of transitions into that eigenstate minus the rate of transitions out of that eigenstate. 
This means the evolution equation for the occupation 
of the four many-body eigenstates
is given by
\begin{eqnarray}
{\rmd \over \rmd t} 
\left(\!
\begin{array}{c} 
P_{\rm 0\tilde{c}} \\ 
P_{\rm 0\tilde{o}} \\
P_{\rm 1\tilde{o}} \\
P_{\rm 1\tilde{c}}
\end{array}\!\right) &=& 
\left(
\begin{array}{cccc}
\mathtt{A}   & 0  & \mathtt{B}  & \mathtt{C}  
\\
0 &  \mathtt{D} &  \mathtt{E} & \mathtt{F} 
\\
\mathtt{G}  &  \mathtt{H}  &  \mathtt{I}  &  0  
\\
\mathtt{J}  & \mathtt{K}  & 0  & \mathtt{L} 
\end{array}\right) \ 
\left(\!
\begin{array}{c} 
P_{\rm 0\tilde{c}} \\ 
P_{\rm 0\tilde{o}} \\
P_{\rm 1\tilde{o}} \\
P_{\rm 1\tilde{c}}
\end{array}\!\right),\qquad
\label{Eq:P-evolution-Eq}
\end{eqnarray}
with the matrix elements given by the transition rates as follows
\begin{subequations}
\label{Eq:trapdoor-matrix-elements}
\begin{eqnarray}
\mathtt{A} &=& -\mathtt{G} -\mathtt{J}
\\ 
\mathtt{B} &=&
\Gamma^{\rm (L)}_{\rm 0\tilde{c};1\tilde{o}} 
+\Gamma^{\rm (R)}_{\rm 0\tilde{c};1\tilde{o}} 
\\ 
\mathtt{C} &=&
\Gamma^{\rm (L)}_{\rm 0\tilde{c};1\tilde{c}} 
+\Gamma^{\rm (R)}_{\rm 0\tilde{c};1\tilde{c}} 
\\ 
\mathtt{D} &=& -\mathtt{H} -\mathtt{K}
\\
\mathtt{E} &=&
\Gamma^{\rm (L)}_{\rm 0\tilde{o};1\tilde{o}} 
+\Gamma^{\rm (R)}_{\rm 0\tilde{o};1\tilde{o}} 
\\ 
\mathtt{F} &=&
\Gamma^{\rm (L)}_{\rm 0\tilde{o};1\tilde{c}} 
+\Gamma^{\rm (R)}_{\rm 0\tilde{o};1\tilde{c}} 
\\
\mathtt{G} &=&
\Gamma^{\rm (L)}_{\rm 1\tilde{o};0\tilde{c}} 
+\Gamma^{\rm (R)}_{\rm 1\tilde{o};0\tilde{c}} 
\\
\mathtt{H} &=&
\Gamma^{\rm (L)}_{\rm 1\tilde{o};0\tilde{o}} 
+\Gamma^{\rm (R)}_{\rm 1\tilde{o};0\tilde{o}} 
\\ 
\mathtt{I} &=& -\mathtt{B} -\mathtt{E}
\\
\mathtt{J} &=&
\Gamma^{\rm (L)}_{\rm 1\tilde{c};0\tilde{c}} 
+\Gamma^{\rm (R)}_{\rm 1\tilde{c};0\tilde{c}} \\
\mathtt{K} &=&
\Gamma^{\rm (L)}_{\rm 1\tilde{c};0\tilde{o}}
+\Gamma^{\rm (R)}_{\rm 1\tilde{c};0\tilde{o}}
\\ 
\mathtt{L} &=& -\mathtt{C} -\mathtt{F}
\end{eqnarray}
\end{subequations}
Note that these matrix elements have the property that each column in the matrix sums to zero.  This has a simple physical origin in the fact that probabilities must sum to one, thus if the probability for the system 
to be in a given state is increasing, there must be an associated reduction
of the probability to be in another state, so that 
${\rmd \over \rmd t}(P_{\rm 0\tilde{c}} +
P_{\rm 0\tilde{o}} +
P_{\rm 1\tilde{o}} +
P_{\rm 1\tilde{c}}) =0$ irrespective of the values of 
$P_{\rm 0\tilde{c}}, \cdots, P_{\rm 1\tilde{c}}$.

If one prepares the system in a given state,
the occupations of the eigenstates will evolve until they reach a 
steady-state.  The decay to the steady-state is typically exponential on a timescale given by the slowest transition rate. 
It is this steady-state that interests us, it is given when the 
left-hand-side of Eq.~(\ref{Eq:P-evolution-Eq}) is zero.
This gives a matrix equation for the steady-state occupation probability, which is solved in Appendix~\ref{Sect:matrix-equation}.
The result cannot usefully be written in a more compact manner than 
Eq.~(\ref{Eq:basic-matrix-solution}) with
\begin{eqnarray}
\left.
\begin{array}{l}
P^\steady_{\rm 0\tilde{c}} \hbox{ given by } P_1 \\ 
P^\steady_{\rm 0\tilde{o}} \hbox{ given by } P_2 \\
P^\steady_{\rm 1\tilde{o}} \hbox{ given by } P_3 \\ 
P^\steady_{\rm 1\tilde{c}} \hbox{ given by } P_4 
\end{array}\right\} 
\!\!
\begin{array}{r}
\\
\hbox{in Eq.~(\ref{Eq:basic-matrix-solution}) 
with $\mathtt{A}$ to $\mathtt{L}$\phantom{.}}
\\ \hskip 3mm
\hbox{given by Eqs.~(\ref{Eq:trapdoor-matrix-elements}).}
\end{array}\nonumber
\\
\label{Eq:trapdoor_P_steady}
\end{eqnarray}
This steady-state of the trapdoor system determines all it steady-state properties,
specifically the currents through the trapdoor system.

\section{Particle and energy currents}

There are four transitions between eigenstates of the trapdoor system that involve an electron going from the trapdoor system to reservoir R, see Fig.~\ref{fig:trajectories}b.  The four transitions are $|1\beta\rangle$ to $|0\alpha\rangle$ where $\alpha$ and $\beta$ can each be $\widetilde{{\rm c}}$ or $\widetilde{o}$. 
At time $t$, the rate of particle flow associated with the transition $|1\beta\rangle \ \to \ |0\alpha\rangle$ is  given by the probability that the system is in state $|1\beta\rangle$ at time $t$ multiplied by the transition rate given in Eq.~(\ref{Eq:golden-rule-rates}); i.e. it is $\Gamma_{0\alpha;1\beta}^{(R)}  \,P_{1\beta}(t)$.  The rate of energy flow associated with that transition equals the particle flow multiplied by the energy carried per particle,
which equals the energy difference between state  $|1\beta\rangle$ and $|0\alpha\rangle$; i.e. it is $(E_{0\alpha} - E_{1\beta})\Gamma_{0\alpha;1\beta}^{(R)}  \,P_{1\beta}(t)$.  Finally the rate of heat flow associated with that transition,
equals the particle flow multiplied by the energy carried per particle measured
from the electro-chemical potential of reservoir R; i.e. it is given by 
$(E_{0\alpha} - E_{1\beta}-\mu_{\rm R})\Gamma_{0\alpha;1\beta}^{(R)}  \,P_{1\beta}(t)$.

For every transition listed above in which an electron flows from
the trapdoor system to reservoir R,  there is the reverse process in which
an electron flows from reservoir R to the trapdoor system. These transitions are from $|0\beta\rangle$ to $|1\alpha\rangle$, where again $\alpha$ and $\beta$ can each be $\widetilde{{\rm c}}$ or $\widetilde{{\rm o}}$.
These transition contribute negatively to the rate of particle flow.

Summing all these transitions, we arrive at the average particle current
into reservoir $j$ at time $t$ \cite{Benenti2017Jun},
\begin{eqnarray}
\big\langle I^{(j)}(t)\big\rangle \!\!&=&\!\!\! 
\sum_{\alpha,\beta \in \{ {\rm \tilde{c},\tilde{o}}\}}
\big(\Gamma_{0\alpha;1\beta}^{(j)}  \,P_{1\beta}(t)
- \Gamma_{1\alpha;0\beta}^{(j)}\,P_{0\beta}(t) \big),
\nonumber \\
\label{Eq:particle-current}
\end{eqnarray}
Similarly, the average energy current into reservoir $j$ at time $t$ is
\begin{eqnarray}
\big\langle \dot{E}^{(j)}(t)\big\rangle \!\!&=&\!\!\! 
\sum_{\alpha,\beta \in \{ {\rm \tilde{c},\tilde{o}}\}} 
\big(\Gamma_{0\alpha;1\beta}^{(j)}  \,P_{1\beta}(t)
- \Gamma_{1\alpha;0\beta}^{(j)}\,P_{0\beta}(t) \big)
\nonumber \\
& &\qquad \qquad \times
\big(E_{1\beta}-E_{0\alpha}\big),\qquad
\label{Eq:energy-current}
\end{eqnarray}
An astute reader will spot that this is ill-defined unless we define
a zero of energy, because the energy current depends on the value of dot A's energy, $E_{\rm A}$, which in turn depends on what we define as the zero of energy \footnote{In this specific model the energy currents do not depend on the value of $E_{\rm o}$ or $E_{\rm c}$, because only the difference between them appears in the equations.}.
There are multiple possibilities commonly used in the literature.
Some take the zero of energy as the electro-chemical potential of the one of the reservoirs, by defining $\mu_{\rm L}=0$ or $\mu_{\rm R}=0$, 
so $E_{\rm A}$ is measured from the electro-chemical potential of the chosen 
reservoir.  
Others take the zero of energy as the mid-point between the two electro-chemical potentials, by defining $\mu_{\rm L}+ \mu_{\rm R}=0$.  This is the choice I make here
for the numerical calculations.

\section{Power and heat currents}

One of the easiest way to define the heat current is to split the energy current 
into work and heat, and define the heat as everything that is not work.

The work done to place a particle entering a reservoir $j$ is equal to its electrochemical potential.
Thus if particles enter reservoir $j$ at a rate given by Eq.~\ref{Eq:particle-current},
it gives the power being generated in reservoir $j$ as;
\begin{eqnarray}
\big\langle \dot{W}^{(j)}(t)\big\rangle \!\!&=&\!\!\! \mu_j
\sum_{\alpha,\beta \in \{ {\rm \tilde{c},\tilde{o}}\}}
\big(\Gamma_{0\alpha;1\beta}^{(j)}  \,P_{1\beta}(t)
- \Gamma_{1\alpha;0\beta}^{(j)}\,P_{0\beta}(t) \big),
\nonumber \\
\end{eqnarray}
One can then identify any other energy flowing into the reservoir as heat,
so 
\begin{eqnarray}
\big\langle \dot{Q}^{(j)}(t)\big\rangle \!\!&=&\!\!\! 
\big\langle \dot{E}^{(j)}(t)\big\rangle - \big\langle \dot{W}^{(j)}(t)\big\rangle
\nonumber \\
&=&\!\!\! 
\sum_{\alpha,\beta \in \{ {\rm \tilde{c},\tilde{o}}\}} 
\big(\Gamma_{0\alpha;1\beta}^{(j)}  \,P_{1\beta}(t)
- \Gamma_{1\alpha;0\beta}^{(j)}\,P_{0\beta}(t) \big)
\nonumber \\
& &\qquad \qquad \times
\big(E_{1\beta}-E_{0\alpha} -\mu_j\big),\qquad
\label{Eq:heat-current}
\end{eqnarray}
where I have used Eq.~(\ref{Eq:energy-current})

If one prepares the system in a given state, this current will depend on time
as the system decays to its steady-state, after which it will be time-independent.
The decay to the steady-state typically occurs on the time-scale of the tunnelling rates, and thus involves only the flow of a few electrons between reservoirs.
After that the flow between reservoirs will be determined by the system's steady-state for as long as the tunnelling occurs; hence this can involve the flow of a macroscopic number of electrons between reservoirs. 
Thus, we are interested in the average steady-state current, 
for which we substitute the steady-state level occupations given in Eqs.~(\ref{Eq:trapdoor_P_steady}) into all the equations in this section;
Eqs.~(\ref{Eq:particle-current}-\ref{Eq:heat-current}).

\section{Particle and energy conservation and the first law}

By construction, particles and energies are conserved at every step in the evolution. This conservation has clear consequences in the steady-state limit, in which the average energy and particle number in the trapdoor system is time independent.
It directly implies that the particle and energy currents into 
reservoir L are opposite to those into reservoir L;
\begin{eqnarray}
\Big\langle I^{(L)}_\steady\Big\rangle = - \Big\langle I^{(R)}_\steady\Big\rangle
\\
\Big\langle \dot{E}^{(L)}_\steady\Big\rangle = - \Big\langle \dot{E}^{(R)}_\steady\Big\rangle
\end{eqnarray}
Writing the second equality in terms of heat and work gives
\begin{eqnarray}
\Big\langle \dot{W}^{(L)}_\steady + \dot{W}^{(R)}_\steady\Big\rangle &=& -
\Big\langle \dot{Q}^{(L)}_\steady+ \dot{Q}^{(R)}_\steady\Big\rangle
\nonumber \\
\label{Eq:rate-eq-1stlaw}
\end{eqnarray}
so the total work production rate is equal and opposite to the total rate of change of heat.
This is the first law of thermodynamics, saying that the sum of heat and work is conserved. 

The total work production rate is
\begin{eqnarray}
\Big\langle \dot{W}^{\rm total}_\steady\Big\rangle &\equiv&
\Big\langle \dot{W}^{(L)}_\steady+ \dot{W}^{(R)}_\steady\Big\rangle 
\nonumber\\
&=& (\mu_{\rm R}-\mu_{\rm L})\,\Big\langle I^{(R)}_\steady\Big\rangle. \qquad
\end{eqnarray}
So when reservoir R has a higher electrochemical potential than reservoir L, a particle current into reservoir R is tuning heat into work.  In contrast, a particle current flowing  
out of reservoir R is turning work into Joule heating.
In the latter case, the quantum system between reservoirs L and R is acting as a resistive element.

\section{Second law of thermodynamics}
\label{Sect:rate-eqn-2nd law}

In the steady-state, the entropy of the trapdoor system does not change on average, only the average entropies in the reservoirs change.
Using the Clausius form for the entropy in each reservoir,
the average total entropy change in the steady-state  is 
\begin{eqnarray}
\Big\langle \dot{S}^{\rm total}_\steady \Big\rangle
&=& \frac{1}{T_{\rm L}}\Big\langle \dot{Q}^{(L)}_\steady\Big\rangle \,+\, \frac{1}{T_{\rm R}} \Big\langle \dot{Q}^{(R)}_\steady\Big\rangle. \qquad
\label{Eq:rate-eqn-entropychange}
\end{eqnarray}

The remarkable thing is that, one can perform an analytic proof that the system will always obey the second law of thermodynamics, whenever
\begin{itemize}
\item[(i)] the reservoir entropy changes are assumed to take the Clausius law, as in Eq.~(\ref{Eq:rate-eqn-entropychange}), and 
\item[(ii)]  heat currents are given by a set of rate-equations with rates given by the local detailed balance formula in Eq.~\ref{Eq:local-detailed-balance}
\end{itemize}
This proof was given in Ref.~\cite{VandenBroeck2013}, and reproduced in section 8.8 of Ref.~\cite{Benenti2017Jun}, so I do not give it here.

In the case of interest to us, where reservoirs L and R are at the same temperature, $T_0$, we can use Eqs.~(\ref{Eq:rate-eq-1stlaw},\ref{Eq:rate-eqn-entropychange}) to write
\begin{eqnarray}
\Big\langle \dot{W}^{\rm total}_\steady \Big\rangle = - T_0 
\Big\langle \dot{S}^{\rm total}_\steady \Big\rangle
\end{eqnarray}
Thus, now we know the second law is obeyed, we see that the total work production rate can never be positive.

A second remarkable thing about the proof that these rate equations always obey the second law, is how transparent it is when formulated in terms of trajectories in 
stochastic thermodynamics.  This is noticeable in Ref.~\cite{VandenBroeck2013}, 
which gave the more pedestrian proof (reproduced in section 8.8 of Ref.~\cite{Benenti2017Jun}) followed by Schmiedl and Seifert's \cite{Schmiedl2007Jan} proof using stochastic thermodynamics
(reproduced in section 8.10.5 of Ref.~\cite{Benenti2017Jun}).  In the pedestrian proof one is making algebraic manipulations without much intuitive meaning. In contrast, in the stochastic thermodynamics proof, one is comparing probabilities of physical processes
with the probabilities of their time-reverse, using the local detail balance formula in Eq.~(\ref{Eq:local-detailed-balance}).
This makes complete intuitive sense when one recalls that entropy is a measure of irreversibility, so should clearly be related to the difference in probabilities between a processes and its time-reverse.
With rather little algebra, it leads to the integral fluctuation theorem, from which the second law automatically follows.
Thus the physical origin of the second law is much more transparent if one is willing to make the effort of learning to handle the trajectories of stochastic thermodynamics. 

\section{The trapdoor as a diode}

\begin{figure}
\centerline{\includegraphics[width=0.95\columnwidth]{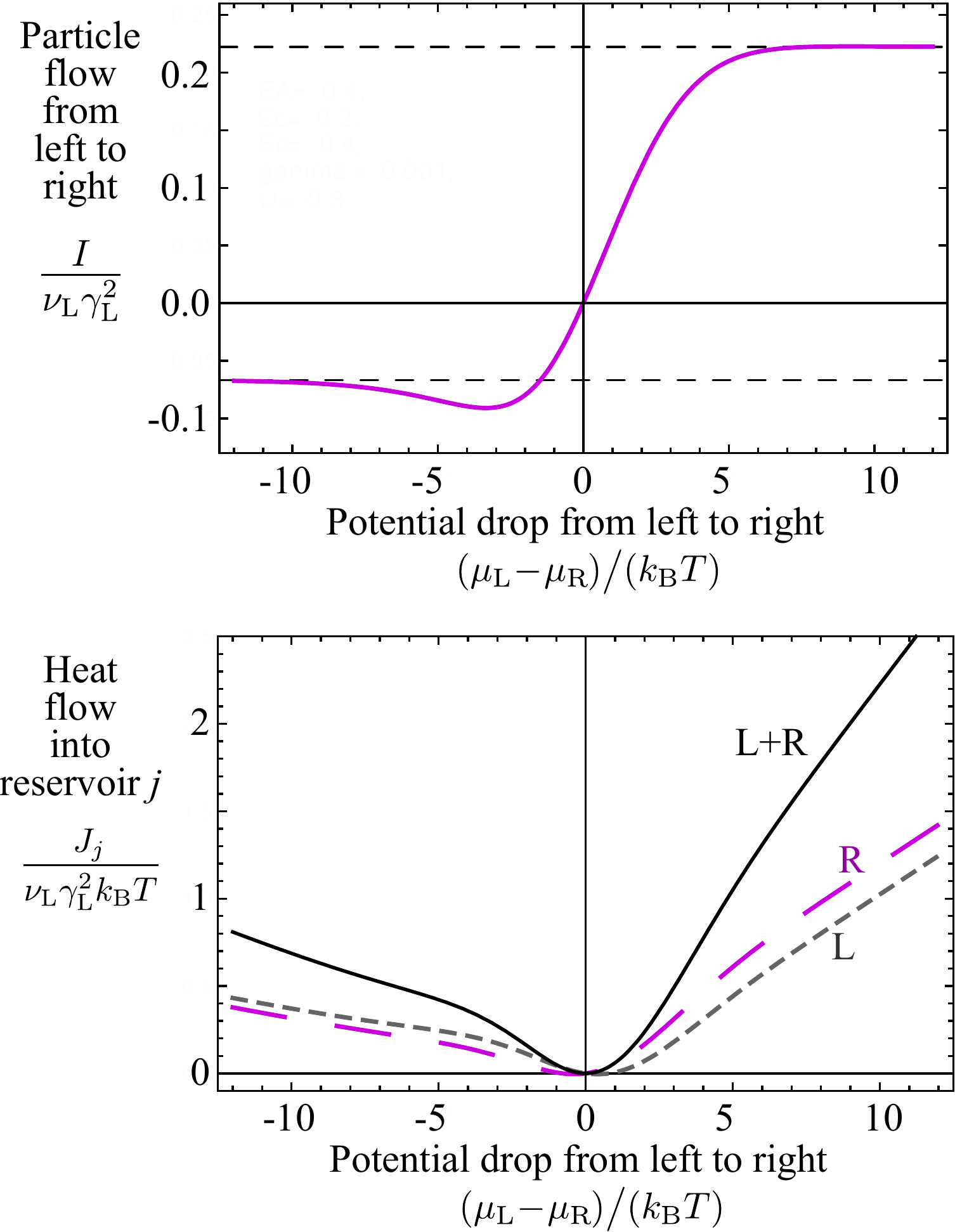}}
\caption{\label{fig:trapdoor-as-diode} 
An example showing that the particle current (a) flows down hill through the trapdoor (from the reservoir with higher electro-chemical potential the one with lower electro-chemical potential), and (b) that this injects heat into reservoirs L and R.  The trapdoor is not acting as a demon, but it is acting as a diode, in which the flow from left to right is stronger than flow from right to left. 
The heat injected into L (dashed grey curve) is different from that generated   in R (long-dashed red curve), but their sum (black solid curve) is simply equal to the Joule heating $(\mu_{\rm L}-\mu_{\rm R})I$.
Here each reservoir's tunnel coupling to dot A has the same magnitude $\gamma_{\rm L}^2\nu_{\rm L}=\gamma_{\rm R}^2\nu_{\rm R}$, and 
$\gamma_{\rm L}$, $\gamma_{\rm R}$, $\nu_{\rm L}$ and $\nu_{\rm R}$ are energy independent.  In this example, I take the tunnelling between levels c and o, $\gamma$, to be much smaller than all other parameters in the trapdoor system; with
$\gamma=0.001 \,k_{\rm B}T$, $E_{\rm A}=0.4 \,k_{\rm B}T$, $E_{\rm c}=0.2\,k_{\rm B}T$, $E_{\rm c}=0.4\,k_{\rm B}T$, $U=0.3\,k_{\rm B}T$. 
} 
\end{figure}

Having explained that the trapdoor system cannot be a perpetual-motion machine, one can ask if it has any function at all.
We recall that the trapdoor was supposed to be an example of the non-return valve imagined by Maxwell (see section~\ref{Sect:intro-Maxwell}).  
A non-return valve for electrical flows is called a diode or rectifier.  So does the trapdoor act as a diode, even if it cannot be a perpetual-motion machine?

The answer is that it is indeed a simple example of a diode, arguably about the  simplest example of a diode.  It is typically a rather poor quality diode,
in which there is not a particularly large asymmetry between flows in the two directions.
That is to say that the current always goes in the direction induced by the potential drop, so electrical power is always dissipated as Joule heating, but the current is larger when the potential drops from left to right, than when it drops from right to left. 

The difference between rectifiers and Maxwell demons, was address by Brillouin \cite{Brillouin1950Jun} who argued that a rectifier could not extract work from the thermal (Nyquist) noise in a resistor. 
Similarly, Feynman \cite{FeynmanLecturesVolOne} showed that his ratchet could act as a rectifier, even though it could not extract work from the thermal fluctuations of a gas.

An example of our two-level trapdoor acting as a diode is shown in 
Fig.~\ref{fig:trapdoor-as-diode}a. There one sees that the particle current from reservoir L to reservoir R
always has the same sign as $\mu_{\rm L}-\mu_{\rm R}$, so the flow is always down hill.  However, the magnitude of the current is larger when $\mu_{\rm L}>\mu_{\rm R}$ than when $\mu_{\rm L}<\mu_{\rm R}$.  
In Fig.~\ref{fig:trapdoor-as-diode}b, one sees that this process heats both reservoirs L and R, and the total heat production is always positive.
This is as it must be, because it is a consequence of the second law.

\section{Getting work by damping out fluctuations}
\label{Sect:damped-trapdoor}

If fluctuations of the two-level trapdoor stop it from working as a perpetual-motion machine, what happens if these fluctuations are damped out?

In the above model, the only damping of the trapdoor's motion comes from
its interaction with the electrons flowing through dot A; the same electrons that are exciting the trapdoor's motion.  Thus here I ask what happens if we add another mechanism to damp the trapdoor's motion.  

I show that this extra damping mechanism can reduce the oscillations of the trapdoor. Then more electrons to go from left to right than the reverse, even if the left reservoir has a lower electrochemical potential than the right one. So it is converting heat in the electronic reservoirs into work (in the form of electrical power).

Is this a perpetual-motion machine? No, the extra damping mechanism requires that the trapdoor is coupled to a reservoir which is colder than the electrons.  So that each time the trapdoor is excited by the electrons, it dissipates this heat to this cold reservoir. This means there is a steady-state heat flow into the cold reservoir. Once this heat flow s taken into account, we see that the second law is restored. The trapdoor system is simply acting as a heat engine, extracting useful work from a heat flow from hot reservoirs (the electron reservoirs) to new the cold reservoir (introduced to damp the trapdoor's motion), without violating any law of thermodynamics.

\subsection{Damping the trapdoor with phonons}
\label{Sect:trapdoor+phonons}

The additional damping mechanism taken here is a weak coupling of the two-level trapdoor to phonons in some reservoir that can be at a lower temperature from the electron reservoirs.
This is most of a thought experiment, as it could be hard to implement experimentally. It might be achieved by have the phonon reservoir being that of the  substrate, and the electrons in the reservoirs being heated to a higher temperature than the substrate by Joule heating (by passing currents through these reservoirs).
At low enough temperatures, the electron-phonon coupling in the reservoirs
would be weak-enough to allow the phonons to remain colder than the electrons.
However, it would then be challenging to get a strong enough coupling between the phonons and the electron in the two-level trapdoor.

The physical mechanism that I have in mind is 
the following; the two-level trapdoor can hop between states by emitting or absorbing a phonon.  If those two levels were not coupling the anything else
(i.e. not coupled to dot A), then this coupling would make the occupation of the two-level system relax to a thermal state in equilibrium with the phonon reservoir.
I model this effect phenomenological by taking the 
Hamiltonian ${\cal H}_{\rm sys}+{\cal H}_{\rm res}$ 
given by Eqs.~(\ref{Eq:Hsys},\ref{Eq:Hres}) and 
adding the following terms
\begin{eqnarray}
 {\cal H}_{\rm phonon} \!\!&=&\!\!\! 
 \sum_k \omega_k a_{{\rm ph};k}^\dagger a_{{\rm ph};k} +\ 
 \gamma_{\rm ph} \left(a_{{\rm ph};k}^\dagger + a_{{\rm ph};k}\right)
 \ \  \nonumber
 \\
 & &\hskip 3.8cm
 \times\left(\hat{d}_{\rm o}^\dagger\hat{d}_{\rm c} - \hat{d}_{\rm c}^\dagger\hat{d}_{\rm o} \right)
\nonumber
\\
\end{eqnarray}
where the first term in the $k$-sum is the Hamiltonian for the phonon reservoir, whose $k$th mode has energy $\omega_k$, and the second term is the coupling to the difference in the charge distribution of levels c and o.

\begin{figure}[t]
\centerline{\includegraphics[width=0.75\columnwidth]{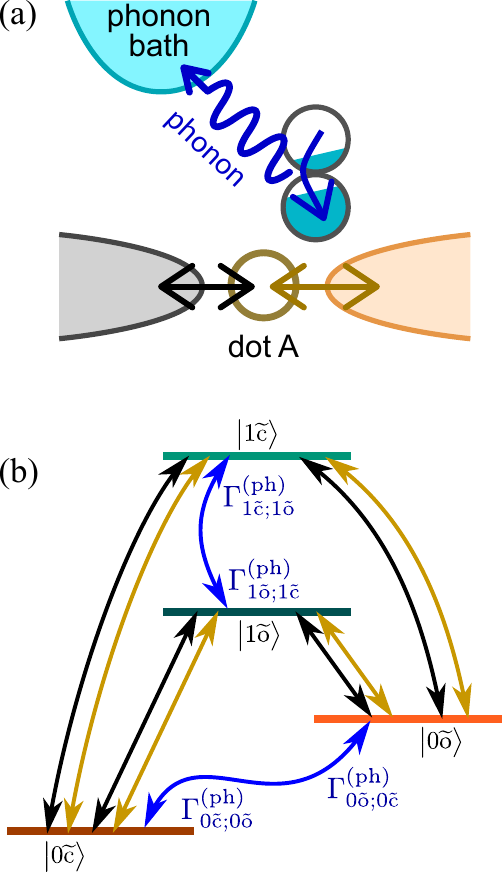}}
\caption{\label{fig:trajectories-with-phonons} 
(a) The two-level trapdoor with an extra bath of phonons to provide a relaxation mechanism to damp out the fluctuations of the trapdoor. 
(b) Blue arrows show the effect of the phonon reservoir, inducing transitions between the many body-eigenstates.
The black and brown arrows are those already shown in Fig.~\ref{fig:trajectories} for transitions induced by the coupling to the L and R electron reservoirs.
Each phonon-induced transition occurs at different rates in each direction; the rate in a given direction is noted next to the relevant arrowhead.
The rates in the two directions on a given arrow obey local detailed balance, as in Eq.~(\ref{Eq:local-detailed-balance-phonons}). 
} 
\end{figure}

%

If this trap-door system's coupling to this phonon reservoir is treated in the same weak-coupling approximation 
as its coupling to the electronic reservoirs, then 
it adds the transitions between many-body eigenstates shown as blue arrows in
Fig. ~\ref{fig:trajectories-with-phonons} (where the black and brown arrows are transitions induced by the electron reservoirs already shown in Fig.~\ref{fig:trajectories}).
This is equivalent to adding a term to the right of the evolution equation in 
Eq.~(\ref{Eq:P-evolution-Eq}) of the form
\begin{eqnarray}
+
\left(\!
\begin{array}{cccc}
\!-\Gamma_{\rm 0\tilde{o};0\tilde{c}}^{\rm (ph)}\!  & \!\Gamma_{\rm 0\tilde{c};0\tilde{o}}^{\rm (ph)}  & 0  & 0  
\\
\Gamma_{\rm 0\tilde{o};0\tilde{c}}^{\rm (ph)}\!  &  \!\!-\Gamma_{\rm 0\tilde{c};0\tilde{o}}^{\rm (ph)}  & 0 & 0 
\\
0  & 0  &  \!\!- \Gamma_{\rm 1\tilde{c};1\tilde{o}}^{\rm (ph)}\!   & \!\Gamma_{\rm 1\tilde{o};1\tilde{c}}^{\rm (ph)} 
\\
0  & 0  & \!\Gamma_{\rm 1\tilde{c};1\tilde{o}}^{\rm (ph)} \!  & \!\!-\Gamma_{\rm 1\tilde{o};1\tilde{c}}^{\rm (ph)} 
\end{array}\!\right) 
\left(\!
\begin{array}{c} 
P_{\rm 0\tilde{c}} \\ 
P_{\rm 0\tilde{o}} \\
P_{\rm 1\tilde{o}} \\
P_{\rm 1\tilde{c}}
\end{array}\!\right),
\nonumber \\
\label{Eq:extra-phonon-term}
\end{eqnarray}
The transition rates are the following
\begin{subequations}
\begin{eqnarray}
\Gamma_{\rm 0\tilde{o};0\tilde{c}}^{\rm (ph)}
 &=& \big|\gamma_{\rm ph}\big|^2 
\cos^2 \theta_0 \ \nu(E_{\rm 0\tilde{o}}-E_{\rm 0\tilde{c}}) 
\nonumber \\
& & \hskip 1cm \times 
n_{\rm ph}(E_{\rm 0\tilde{o}}-E_{\rm 0\tilde{c}})
\\
\Gamma_{\rm 0\tilde{c};0\tilde{o}}^{\rm (ph)}
 &=& \big|\gamma_{\rm ph}\big|^2 
\cos^2 \theta_0 \ \nu(E_{\rm 0\tilde{o}}-E_{\rm 0\tilde{c}}) 
\nonumber \\
& & \hskip 1cm \times 
\left(n_{\rm ph}(E_{\rm 0\tilde{o}}-E_{\rm 0\tilde{c}}) +1 \right)\qquad
\\
\Gamma_{\rm 1\tilde{c};1\tilde{o}}^{\rm (ph)} 
&=& \big|\gamma_{\rm ph}\big|^2 
\cos^2 \theta_0 \ \nu(E_{\rm 1\tilde{c}}-E_{\rm 1\tilde{o}}) 
\nonumber 
\\ 
& & \hskip 1cm \times n_{\rm ph}(E_{\rm 1\tilde{c}}-E_{\rm 1\tilde{o}})
\\
\Gamma_{\rm 1\tilde{c};1\tilde{o}}^{\rm (ph)} 
&=& \big|\gamma_{\rm ph}\big|^2 
\cos^2 \theta_0 \ \nu(E_{\rm 1\tilde{c}}-E_{\rm 1\tilde{o}}) 
\nonumber 
\\ 
& & \hskip 1cm \times \left(n_{\rm ph}(E_{\rm 1\tilde{c}}-E_{\rm 1\tilde{o}})
+1\right) 
\end{eqnarray}
\end{subequations}
where $\gamma_{\rm ph}$ is the electron-phonon coupling, and $\nu(E)$ is the phonon density of states at energy $E$. Here, $n_{\rm ph}(E)$ is the number of phonons at energy $E$, given by the Bose distribution,
\begin{eqnarray}
n_{\rm ph}(E)= \Big(\exp\big[E\big/T_{\rm ph}\big]-1\Big)^{-1}
\end{eqnarray}
This $n_{\rm ph}(E)$ appears in the transition rates for the trapdoor system to absorb a phonon,
while $\big(n_{\rm ph}(E)+1\big)$ appears in transition rates for the trapdoor system to emit a photon. The $n_{\rm ph}(E)$  in this factor of \mbox{$\big(n_{\rm ph}(E)+1\big)$} comes from stimulated emission, while the +1 comes from spontaneous emission. This means that there is always one more way for a system to emit a phonon than to absorb a phonon.
As $\big(n_{\rm ph}(E)+1\big)$ is equal to $n_{\rm ph}(E)\exp\big[E\big/T_{\rm ph}\big]$,
we have local detailed balance for the phonons
\begin{eqnarray}
\Gamma^{\rm (ph)}_{\rm \alpha\tilde{o};\alpha\tilde{c}}   
\ =\ 
\Gamma^{\rm (ph)}_{\rm \alpha\tilde{c};\alpha\tilde{o}}   
\ \exp\left[-\frac{E_{\rm \alpha\tilde{o}}-E_{\rm \alpha\tilde{c}}}{T_{\rm ph}} \right],
\label{Eq:local-detailed-balance-phonons}
\end{eqnarray}
for $\alpha =0,1$.
This is the phonon equivalent of the relation for electrons in Eq.~(\ref{Eq:local-detailed-balance}).

\begin{figure}
\centerline{\includegraphics[width=0.95\columnwidth]{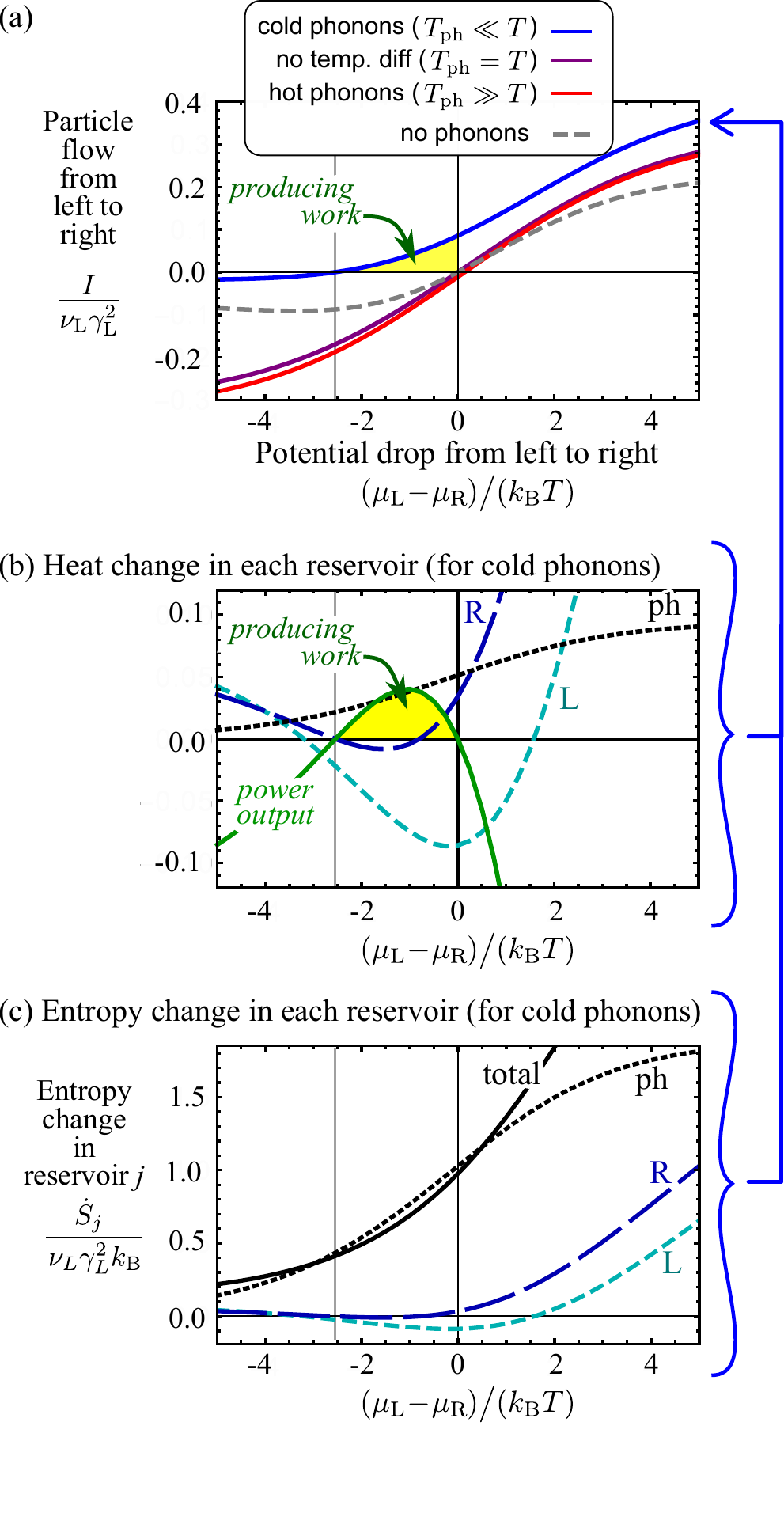}}
\caption{\label{fig:trapdoor-phonons} 
(a) The particle current when the two-level trapdoor is coupled to phonon reservoir at temperature $T_{\rm ph}$, while the electron reservoirs are at temperature $T$.  For $T_{\rm ph}=0.05T$ (blue curve), the electron-flow is producing work in the region marked in yellow.  
For $T_{\rm ph}=T$ (purple curve), no work is produced. 
For $T_{\rm ph}=30T$ (red curve), there is a small region (not really visible in this plot) where the electron-flow produces work, because the curve passes slightly below the origin.
Here the coupling to the phonon reservoir, $\gamma_{\rm ph}=3\gamma_{\rm L}$, and the other parameters are as in Fig.~\ref{fig:trapdoor-as-diode}.
(b) The heat change in the reservoirs for $T_{\rm ph}=0.05T$, and the power output (green).
(c) The entropy changes in the reservoirs for $T_{\rm ph}=0.05T$. It shows that the region where work is produced involves the entropy in the electrons reservoirs going down, but the entropy in the phonons is going up by a larger amount, so the total entropy change is positive.
} 
\end{figure}

The blue curve in Fig.~\ref{fig:trapdoor-phonons}a corresponds to dynamics in which the trapdoor's motion is damped by the cold phonon reservoir, we see it has 
a region (marked in yellow) where it is producing work as the electron current flows against the bias (electrons flowing up-hill). Considering Fig.~\ref{fig:trapdoor-phonons}a alone, it looks like adding a relaxation process is enough to cause work to be produced {\it from nothing!}  
However, the relaxation dynamics come because the trapdoor is coupled to a phonon reservoir that is colder than the electrons.
We can see in Fig.~\ref{fig:trapdoor-phonons}b that heat is flowing into the cold phonon reservoir, and out of one (or both) of the hot electron reservoirs.
From the first law, we have that the power output plus the heat output into each wire is zero, and we see by eye that this is the case. 
Fig.~\ref{fig:trapdoor-phonons}c plots the rate of entropy change in each reservoir; given by
$\dot{S}_j=J_j/T_j$ where $J_j$ is the heat flow into reservoir $j$, 
and $T_j$ is the temperature of reservoir $j$. 
It shows that the region where work is produced is a region where the entropy in the electrons (sum of entropy in L and R reservoirs) is going down, but the entropy in the phonons is going up by a larger amount. 
So the total entropy change is positive.
We see that the machine is below Carnot efficiency
because  total entropy change is non-zero everywhere in the region of positive power output.

If we then look at the case where the phonon reservoirs is 
at exactly the same temperature as the electronic reservoirs, 
the purple curve goes exactly through the origin in the current-bias plane in Fig.~\ref{fig:trapdoor-phonons}a, meaning the current is never against the bias, so electrons only flow down-hill.  Thus no work production is possible. 

When the phonons are hotter than the electrons, then there is a heat flow
from phonons to electrons, and the trapdoor system can act as a heat engine turning heat into work. We could see this if we zoomed on the red-curve
in Fig.~\ref{fig:trapdoor-phonons}a; it passes slightly below the origin,
which means that there is a small region---at positive but small $(\mu_{\rm L}-\mu_{\rm R})$---in which the electrons flow up-hill. In other words,
the current is against the bias.  We note from Fig.~\ref{fig:trapdoor-phonons} that the trapdoor does not function very well as this sort of heat engine.  This is because the parameters were chosen to demonstrate a significant power output when the phonons are cold, other parameter regimes are likely to show more power output when the phonons are hot.

\subsection{Maxwell demon or heat engine?}
\label{Sect:demon-or-heatengine}

I conclude this chapter by asking if the trapdoor with coupling to a cold phonon reservoir is a Maxwell demon or a heat-engine.
This raises the question of whether one can make a clear distinction between 
this Maxwell demon and a heat engine.

The trapdoor with its cold phonon reservoir is similar to the Feynman ratchet in which 
the pawl is allowed to relax by being thermally coupled to a gas at a colder temperature than that of the gas causing the fluctuations.
Feynman described this as a heat engine \cite{FeynmanLecturesVolOne}, and this seems completely reasonable at the 
macroscopic level.
One could equally call it a cold engine, because the resource it is exploiting is the ``cold'' of the cold reservoir\footnote{For example, the fuel could be a block of ice, with the engine generating power by heating-up the ice. Once the ice has melted and reached ambient temperature, the cold resource has been consumed and no more power can be generated without replacing it with a new block of ice.}. This changes the thermodynamic formulas a bit compared to a heat engine (see section 9.6 of Ref.~\cite{Benenti2017Jun}), but it remains conceptually the same as a heat engine.

However, when you look at the system microscopically, it seem to have the 
requirements for a Maxwell demon. The system is measuring particle by particle, and opening or closing a trapdoor depending on the results of the measurement.
This view point comes directly from Ref.~\cite{Strasberg2013Jan}, in the context of a slightly different autonomous Maxwell demon that I discuss in Chapter~\ref{Chap:experiments}.
So maybe it is best to say that this is simply a heat-engine whose microscopic mechanism corresponds to a Maxwell demon.

However, it was recently proposed \cite{Freitas2021Mar} that we could 
clarify the situation be tightening the definition of the Maxwell demon.
They proposed that a system can only be categorized as a ``true'' or ``strict'' Maxwell demon, if it {\it never} exchanges energy with the system it acts upon. 
This is a very logical definition, given that the ideal version of Maxwell's original demon
does not change the energy of the particles in the working fluid when it observes them, nor when it opens and closes the door.  
We have seen that it is not fulfilled by the two-level trapdoor system, when it is generating power (since it is extracting energy from the electrons and dumping it in the cold phonon reservoir). 
Within that categorization, the two-level trapdoor system is not a ``true'' Maxwell demon, it is simply a heat-engine. 
Indeed, this is the case for most experimental systems described as Maxwell demons in the literature (including the one in  Ref.~\cite{Strasberg2013Jan}), as few satisfy the condition of {\it never} exchanging energy with the system it acts upon. 
We will discuss this elegant, but experimentally challenging, definition more fully in section~\ref{Sect:Freitas}.

\chapter{Why a trapdoor alone creates no asymmetry}
\label{Chap:Understanding-trapdoor}

In the previous chapter, we took the two-level trapdoor model 
and solved its quantum dynamics, proving that
when the two-electronic reservoirs are at the same temperature, the two-level trapdoor
equilibrates in a thermal state at that temperature. Then every transition
in the trapdoor system occurs with the same probability as its time-reverse.
This means we have proven that despite is asymmetric nature, it causes strictly no asymmetry
between the probability of electron flow from left to right and the probability of electron flow from right to left.
This is why it cannot violate the second-law of thermodynamics.

\begin{figure}
\centerline{\includegraphics[width=0.35\columnwidth]{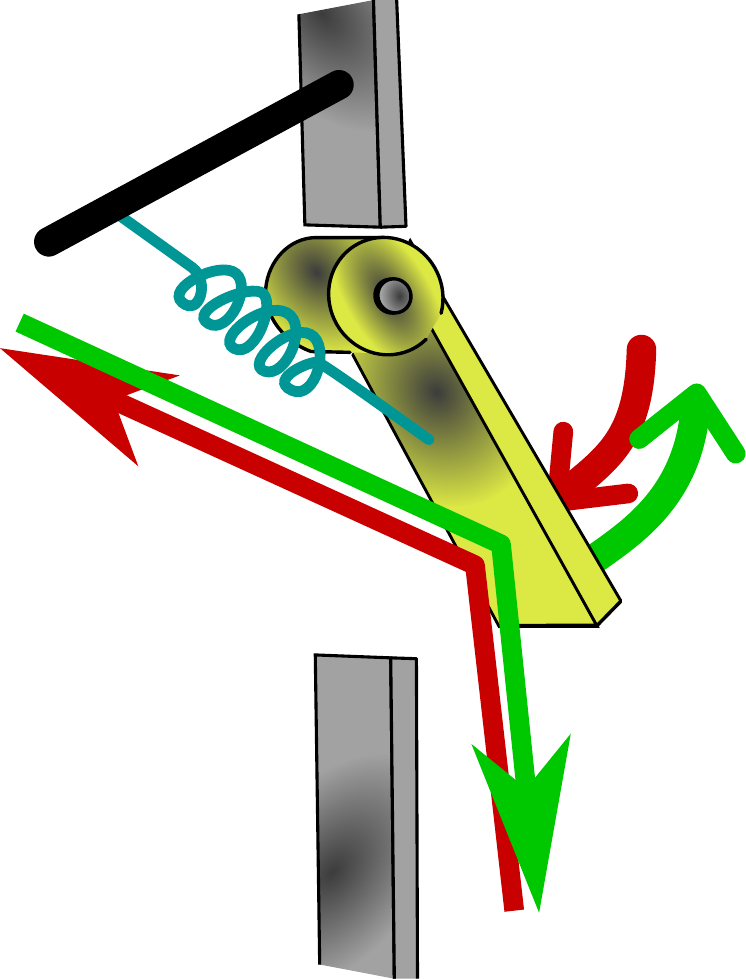}}
\caption{\label{fig:trapdoor-why-it-does-not-work} 
To understand why the trapdoor does not act as a demon in the manner sketched in Fig.~\ref{fig:trapdoor}, we note that for every trajectory marked by green arrows, there is its time-reverse marked by red arrows. 
The green arrows show a particle going from left to right, by pushing the trapdoor open.
The red arrows show the trapdoor pushing a particle from right to left. The existence of trajectories like the red arrows tells us that the trapdoor not only stops some particles going from right to left (as was sketched in Fig.~\ref{fig:trapdoor}a), its thermal motion also means it sometimes pushes particle
from right to left. We will argue here that this stops the trapdoor inducing an asymmetry between the probability of a particle flowing from left to right, and the probability of a particle flowing from right to left.} 
\end{figure}

To get a good intuitive feel for this statement, it is worthwhile making the analogy back to 
the original trapdoor considered by Maxwell and Smoluchowski (see section~\ref{sect:intro-trapdoor}).  We will only make this analogy at a handwaving level here, using the sketch in Fig.~\ref{fig:trapdoor-why-it-does-not-work}, although we expect the analogy could be made rigorous with a suitable (classical or quantum) calculation for a trapdoor of the type in Fig.~\ref{fig:trapdoor-why-it-does-not-work}.  

I can summarize the handwaving analogy between our proof for the two-level
trapdoor, and a trapdoor of the type in  Fig.~\ref{fig:trapdoor-why-it-does-not-work} as follows. When the gases to the left and right of the trapdoor are at the same temperature,
then the trapdoor will equilibrate at that temperature. Then we can expect that
there is the same probability for any process and its time-reverse. This is why the trapdoor will not create an asymmetry in the probability of a particle flowing from left to right and the probability of a particle flowing from right to left. This is why it will not violate the second law.

The examples of the green and red processes sketched in 
 Fig.~\ref{fig:trapdoor-why-it-does-not-work}, are particularly enlightening.
The original intent of the trapdoor was create an asymmetry by stopping some particles going from right to left, however we see that its thermal motion also sometimes pushes particles from right to left (the red arrow). The above argument tells us the simplest way to see the trapdoor's net effect is {\it not} to 
try to calculate the probability it stops particles coming from the right (like the trajectory in Fig.~\ref{fig:trapdoor}a), and compare that to the probability it actively pushes particle from right to left (like the red trajectory in  Fig.~\ref{fig:trapdoor-why-it-does-not-work}).
Instead, we should compare trajectories with their time-reverses.  If every trajectory involving a particle going from left to right occurs with the same probability as the time-reverse trajectory, then it is clear that the trapdoor causes no asymmetry on the flow, and the second law cannot be violated.

If, as I argue, the probability of the green and red trajectories in Fig.~\ref{fig:trapdoor-why-it-does-not-work} are equal in equilibrium, then if we slightly increase the pressure of the gas on the left,
then there will be a slightly higher chance of a particle arriving from the left along the trajectory marked in green, than a particle arriving from the right along the trajectory marked in red. Then the net flow will be from right to left, meaning that the flow is always from higher to lower pressure in agreement with the second law.

\chapter{Demonic experiments}
\label{Chap:experiments}

If we take the trapdoor system with the phonon bath described at the end of the previous chapter (shown in Fig.~\ref{fig:trajectories-with-phonons}a), and simplify it a little bit, then we get a system that has been studied experimentally, both as a demon \cite{Koski2015Dec}
and as a heat-engine \cite{Roche2015Apr,Hartmann2015Apr,Thierschmann2015Aug,Thierschmann2016Dec}.
These experiments were extremely challenging, because the control of heat flow and the measurement of temperatures at the nanoscale are both extremely difficult.  The experimentalists used clever thermometry techniques like noise thermometry, and often added superconducting elements to block heat flows. However, here I will (unfairly) gloss over these issues to concentrate on the demonic physics they observe. 

\begin{figure}[t]
\centerline{\includegraphics[width=0.75\columnwidth]{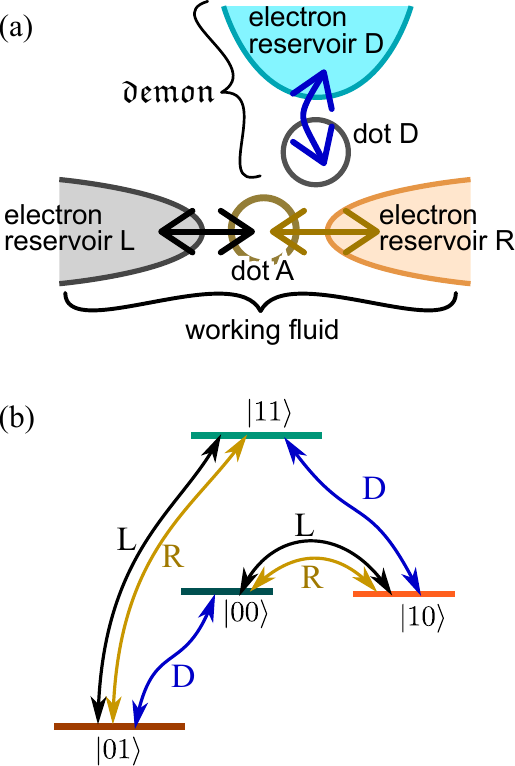}}
\caption{\label{Fig:3term} 
(a) A simplified model of the three terminal experimental devices, studied independently in four different teams \cite{Koski2015Dec, Roche2015Apr,Hartmann2015Apr,Thierschmann2015Aug,Thierschmann2016Dec}.
(b) The transitions in the theoretical model of Refs.~\cite{Sanchez2011Feb,Strasberg2013Jan}, with the levels labelled $|n_{\rm A}\ n_{\rm D}\rangle$, where $n_{\rm A}$ and $n_{\rm D}$ are the occupations of dots A and D. The experimental systems are similar to these models, but are more complicated because they involved a pair of metallic grains (rather than quantum dots or molecules) each with many levels rather than just one (see section~\ref{sect:metal}).
} 
\end{figure}

These experimental systems are analogous to taking the trapdoor in Fig.~\ref{fig:trajectories-with-phonons}a, and replacing the double-dot and phonon reservoir with a single dot coupled to an electronic reservoir, as sketched in Fig.~\ref{Fig:3term}a.  This slightly simplifies the physics \cite{Sanchez2011Feb,Sothmann2012May,Strasberg2013Jan}. In the two-level trapdoor case, the two-level trapdoor exhibited damped coherent oscillations (with the damping coming from the phonon bath).  In this new three-terminal set-up, there are no coherent oscillations, if dot D is put is some arbitrary state, it simply decays back to equilibrium with reservoir D without coherent oscillations.

Like with the system in Fig.~\ref{fig:trajectories-with-phonons}a, there is no effect if all three reservoirs are in equilibrium with each other. 
However, Ref.~\cite{Strasberg2013Jan} showed theoretically that if reservoir D is colder than the others, then it allows dot D and reservoir D in combination to act as a demon, detecting if there is an electron in dot A and modify the tunnelling between dot A and reservoir R in consequence.
It can thereby extract entropy from the working fluid (made of dot A and reservoirs L,R),
while generating electrical power by driving a particle current from L to R against its natural direction of flow (from reservoir L with lower electrochemical potential to reservoir R with higher electrochemical potential). 
This system was experimentally studied in Ref.~\cite{Koski2015Dec}, but they did not follow the exact protocol given in Ref.~\cite{Strasberg2013Jan}. 
Instead of observing a power production, they observed a sort of demonic cooling which we will discuss in section~\ref{Sect:Pekola-expt} below.

\begin{figure}[t]
\centerline{\includegraphics[width=0.9\columnwidth]{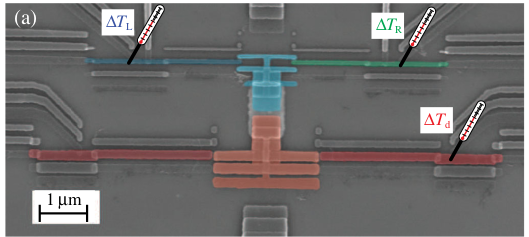}}
\centerline{\includegraphics[width=0.9\columnwidth]{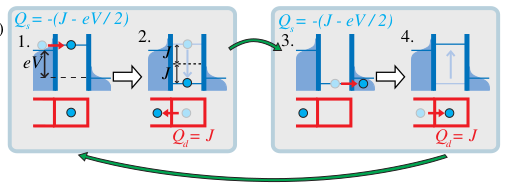}}
\caption{\label{fig:Koski-Pekola} 
Taken from Ref.~\cite{Koski2015Dec}. (a) A scanning
electron micrograph of the experimental structure. (b) Their diagram of how it works
as a demon (we provide a more detailed explanation of this below).
} 
\end{figure}

The same type of set-up was also shown experimentally to generate electrical power when reservoir D is hotter than the other reservoirs (much like the two-level trapdoor with phonons in Section~\ref{Sect:trapdoor+phonons}), so it acts as a heat engine.  This power production was observed simultaneously in three groups \cite{Roche2015Apr,Hartmann2015Apr,Thierschmann2015Aug,Thierschmann2016Dec}.\footnote{In Ref.~\cite{Hartmann2015Apr} the heating of reservoir D was replaced by a randomly fluctuating voltage, so it should also be considered a rectifiers that convert fluctuating voltages into a dc power output.}   
In principle, one simply needs to reverse the temperature gradient (keep reservoir D cold, while heating reservoirs L and R) to observe the demonic action producing power.
Of course, this raises the question of whether this system would be a demon or simply a thermodynamic engine that exploits cold rather than heat. We addressed this question in 
section~\ref{Sect:demon-or-heatengine} in the context of the two-level trapdoor, and it is the same here.

\begin{figure}[t]
\centerline{\includegraphics[width=0.6\columnwidth]{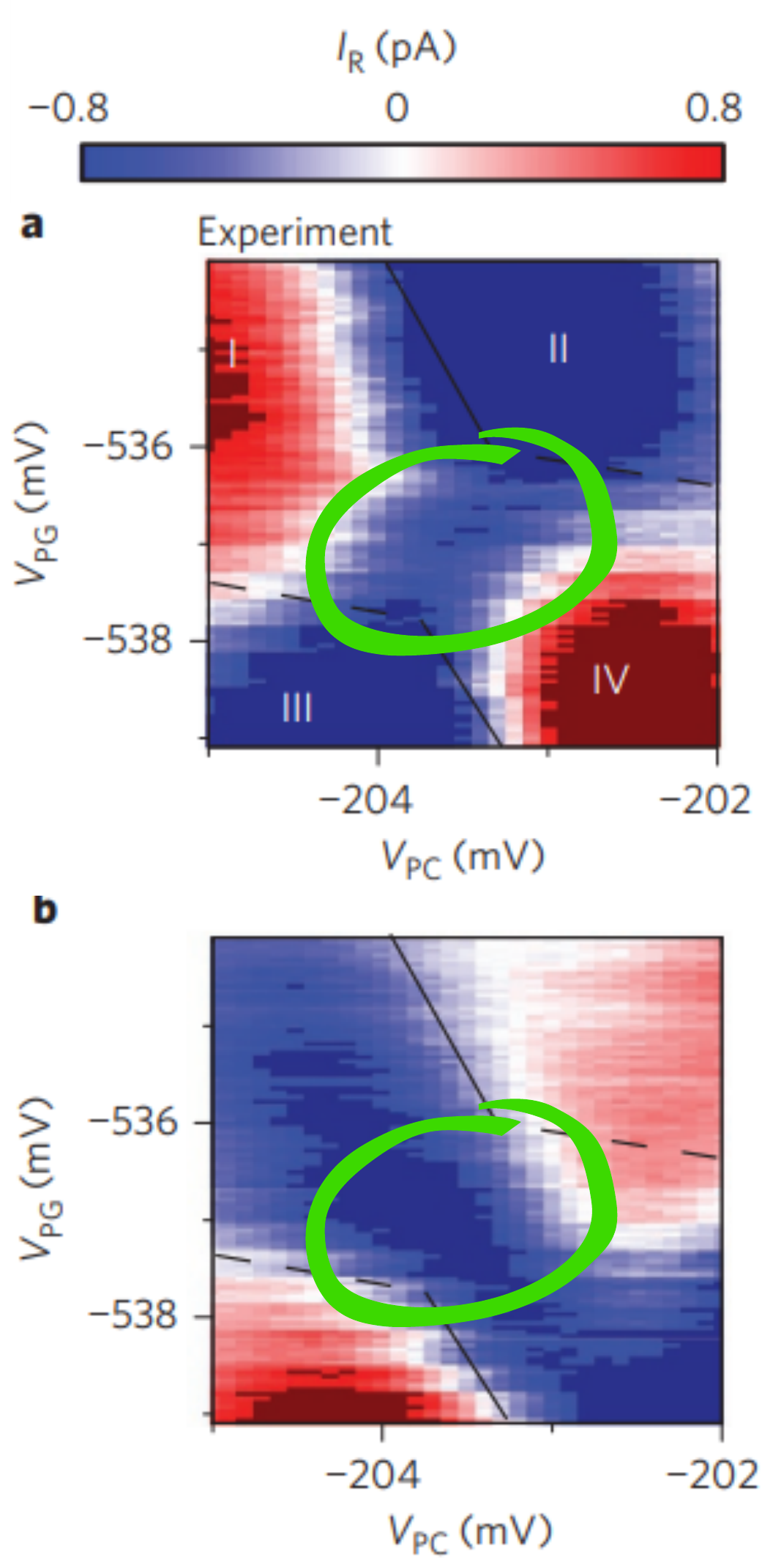}}
\caption{\label{Fig:Thierschmann-expt} 
Taken from Ref.~\cite{Thierschmann2015Aug}. The two plots are the current from reservoir L to reservoir R, for (a) positive and (b) negative biases applied between L and R.
In much of the parameter regime, one sees a standard behaviour, where the current flows from high to low bias, and thus changes sign when the bias changes sign.
However in the parameter regime highlighted in green, the current is negative, irrespective of the sign of the bias. Thus in one of the cases (case a) the current is against the bias, generating electrical power. 
} 
\end{figure}

\section{Theoretical model}
\label{Sect:3term-model}
The story of such systems started with two theoretical works \cite{Sanchez2011Feb,Strasberg2013Jan}, which modelled the system sketched in Fig.~\ref{Fig:3term} using rate equations.
The transitions modelled by the rate equation are those shown in Fig.~\ref{Fig:3term}.
We reviewed this in detail in section 9.4 of Ref.~\cite{Benenti2017Jun}, so I will not go into much details here.
I simply note that if you look at the rate equation for this system from 
section 9.4 of Ref.~\cite{Benenti2017Jun},  we have
\begin{eqnarray}
{\rmd \over \rmd t}
\left(
\begin{array}{c}\!\!
P_{00}  \\ P_{11} \\ P_{01} \\ P_{10}
\end{array}\!\!
\right)
&=& \left(
\begin{array}{cccc}
\mathtt{A}   & 0  & \mathtt{B}  & \mathtt{C}  
\\
0 &  \mathtt{D} &  \mathtt{E} & \mathtt{F} 
\\
\mathtt{G}  &  \mathtt{H}  &  \mathtt{I}  &  0  
\\
\mathtt{J}  & \mathtt{K}  & 0  & \mathtt{L} 
\end{array}
\right)
\left(\!
\begin{array}{c}
P_{00}  \\ P_{11} \\ P_{01} \\ P_{10}
\end{array}\!
\right), \qquad
\end{eqnarray}
where $P_{ij}$ is the probability for the double-dot to be in state $ij$, with $i$ being the state of dot A and $j$ being the state of dot D.
The matrix elements $(\mathtt{A},\mathtt{B},\cdots,\mathtt{L})$ depend on details (reservoir couplings, temperature, biases, etc), as explained in section 9.4 of Ref.~\cite{Benenti2017Jun}, however the crucial point is that
the mathematical structures is the same as for the two-level trapdoor in the previous chapter. Thus we have already done the mathematics to get an exact algebraic expression for the steady-state in Appendix~\ref{Sect:matrix-equation} here. 
The rest of the analysis proceeds as for the two-level trapdoor in the previous
chapter, as explained in  section 9.4 of Ref.~\cite{Benenti2017Jun}.

In the case where reservoir D is hotter than the other reservoirs, the experimental system generates electrical power as shown in Fig.~\ref{Fig:Thierschmann-expt}.
The theory of Ref.~\cite{Strasberg2013Jan} 
shows it should also generate electrical power when 
reservoir D is colder than the other reservoirs, then this power generation would be interpreted dot D 
detecting if there is an electron in dot A and opening the trapdoor
if there is.  It thereby acts like a Maxwell demon, lowering the entropy of the working fluid, while generating entropy in reservoir D.  This occurs through a net energy flow from reservoirs L and R to reservoir D.
This has not yet been demonstrated experimentally.
As it is challenging to cool electronic reservoirs below the temperature of the cryogenics, the easiest way to demonstrate this would be to leave reservoir D at
the cryostat temperature, and to use Joule heating to heat reservoirs L and R
to a bit above the cryostat temperature.  A technical challenge in doing this would be to
get the temperature of L and R close enough that one could be sure that the power is being generated by the presence of the demon, and not by  
the thermoelectric response of dot A's energy dependent coupling between reservoirs L and R at different temperatures.

\begin{figure}[t]
\centerline{\includegraphics[width=0.65\columnwidth]{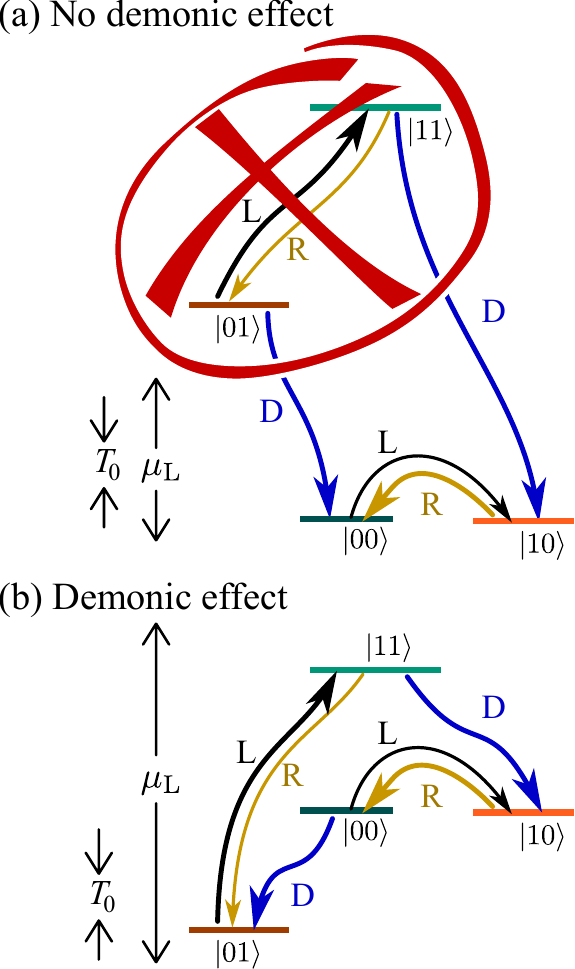}}
\caption{\label{Fig:sketch-Koski-expt} 
Simple theoretical model of transitions in the two regimes of interest to us in the demonic experiment in Ref.~\cite{Koski2015Dec}. The levels are labelled $|n_{\rm A}\ n_{\rm D}\rangle$, where $n_{\rm A}$ and $n_{\rm D}$ are the occupations of dots A and D.
While each transition happens in both directions, we use the arrow head to indicate the net average flow for each transition.   
In (a) the energy-level of dot D are moved too high to be accessible (energy $\gg \kB T_0, \mu_{\rm L}$). Then dot D is always empty and it cannot play the role of a demon.
In (b) the energy-level of the dot D is moved down so it is within the window accessible by the bias between reservoirs L and R. Then a cycle is created in which heat is always flowing into reservoir D.  
} 
\end{figure}

\begin{figure}[t]
\centerline{\includegraphics[width=0.75\columnwidth]{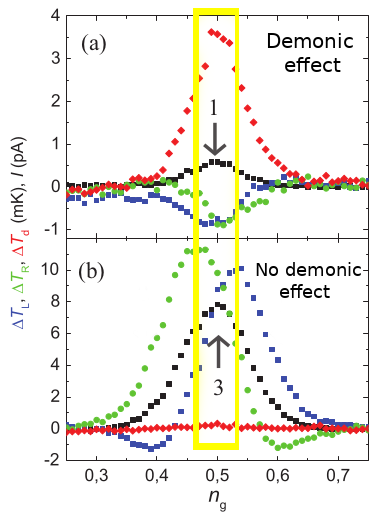}}
\caption{\label{Fig:Koski-expt-results} 
Taken from Ref.~\cite{Koski2015Dec}. For simplicity, we concentrate on the region at $n_{\rm g}=0.5$ highlighted in yellow. 
Looking at (b) first (where the demon is not operating) and we see a resistive heating of reservoirs L and R, leading to a small increase in their temperatures $\Delta T_{\rm L}$ and $\Delta T_{\rm R}$. Now turning to (a) where the demon is operating, we see that 
despite the current flow from L to R generating Joule heating, reservoirs L and R are cooled rather than heated.  
} 
\end{figure}

\section{Demonic experiment of Koski et al.}
\label{Sect:Pekola-expt}

The demonic experiment of Koski et al \cite{Koski2015Dec} is a little different from those discussed thus far in the review.
They adjusted the energy level of dot D to compared two different parameter regimes of the three-terminal system.  These two regimes are sketched in Fig.~\ref{Fig:sketch-Koski-expt}. In both cases, reservoir L has a large bias, $\mu_{\rm L}=eV \gg T_0$ and reservoir R is unbiased (as in Fig.~\ref{fig:Koski-Pekola}b).  
For simplicity, we consider only the case where dot A's gate is tuned so dot A wants to have half an electron in it ($n_g=1/2$ \cite{Koski2015Dec}), so it has the same energy whether it is empty or full.

They first consider the case where there is no demon,  because dot D's level is pushed so high in energy that dot D is always empty and plays no role.  This is shown in Fig.~\ref{Fig:sketch-Koski-expt}a.  
The rates in the rate equation are then such that electrons are more likely to enter dot A from the left and leave to the right (as indicated by the direction of the arrows 
in Fig.~\ref{Fig:sketch-Koski-expt}a).
Then the current flows through dot A in the direction determined by the bias (from L to R), generating a Joule heating, which heats reservoirs L and R. 
This heating of Joule reservoirs L and R can be seen as a small increase in their temperatures (respectively blue and green in Fig.~\ref{Fig:Koski-expt-results}b).
So dot A is acting as a perfectly normal resistive element.

They then add the demon to the set-up, by bringing dot D's level
down in energy, to an energy when it can fill and empty as it interacts with dot A, see Fig.~\ref{Fig:sketch-Koski-expt}b. There is still a large bias between reservoirs L and R, driving a current from L to R, and causing Joule heating. However looking at Fig.~\ref{Fig:sketch-Koski-expt}b, 
we see that there is now a new cycle for current to flow from L to R involving dot D;
this cycle is $|01\rangle \to  |11\rangle \to |10\rangle \to  |00\rangle \to |01\rangle$.
This cycle becomes the dominant one, if dot A's coupling to reservoirs L and R is much weaker than dot D's coupling to reservoir D.
We see that the net flow associated with this cycle involves reservoir D absorbing energy; transitions involving reservoir D have a net flow from higher to lower energy dot states,  so energy conservation means they increase the energy in reservoir D.
Remarkably, the experiment shows that the demon is extracting so much heat from reservoirs L and R that their temperature actually drops rather than increasing
(respectively blue and green in Fig.~\ref{Fig:Koski-expt-results}a).
This heat is now being dissipated in reservoir D, so its temperature increases (red in Fig.~\ref{Fig:Koski-expt-results}a.
Thus the demon's measurement-feedback cycle enables it to use the energy of a 
current being dissipated through a resistive element to actually cool that resistive element!

This result seems so bizarre as to appear almost impossible,
so it is worthwhile making an analogy to show it is not as bizarre as all that.
Suppose you had a normal household refrigerator, but made it running off a battery. 
Then you place the battery inside the refrigerator (but with the battery still connected to wires that go to the pump motor on the back of the refrigerator). The refrigerator will still work, it will take the electrical work from the battery to cool down everything inside the refrigerator, including the battery, while dissipating heat into the room that the refrigerator is in.
No laws of thermodynamics have been broken.
This is a thermodynamic analogy to what is happening in the demonic experiment.
The electrical power supplied by the bias between reservoirs L and R (the two poles of the battery) is being used to cool reservoirs L and R (the battery), at the cost of dissipating heating heat into reservoir D (the room the refrigerator is in). 
The difference in this experiment is that the refrigeration mechanism is a microscopic measurement and feedback cycle.


\chapter{Nonequilibrium reservoirs and N-demons}
\label{Chap:N-demon}

This chapter focuses on using nonequilibrium resources to 
do things that  superficially appear to be thermodynamically impossible.
For example, a system turning its heat into work without a supply of energy,
apparently violating Kelvin's formulation of the second-law.
As you will guess from having read the rest of this review, once again a more profound analysis shows that there is no violations of the laws of thermodynamics.

\begin{figure}[t]
\centerline{\includegraphics[width=0.85\columnwidth]{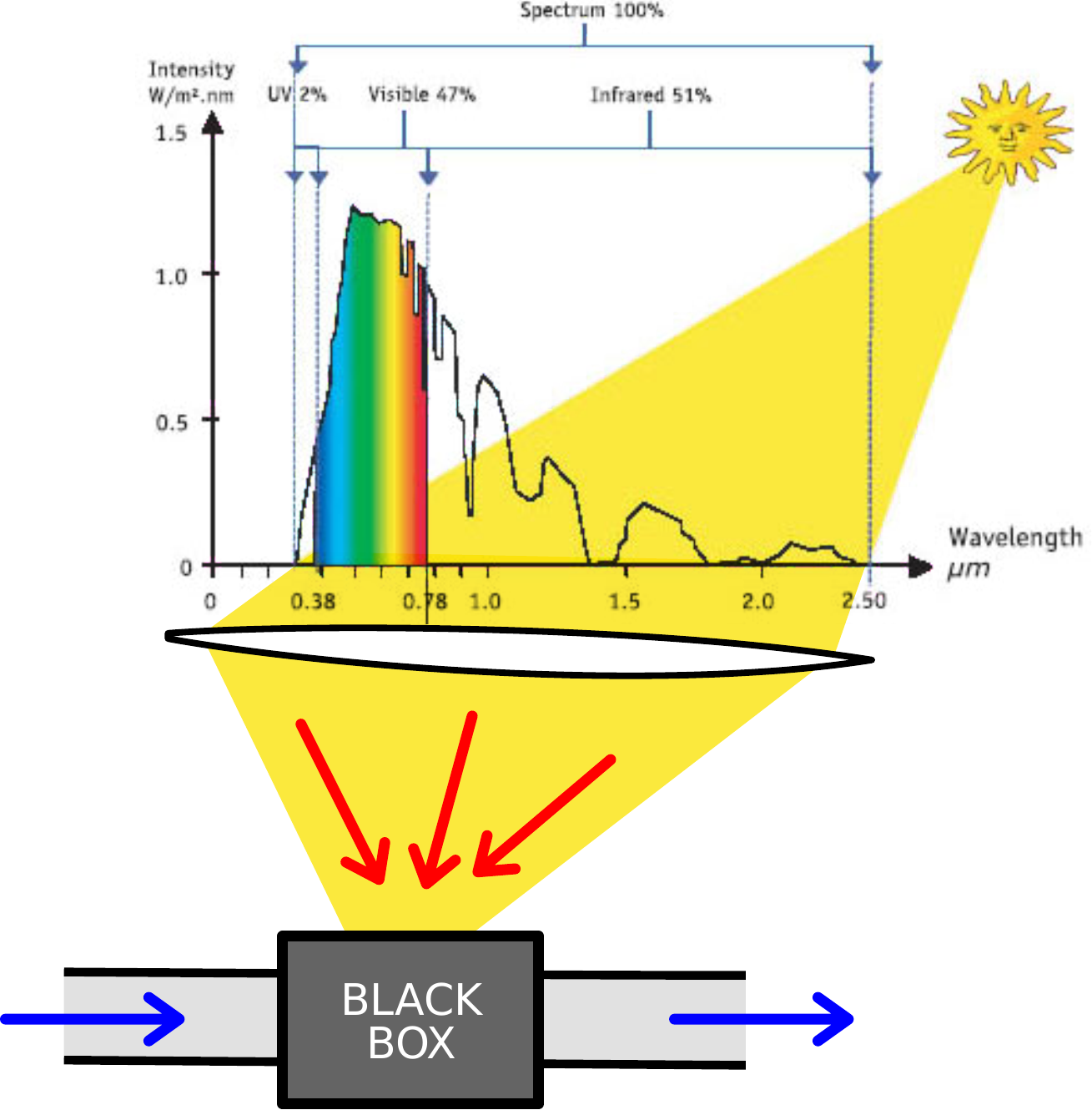}}
\caption{
\label{Fig:sun_spectrum}
The nonequilibrium (i.e. non-thermal) spectrum of sunlight at ground level going into some black box that can exploit it to make useful work (electrical power).  The sunlight's spectrum is nonequilibrium because some wavelengths are absorbed when it traverses the earth's atmosphere. This spectrum is taken from the internet (origin unknown).}
\end{figure}

However this does not make such effects boring.
Nonequilibrium resources are all around us, typically as bi-products
of other processes. One of my favourite examples is sunlight; it leaves the surface of the sun as an almost perfect thermal distribution (a black-body distribution at bit less than 6000\,K). However, when it reaches ground-level it has passed through the atmosphere which has absorbed certain wavelengths, see Fig.~\ref{Fig:sun_spectrum} \footnote{Many people like the image of the solar spectrum that I use in  Fig.~\ref{Fig:sun_spectrum}, it is in a huge number of reports and blogs on the internet, but I have yet to find its origin.}.  Thus it has acquired a nonequilibrium distribution.  So, in principle one could exploit this nonequilibrium-ness to 
extract more useful work from sunlight, than if it were a purely thermal distribution with the same energy density. This is definitely not the simplest example to exploit (because the nonequilibrium parts are mainly in the infrared, which we rarely exploit at all in photovoltaics), a more plausible application is discussed in our recent work~\cite{Tesser2022Sep}.
It was only after working on such nonequilibrium resources for a while that we realised their similarity to Maxwell demons, as we then explained in Ref.~\cite{Sanchez2019Nov}.

\section{\mbox{Non-locality of} \mbox{thermodynamic laws}}

In conventional circumstances, the laws of thermodynamics are local in space.
The entropy of a system is only reduced if some external agent supplies energy directly to that system. However, this is not a requirement of the laws of thermodynamics.
Here we present ``exotic'' situations in which the entropy of a system can go down, because the entropy of another system (some distance away) goes up, even when there is no energy exchanged between the systems.
Just as we found for a system (see below) in  Ref.~\cite{Whitney2016Jan}.
We called this a {\it non-locality of the laws of thermodynamics}, in which there is spatial separation of the first-law of thermodynamics (heat to work conversion) from the second-law of thermodynamics (generation of entropy).
The first law (energy conservation) is operating within one system, with heat being turned into work, thereby generating useful work {\it and} cooling that system, and reducing that systems entropy. 
However, the second law is operating elsewhere, in another system, ensuring that it produces more entropy than the first system's entropy reduction.
All of this occurring without energy exchange between the two systems.

\section{\mbox{N-demon}}
\label{sect:Ndemon}

Imagine a situation where the demon has two thermal reservoirs 
at its disposal (one very cold and one very hot).  Then every time it takes a pair of particles from the working fluid, it does not measure them. Instead it 
just throws that pair of particles into one of its reservoirs (say the very cold reservoir), and replaces them with a pair of particles; one taken from the very cold reservoir and one taken from the very hot reservoir.  

This demon is doing the exact reverse of a normal relaxation process.
In a normal collision between two identical particles, we expect the faster particle to give kinetic energy to the slower one, so the fast particle slows-down and the slow particle speeds-up. The repeated effect of such scatterings is what makes a gas relax to a thermal distribution. As the demon is doing the reverse of this, it is pushing the working fluid {\it away} from a thermal distribution.

Compared to the demons in previous chapters, this makes three rather profound changes to 
the demon---profound enough that we think of it as a new kind of demon different from a Maxwell demon. We call it an N-demon, with ``N'' for ``nonequilibrium'', because we will see that it is exploiting
a nonequilibrium resource, rather than a measurement-feedback loop.
In the present case, the nonequilibrium resource is the pair of reservoirs 
(one hot and one cold) that are not in equilibrium with each other, but below we will consider more general nonequilibrium resources.
The three profound changes made to the demon are:
\begin{itemize}
\item There is no longer a measurement at the level of individual particles, and there is no longer a measurement-feedback loop. This means the demon has no need for a memory,
so there is no entropy production associated with resetting this memory. 

\item There are now entropy changes to be calculated in both the demon's reservoirs;
particles taken from the working fluid are being injected into these reservoirs, and particles taken from these reservoirs are being injected into the working fluid.
This implies heat flows and hence entropy changes.

\item The temperatures of the two reservoirs used by the demon can always be adjusted
to fix the average energy of the particles it injects into the working fluid, but as those particles are taken from a distribution, the energy will be different each time the demon injects such a pair of particles.  Similarly, the average energy of the particle extracted from the working fluid may be fixed, but it will be different each time a pair is extracted.  Thus even if everything is adjusted so the average energy of injected particles equals that of extracted particles, there will be energy fluctuations.
Thus the demon exchanges no energy with the working fluid on average. There always likely to be a fluctuating energy between the two (see section~\ref{Sect:Freitas} for more on this).
\end{itemize}

\begin{figure}
\centerline{\includegraphics[width=0.95\columnwidth]{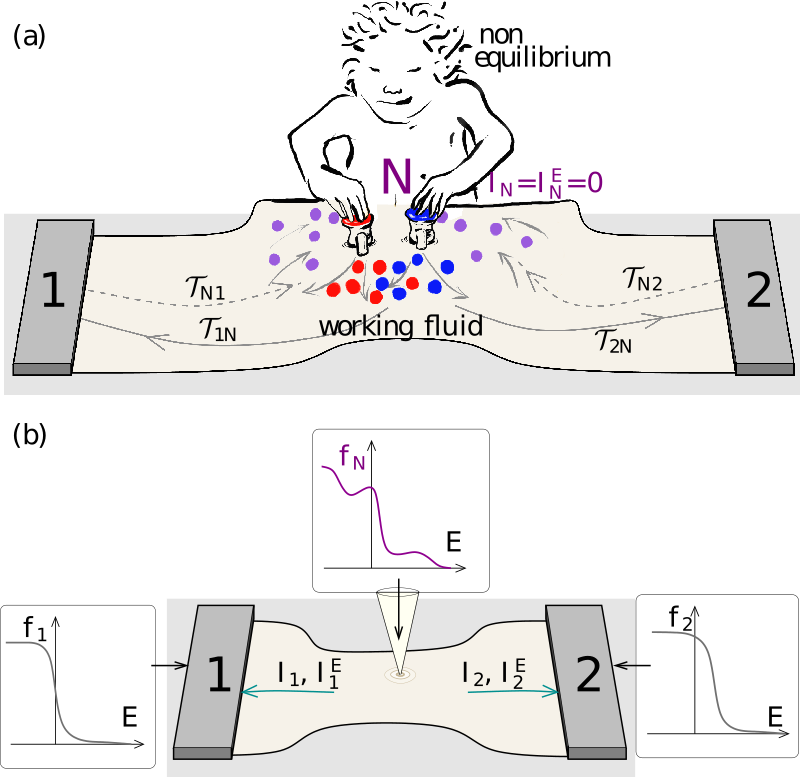}}
\caption{
\label{Fig:Ndemon}
Sketch by Rafael Sanchez taken from our article \cite{Sanchez2019Nov}.
(a) An N-demon supplies no heat or work, instead it supplies a nonequilibrium distribution to the working fluid containing equilibrium reservoirs 1 and 2.  This nonequilibrium distribution could be a non-thermalized mixture of hot (red) and cold (blue) particles.
(b) A physical implementation in which the nonequilibrium distribution $f_{\rm N}$ is injected locally into the working fluid.}
\end{figure}

In the spirit of the last point in the list, we do not {\it need} the demon
to extract particles in pairs, or inject them in pairs.  It is enough
that it injects the same number of particles as it extracts  {\it on average over time}, and the injected particles have the same energy as those extracted  {\it on average over time}.
This enabled us to propose the N-demon sketched in Fig.~\ref{Fig:Ndemon}.

The N-demon's basic property is to inject a nonequilibrium distribution of particles into the working fluid that the N-demon is acting on.
There are numerous ways such a nonequilibrium distribution could be achieved, 
however for concreteness we consider only one here.
We assume the N-demon's nonequilibrium distribution is constructed from multiple reservoirs (at least two) each of which contains a thermal distribution, but which are not in equilibrium with each other. 
They should be coupled to each other to ensure that the 
proportion of particles entering the working fluid each distributions (each reservoirs in the N-demon) is different at different energies, This ensure a non-equilibrium distribution of particles entering the working fluid from the N-demon.  

\begin{figure*}
\centerline{\includegraphics[width=0.85\textwidth]{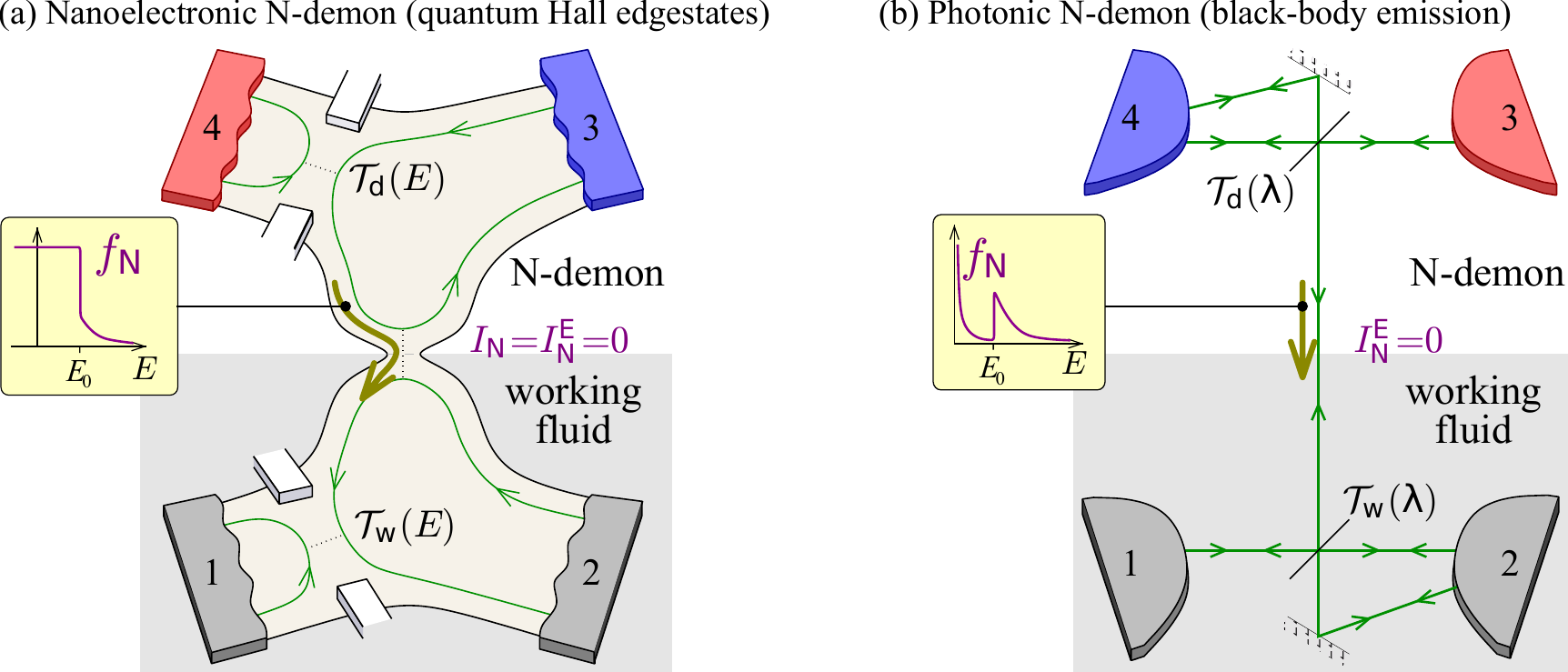}}
\caption{
\label{Fig:Ndemon2}
Two implementations of the N-demon, taken from our article \cite{Sanchez2019Nov}.
(a) An N-demon made with quantum Hall edge-states.
(b) An N-demon made with light flowing between black-bodies.
In both cases $f_{\rm N}$ is a on-equilibrium distribution injected into the lower part (the working fluid), from the upper-part (the N-demon).  This nonequilibrium distribution is injected at a point, but is built out of the hot and cold thermal distributions in reservoirs 3 and 4.  It is chose so that there is no net flow of particles or energy (on average) between the N-demon and the working fluid, 
i.e.~$I_{\rm N}=0$ and $I_{\rm N}^{\rm E}=0$. 
Yet in (a) the working fluid's heat reduces as it produces electrical work between reservoirs 1 and 2, while in (b) heat flows against a temperature difference between 
reservoirs 1 and 2.  In both cases, the entropy of the lower part (working fluid plus reservoirs 1 and 2) drops, but the entropy of the upper part (N-demon, reservoirs 3 and 4) grows by more.}
\end{figure*}

The simplest example of this is shown in Fig.~\ref{Fig:Ndemon2}b, where the nonequilibrium distribution produced by the demon, $f_{\rm N}$, is generated by a 
partially-silvered mirror between two thermal reservoirs 3 and 4, which are two black bodies at different temperature. The partially-silvered mirror has a frequency-dependent transmission coefficient (i.e. wavelength dependent), ${\cal T}(\lambda)$,
which reflects all photons with energy above a certain threshold, $E_0$, and does not reflect any photons with energy below $E_0$.
This results in a nonequilibrium injected downwards into the working fluid, where all the photons below $E_0$ come from the cold reservoir 4, and all those above $E_0$ come from the hot reservoir 3.  This results in the nonequilibrium distribution, $f_{\rm N}$ sketched in Fig.~\ref{Fig:Ndemon2}b. 

A similar example in nanoelectronics using quantum Hall edge-states is shown 
in Fig.~\ref{Fig:Ndemon2}a.  There the demon is injecting a distribution of fermions
(electrons), $f_{\rm n}$, in which energy states below some threshold, defined by the transmission at the point contact marked ${\cal T}_{\rm d}(E)$,  come from reservoir 3.
Reservoir 3 is cold and biased so that all fermionic states are approximately full up to $E_0$, so $f_{\rm N}(E \leq E_0)=1$.  In contrast states above $E_0$ come from the hot unbiased reservoir 4, so  $f_{\rm N}(E > E_0) < 0.5$ and decays as $E$ increases.
This is the nonequilibrium distribution $f_{\rm N}(E)$ injected into the working fluid, sketched in Fig.~\ref{Fig:Ndemon2}a.

In the case of the nanoelectronic N-demon in Fig.~\ref{Fig:Ndemon2}a, the N-demon drives electrons from reservoir 1 to reservoir 2 against a potential
(we assume the electrochemical potential of reservoir 2 is higher than reservoir 1), meaning it is doing electrical work.   This is achieved by placing an energy barrier in the working fluid. 
This reduces the entropy of the working fluid, as it converts heat in the working fluid into work (in the form of electrical power).

In the case of the photonic N-demon in  Fig.~\ref{Fig:Ndemon2}a, the N-demon drives heat against its natural direction of flow,
from a colder reservoir 1 to a hotter reservoir 2.  Again this is achieved by placing an energy barrier in the working fluid. 

It is easy to imagine making the classical limit of the photonic N-demon, with an effectively infinite number of photons flowing out of each black-body.  Then there will be fluctuations in the energy flows, but they will be negligible compared to the flows themselves; so they will be negligible compared to the heat flow causing the entropy reduction in the working fluid (the heat flow from colder reservoir 1 to hotter reservoir 2).  Thus while we are careful to say the energy and heat flow between the N-demon and the working fluid is only zero on average, the small fluctuations about zero are small enough to be neglected in this N-demon, and so cannot be the cause of the entropy reduction in the working fluid.

\subsection{Why this confuses our intuition}

The N-demon confuses our intuition, when we see that it reduces the working fluid's entropy, without injecting energy, heat or work.  This is because our intuition is based on everyday experiences of heat flow, and we are most familiar with flows that thermalized extremely quickly.  
We expect that if we inject a non-equilibrium fluid into one end of a pipe (for example by injecting hot and cold water from two different water tanks) it typically emerge at the other end as an equilibrated distribution, in which each particle has forgotten which water tank it came from.
At each point in space this equilibrated distribution in entirely defined by three parameters; local net velocity, local pressure and local temperature.  If we then use this fluid in a thermodynamic machine, there are no surprises; its velocity and pressure provide work sources, and its temperature provides a heat source heat.  If there is no difference in this velocity, pressure or temperature between it and a working fluid that we connect it to, then there is no flow of energy, work or heat, and it cannot reduce the entropy of that working fluid.
This is because there is non-equilibrium-ness left to exploit.

In contrast, the machines proposed in Fig.~\ref{Fig:Ndemon2} work because the particles flowing from the N-demon are still in a non-equilibrium distribution when they reach the working fluid.
It is this source of non-equilibrium-ness that can reduces the entropy of the working fluid.

If there is a process of relaxation between the non-equilibrium particles as they flowed from the N-demon towards the 
working fluid, it would suppress the non-equilibrium-ness of the distribution, and reduce the effectiveness of the machine (such a process was modelled in Refs.~\cite{Sanchez2019Nov,Hajiloo2020Oct}).
If this relaxation is strong, then the particles flowing down into the working fluid 
would arrive at the working fluid in a thermal distribution. As it is established that this flow  provides no heat or work the working fluid, this means the thermal distribution of these particles is the same as those of the working fluid (same temperature),  Hence, it is  incapable of reducing the entropy of the working fluid.
Thus to build a working N-demon, its particles must not relax too quickly to a thermal state.  This constraint
is discussed more in Section~\ref{Sect:where-to-find-Ndemons}.

In many ways the N-demon works in a similar manner to 
a conventional machine driven by a coupling to two reservoirs at {\it two different points in space},
if these reservoir are both in internal equilibrium, but are not in equilibrium with each other. For example, suppose that one of these reservoir is hotter than the working fluid and the other is colder. Then it is easy to 
imagine a situation where the hot reservoir supplies heat to the working substance, while the cold reservoir removes the same amount heat (upon on averaging over thermal fluctuations).  Then one can engineer a pair of heat engines that exploit these two heat flows independently to both generate work in the working fluid \cite{Freitas2021Mar}.
Then what clearly matters is the heat flow into (or out of) each reservoir individually; it is no real relevance that the sum of the two is zero.
The surprise of the N-demon is that one can have similar physics to this, when particles from both reservoirs enter the working substance as a single non-equilibrium distribution at {\it a single point in space}, so long as those particles do not thermalize with each other before their non-equilibrium distribution can be exploited.  Then this non-equilibrium flow carries no energy or heat on average, yet its non-equilibrium nature can be exploited.

\section{\mbox{Quantifying nonequilibrium} \mbox{resources using free-energy}}

In general, a given resource could be a source of a combination of heat, work and nonequilibrium-ness.  Such a combined resource can be quantified via a \textit{free-energy} \cite{Strasberg2017Apr,Hajiloo2020Oct}, $F_\text{neq}$, 
which takes the form of a Helmholtz potential \cite{Callen1985Sep}, 
\begin{eqnarray}
F_\text{neq}=U_\text{neq}-T'S_\text{neq},
\end{eqnarray}
where $U_\text{neq}$ and $S_\text{neq}$ are the nonequilibrium resource's internal energy and entropy. The critical point is to identify which temperature $T'$ appears in this formula; we argued in Ref.~\cite{Hajiloo2020Oct} that 
it is the ``ambient'' temperature; i.e. the temperature that all reservoirs would have if there were no mechanisms keeping them hot, cold or out-of-equilibrium.

The power produced from such a resource is bounded by 
\begin{eqnarray}
P \leq -\dot{F}_\text{neq},
\label{Eq:Free-energy-inequality}
\end{eqnarray}
where $-\dot{F}_\text{neq}$ is the rate of reduction of the non-equilibrium resource's free energy. 

For a standard heat-engine, the resource is a hot thermal reservoir,
in internal equilibrium at temperature $T_{\rm hot}$, with no bias ($\mu_{\rm hot}=0$).
Then
$\dot{S}_\text{hot} =  \dot{U}_\text{hot}/T_\text{hot}$, where 
$\dot{U}_\text{hot}$ is the heat flow {\it into} the hot reservoir.
As a result
\begin{eqnarray}
\dot{F}_\text{hot}&=& \dot{U}_\text{hot} \ \left(1 -T_0\big/T_\text{hot}\right),
\end{eqnarray}
where $T_0$ is the ambient temperature.
Then Eq.~(\ref{Eq:Free-energy-inequality}) reduces to Carnot's well-known result for heat-engines
\begin{eqnarray}
P/Q &\leq& \left(1 -T_0\big/T_\text{hot}\right)
\end{eqnarray}
where the heat flow {\it out of} the hot reservoir, $Q \equiv - \dot{U}_\text{hot}$. 

If the resource is a reservoir at ambient temperature, but it is biased with a shift of its electrochemical potential of $\mu_{\rm bias}$, then  $\dot{S}_{\rm bias}= (\dot{U}_{\rm bias}-\dot{N}_{\rm bias}\,\mu_{\rm bias})/T_0$, where 
$\dot{N}_{\rm bias}$ and 
$\dot{U}_{\rm bias}$ are  the rate of particle number and energy in the reservoir.
Thus the free energy is 
\begin{eqnarray}
\dot{F}_\text{hot}&=& \dot{N}_\text{bias}\,\mu_{\rm bias}  \  .
\end{eqnarray}
Which means the useful power is 
\begin{eqnarray}
P &\leq& I_\text{bias}\,\mu_{\rm bias} 
\end{eqnarray}
where the particle current {\it out of} the reservoir is 
$I_\text{bias}\equiv - \dot{N}_\text{bias}$.
This gives the trivial result that under ideal conditions, the power in the particles flowing out of this reservoir can all be turned into useful work, (for example, turned into mechanical work by a motor). However,  some of it is usually lost as heat (resistance in wires etc), and then the useful power $P$ is less than $I_\text{bias}\,\mu_{\rm bias}$.

If the resource is a N-demon, which supplies nonequilibrium-ness, 
but no energy or heat (what we call a strict N-demon in Ref.~\cite{Hajiloo2020Oct}),
then $\dot{U}_\text{neq}=0$, and the free energy change reduces to \cite{Hajiloo2020Oct},
\begin{eqnarray}
\dot{F}_\text{N-demon}&=& -T_0 \dot{S}_\text{N-demon}  \  .
\end{eqnarray}
leading directly to the inequality \cite{Sanchez2019Nov}
\begin{eqnarray}
P \leq T_0 \dot{S}_\text{N-demon} \ .
\end{eqnarray}
A different case is when the N-demon is not used to produce power, but rather to further cool the colder of two-reservoirs.
This is the case the the N-demon made of light emission from black-bodies in Fig.~\ref{Fig:Ndemon2}b. Then, let us assume the colder of the two reservoirs is at temperature $T_{\rm cold}$ and the other is at ambient temperature $T_0$, then
the limit on cooling power is given by\cite{Hajiloo2020Oct}
\begin{eqnarray}
J_{\rm cold} \leq {T_{\rm cold}T_0 \over T_0-T_{\rm cold}} \dot{S}_\text{N-demon} 
\end{eqnarray}
where $J_{\rm cold}$ is the rate of heat-flow out of the cold reservoir (and into the ambient reservoir) induced by the N-demon. 

Ref.~\cite{Hajiloo2020Oct} contains numerous examples of concrete implementations of such 
systems in the context of quantum Hall edge states. In particular, it shows the crossover from an N-demon that exploits only nonequilibrium-ness of a resource, to a system that exploits both the nonequlibrium-ness and the heat in a resource.

\section{\mbox{Where to look for} \mbox{nonequilibrium reservoirs}}
\label{Sect:where-to-find-Ndemons}

At the beginning of this chapter we argued that nonequilibrium resources are all around us, and gave the example of sunlight (Fig.~\ref{Fig:sun_spectrum}). 
However there are a few things to keep in mind when looking for 
physical systems that could play the role of nonequilibrium reservoirs.

Firstly, there must be a mechanism that drives the reservoir out of equilibrium.
In the examples from Refs.~\cite{Sanchez2019Nov,Hajiloo2020Oct}, shown in 
Fig.~\ref{Fig:Ndemon2}, the nonequilibrium reservoir is bi-partite, made of one hot and one cold reservoir.  It is implicitly assumed that there are mechanisms that keeps the hot reservoir hot and keeps the cold reservoir cold, otherwise they would thermalize with each other and the nonequilibrium reservoir would cease to be nonequilibrium.
However, the reservoir does not need to be bipartite, it is enough that there is some  mechanism that drives the reservoir away from equilibrium in some manner. 

Secondly, there must be certain states in the reservoir that relax slowly, so that when they are driven out of equilibrium, they do not immediately relax back to an equilibrium state.  These are the only reservoir modes that re likely to be sufficiently out of equilibrium to exploit as a nonequilibrium reservoir.  In this context, photons (like sun-light) are more accessible than electrons, because photons relax very slowly 
(they do not relax at all in vacuum or in linear media).  
This is why the N-demon involving light emission by black-bodies in Fig.~\ref{Fig:Ndemon2}b could be macroscopic. Indeed, the upper two reservoirs could be kilometres from the lower two reservoirs, so long as the photons flowing from top to bottom do not thermalize amongst themselves.

In contrast, electrons tend to thermalize fast due to electron-electron and electron phonon interactions, this relaxation is much slower at sub-Kelvin temperatures, but one typically still has to go to the nanoscale to get a device small enough to ``catch'' electrons that have been driven out of equilibrium before they thermalize back to an equilibrium distribution. 
The techniques to do this for electrons in high magnetic field, where they follow edge-states (Fig.~\ref{Fig:Ndemon2}a) are well-known, and relatively straightforward theoretically and experimentally.  However, it could also be done will normal electron flows (in the absence of a magnetic field).  

It is worth mentioning that this difficulty of ``catching'' electrons before they thermalize is the reason why it is so challenging to build hot-carrier photovoltaics \cite{Green2003}.  
However, recent experimental progress has been made using nanostructures \cite{Green2003,Hirst2014Jun,Chen2020Jun,Fast2021Jun,Fast2022Jun}, and we recently showed how the basic operating principle of such a hot-carrier photovoltaic is similar to that of an N-demon \cite{Tesser2022Sep}.

There is one final thing to look for if one wishes to practically exploit nonequilibrium resources. The process that generates the nonequilibrium state should be of little cost to us.  There is not much practical interest if we are expending the energy necessary to maintain the N-demon's nonequilibrium state;
for example shining a laser that drive electrons out of equilibrium.    Such reservoirs are ideal for proofs of principle, but not the place to look for practical applications.  Instead, the ideal case is when the nonequilibrium distribution is a by-product, or even waste-product, of some other process. Then it is effectively free (or cheap) for us, and it makes sense to use its nonequilibrium nature as a resource from which to extract useful work.

In many cases, the waste distributions that come from other processes will be hot as well as nonequilibrium.  It is natural to try to recover useful work from its waste heat, we argue that it is equally natural to try to recover useful work from its waste nonequilibrium-ness.

\section{\mbox{Scattering theory of} \mbox{nonequilibrium reservoirs}}

Having seen how important the entropy change is for a nonequilibrium reservoir, it is important to develop methods of calculating it.
Ref.~\cite{Deghi2020Jul} calculated this within scattering theory.
This applies if the electrons in the nano-structure are independent, and the electrons in the 
resource are completely defined by a nonequilibrium distribution $p(E)$ as a function of energy.  Note that this assumes that the resource is a macroscopic reservoir incoherent electrons with no classical correlations or quantum entanglement between them.
Then \cite{Deghi2020Jul}
\begin{eqnarray}
{\rmd S^{(\alpha)} \over \rmd t } &=& \kB \int {\rmd E \over h}  \ \ln\left[ {1-p_\alpha(E) \over p_\alpha(E)} \right] 
\nonumber \\ 
& & \qquad \times \sum_\beta
\big[{\cal T}_{\alpha\beta}(E) \, p_\beta (E) -  {\cal T}_{\beta\alpha}(E)\, p_\alpha (E)  \big]. \quad
\nonumber \\
\label{Eq:dotS}
\end{eqnarray}
where $S^{(\alpha)}$ is the entropy in reservoir $\alpha$, 
$p_\alpha(E)$ is the nonequilibrium (or equilibrium) distribution in reservoir $\alpha$,
and ${\cal T}_{\alpha\beta}(E)$ is the transmission from reservoir $\beta$ to reservoir $\alpha$.
Other quantities like the currents take the same form as for reservoirs in internal equilibrium,  with the Fermi distribution replaced by the nonequilibrium distribution.
Hence the rate of change of the particle number and energy in reservoir $\alpha$ are
\begin{eqnarray}
\dot{N}_\alpha &=&  \int {\rmd E \over h}  
\sum_\beta
\big[{\cal T}_{\alpha\beta}(E) \, p_\beta (E) -  {\cal T}_{\beta\alpha}(E)\, p_\alpha (E)  \big]. \quad
\nonumber \\
\dot{U}_\alpha &=&  \int {\rmd E \over h}  E
\sum_\beta
\big[{\cal T}_{\alpha\beta}(E) \, p_\beta (E) -  {\cal T}_{\beta\alpha}(E)\, p_\alpha (E)  \big]. \quad
\nonumber \\
\end{eqnarray}
where the electric current into reservoir $\alpha$ is $I_\alpha=e\dot{N}_{\alpha}$,
and the heat current into reservoir $\alpha$ is 
$J_\alpha = \dot{U}_\alpha-\mu_\alpha\dot{N}_\alpha$.

If reservoir $\alpha$ is in internal equilibrium, then $p_\alpha(E)$ becomes the Fermi distribution in the $\alpha$th reservoir, 
$p_\alpha(E) \to F_\alpha(E) = 1\big/\big(1+\exp[(E-\mu_\alpha)\big/T_\alpha]\big)$.
This means $\ln\left[ (1-p_\alpha(E))\big/ p_\alpha(E) \right]  \to (E-\mu_\alpha)/T_\alpha$.
Then and only then does the entropy change in reservoir $\alpha$ reduces to 
the Clausius relation 
\begin{eqnarray}
\dot{S}^{(\alpha)} \ =\  {J_\alpha\over T_\alpha} \ \equiv\ {\dot{U}_\alpha-\mu_\alpha\dot{N}_\alpha \over T_\alpha}
\label{Eq:Classius-for-scattering}
\end{eqnarray}
This means that if there is no average particle or energy flow into reservoir $\alpha$ (i.e.~$\dot{U}_\alpha=\dot{N}_\alpha=0$), then its entropy cannot increase, so its is forbidden to reduce the entropy of another reservoir by the second law.

In contrast, in any case where $p_\alpha$ is not a Fermi function, the Clausius relation in Eq.~(\ref{Eq:Classius-for-scattering}) does not hold,
and there is no simple relationship between the change of entropy in reservoir $\alpha$ and heat and particle flow into reservoir $\alpha$. Thus its entropy, $S_\alpha$, can increase even when there is no average particle or energy flow into it ($\dot{U}_\alpha=\dot{N}_\alpha=0$).  This means that there is nothing forbidding it being the cause of an entropy reduction in another reservoir.

\section{\mbox{The second law within} \mbox{the scattering theory of} \mbox{non-equilibrium reservoirs}}

Now it is important to prove that the entropy production in the scattering theory for non-equilibrium reservoirs, satisfies the second law of thermodynamics.
For this we simply define
\begin{subequations}
\label{Eq:define-x}
\begin{eqnarray}
x_\alpha(E) &=&   \ln\left[ {1-p_\alpha(E) \over p_\alpha(E)} \right] ,
\end{eqnarray}
which implies that for any non-thermal distribution
\begin{eqnarray}
p_\alpha(E)\ =\ \left(\e^{x_\alpha(E)}+1\right)^{-1} \ \equiv\ f\left( x_\alpha(E) \right).
\end{eqnarray}
\end{subequations}
where $f(x)$ as a Fermi function. 
If we substitute this into Eq.~(\ref{Eq:dotS}) we get
\begin{eqnarray}
{\rmd S_\alpha \over \rmd t} 
&=& \kB \int {\rmd E \over h}\ x_\alpha
\nonumber \\
& & \quad \times\ \sum_\beta
\big[{\cal T}_{\alpha\beta}(E) \, f(x_\beta) -  {\cal T}_{\beta\alpha}(E)\, f(x_\alpha)  \big], \nonumber \\
\label{Eq:dotS-scatter-x}
\end{eqnarray}
where $x_\alpha$ is a function of $E$.
Thus we have a formal mapping of the scattering theory for non-equilibrium distributions onto an object that looks like a scattering theory for equilibrium distributions, with all the complexity hidden in the highly non-trivial relationship between $x_\beta$ and $E$.
This allows us to simply reproduce the derivation of the second law for equilibrium reservoirs
in Ref.~\cite{Nenciu2007Mar}, see also Ref.~\cite{Whitney2013Mar}.

\subsection{Second law for two reservoirs}

Despite the differences,
the proof that this obeys the second-law follows exactly as for systems with equilibrium reservoirs.
This proof is extremely simple for a scatterer between two reservoirs (L and R), where
the symmetries of the two-terminal scattering matrix mean that ${\cal T}_{LR}(E)={\cal T}_{RL}(E)$.
Then Eq.~(\ref{Eq:dotS-scatter-x}) implies that 
\begin{eqnarray}
{\rmd S_{\rm total} \over \rmd t} &=&  
{\rmd S_L \over \rmd t} + {\rmd S_R \over \rmd t} 
\\
&=& -\kB \int {\rmd E \over h} 
\left(x_L -x_R\right) \,{\cal T}_{LR}(E) 
\nonumber \\
& & \qquad \qquad \times\  \big[ f(x_L) -  f(x_R)  \big], \qquad
\end{eqnarray}
Now since $f(x)$ is a monotonically decaying function of $x$, 
we can see that 
\begin{eqnarray}
\left(x_L -x_R\right)\big[ f(x_L) -  f(x_R)  \big]  &\leq& 0
\end{eqnarray}
for all $x_L$ and $x_R$. This means that this inequality holds for all $E$, no matter how complicated the relationship is between $x_L$, $x_R$ and $E$. 
Since ${\cal T}_{LR}(E)\geq 0$ for all $E$,  the integral over $E$ can never be positive.
Thus we see that the second law of thermodynamics is always satisfied by 
Eq.~(\ref{Eq:dotS}) for two-reservoir systems, i.e.\ any such two-reservoir systems will obey
\begin{eqnarray}
{\rmd S_{\rm total} \over \rmd t}   \ \geq 0\ .
\label{Eq:second-law}
\end{eqnarray}

\subsection{Second law for many reservoirs}

Now we turn to the case of an arbitrary number of reservoirs, where the proof is more intricate,
because we do not have such simple symmetries of the scattering matrix.  The proof for equilibrium reservoirs 
was done in Ref.~\cite{Nenciu2007Mar} and reviewed in section 6.4.2 of Ref.~\cite{Benenti2017Jun}.
We start by noting that particle conservation means that ${\cal N}_\alpha = \sum_\beta {\cal T}_{\beta \alpha}$,
where ${\cal N}_\alpha$ is the number of modes in the connection between reservoir $\alpha$ and the scatterer.
Them we can write Eq.~(\ref{Eq:dotS}) as 
\begin{eqnarray}
{\rmd S_\alpha \over \rmd t} \!\!
&=& \!\!\kB \int {\rmd E \over h}\  \ln\left[ {1-p_\alpha(E) \over p_\alpha(E)} \right] 
\nonumber \\
& & \qquad \quad\times\ \sum_\beta
\big[{\cal T}_{\alpha\beta} -\delta_{\alpha\beta}{\cal N}_\alpha\big] p_\beta (E),
\qquad 
\label{Eq:dotS-scatter1}
\end{eqnarray}
Summing over all reservoirs, and substituting in Eqs.~(\ref{Eq:define-x}) gives
\begin{eqnarray}
{\rmd S_{\rm total} \over \rmd t} 
&=&\sum_\alpha {\rmd S_\alpha \over \rmd t} 
\\
&=& \kB \int {\rmd E \over h}  \sum_{\alpha,\beta}  x_\alpha \,
\big[{\cal T}_{\alpha\beta} -\delta_{\alpha\beta}N_\alpha\big] \, f(x_\beta).
\qquad
\nonumber \\
\label{Eq:dotS-scatter1-x}
\end{eqnarray}
If we compare this with the equation in section 6.4.2 of Ref.~\cite{Benenti2017Jun},
we see that it has the same form, but with $\xi_\alpha=(E-\mu_\alpha)/T_\alpha$ replaced by
$x_\alpha$, and $A_{ij}$ in Ref.~\cite{Benenti2017Jun} being
$-\big[{\cal T}_{\alpha\beta} -\delta_{\alpha\beta}N_\alpha\big]$.

Now let us use the fact that $\sum_\alpha \big[{\cal T}_{\alpha\beta} -\delta_{\alpha\beta}N_\alpha\big] =0$
to subtract the following term from the right-hand-side of the above equation (without changing the left-hand-side)
\begin{eqnarray}
 \kB \int {\rmd E \over h}  \sum_{\alpha,\beta}  x_\beta \,
\big[{\cal T}_{\alpha\beta} -\delta_{\alpha\beta}N_\alpha\big] \, f(x_\beta) \ = \ 0.
\qquad
\end{eqnarray}
After which
\begin{eqnarray}
{\rmd S_{\rm total} \over \rmd t} 
&=& \kB \int {\rmd E \over h}  \sum_{\alpha,\beta} \,
\big[{\cal T}_{\alpha\beta} -\delta_{\alpha\beta}N_\alpha\big] \
\nonumber \\
& & \qquad \qquad \times (x_\alpha-x_\beta) \, f(x_\beta).
\qquad
\label{Eq:2nd-law-step1}
\end{eqnarray}
Next, we use Nenciu's trick \cite{Nenciu2007Mar} of define $F(x)$ as the integral of the Fermi function $f(x)$, in other words $F(x) = \int^{x} \rmd x'\,f(x')$.
Since $f(x)$ is a monotonically decaying function of $x$, we know that $F(x)$ is a concave function of $x$.
this means that $F(x)$  is always less than its first order Tailor expansion about any $x_0$,
i.e.
\begin{eqnarray}
F(x) -F_(x_0) &\leq& (x-x_0) f(x_0)  
\end{eqnarray}
Now taking this inequality with $x=x_\alpha$ and $x_0=x_\beta$, we can substitute it into the right-hand-side
of Eq.~(\ref{Eq:2nd-law-step1}) to get
\begin{eqnarray}
{\rmd S_{\rm total} \over \rmd t} 
&\geq& \kB \int {\rmd E \over h}  \sum_{\alpha,\beta} \,
\big[{\cal T}_{\alpha\beta} -\delta_{\alpha\beta}N_\alpha\big] \
\nonumber \\
& & \qquad \qquad \times \left(F(x_\alpha)-F(x_\beta)\right).
\qquad
\label{Eq:2ndlaw-n-terminal-inequality}
\end{eqnarray}
Now the prefactor on the $F(x_\alpha)$ term contains 
$$\sum_\beta 
\big[{\cal T}_{\alpha\beta} -\delta_{\alpha\beta}N_\alpha\big] =0,
$$
and the prefactor on the $F(x_\beta)$ term contains 
$$
\sum_\alpha
\big[{\cal T}_{\alpha\beta} -\delta_{\alpha\beta}N_\alpha\big] =0.
$$
Thus the whole right-hand-side of Eq.~(\ref{Eq:2ndlaw-n-terminal-inequality}) is zero, 
irrespective of how  complicated the relationship is between $x_\alpha$, $x_\beta$ and $E$.
Thus get the second law of thermodynamics
given in Eq.~(\ref{Eq:second-law}), but now for an arbitrary number of non-equilibrium reservoirs.

\section{N-demons in classical electric circuits}

\begin{figure}
\centerline{\includegraphics[width=0.9\columnwidth]{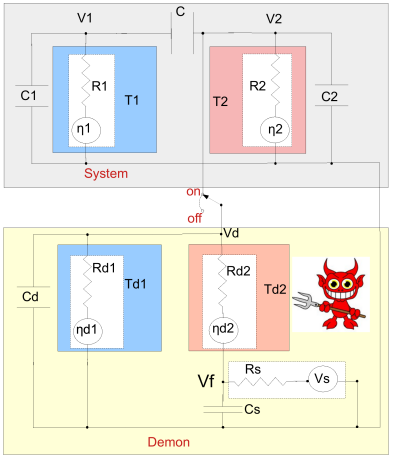}}
\caption{
\label{Fig:Ciliberto-Ndemon}
Taken from Ref.~\cite{Ciliberto2020Nov}. A classical electric circuit in which the lower part  (yellowish beige) acts as a N-demon. It drives heat in the upper part of the circuit (grey) from the colder resistor to the hotter resistor (against its natural flow direction), without the N-demon supplying any energy to the upper part of the circuit. }
\end{figure}

Following our work, Sergio Ciliberto \cite{Ciliberto2020Nov} proposed an N-demon in a classical
electric circuit (calling it an out-of-equilibrium Maxwell demon).
It is an autonomous \cite{autonomous} system that can reverse the natural direction of
the heat flux between two electric circuits maintained at different temperatures, when they are coupled by the electric
thermal noise, see Fig.~\ref{Fig:Ciliberto-Ndemon}. As in the examples that 
we gave above the demon does not process any information. Instead, it operates via
a frequency dependent coupling with the hot and cold reservoirs. 
There is no energy flux
between the demon and the system, but the total entropy production (system+demon) is positive.
Its physics is governed by
equations similar to those of two coupled Brownian particles. 
His work shows that while we discovered the N-demon in the context of nanoelectronic systems, the principles apply equally to macroscopic circuits.

\section{\mbox{Quantum non-equilibrium} \mbox{reservoirs: squeezed states, etc}}

The reservoirs in this chapter are nonequilibrium state, without a uniquely defined temperature, but they are fairly classical in nature.
They contain classical mixtures of particles at different energies.
Others have considered making engineered reservoirs that are quantum in nature, 
such as squeezed-states~\cite{Rossnagel2014Jan,Correa2014Feb,Abah2014May,Manzano2016May,Agarwalla2017Sep,Manzano2018Oct,Niedenzu2018Jan,Ghosh2019Apr} or other quantum correlations~\cite{Scully2003Feb,Francica2017Mar}.
These proposals involved exploiting such quantum reservoirs to demonstrate demonic effects (apparent violations of the second law),
and the authors often attributed these effects to the quantum nature of the reservoirs.
However, the N-demons presented in this chapter makes it clearer that quantum-ness in not the necessary condition for such demonic effects, it the the non-equilibrium-ness that is crucial.
Making a reservoir of squeezed states or with other quantum correlations is but one way to make a non-equilibrium reservoir.  Of course, it can be more difficult to quantify the potential of such a quantum reservoir to do useful work, compared to a reservoir that is in a classical non-equilibrium state. Often this is done via calculations of the quantum reservoir's {\it ergotropy} \cite{Francica2017Mar,Niedenzu2018Jan,Manzano2018Oct,Ghosh2019Apr}.

\chapter{Classifying demons \mbox{\& the definition of strict Maxwell demons}}
\label{Chap:strict-demon}

Often in physics, understanding is more important than terminology,  
however inventing precise terminology can help classify things, and thereby help our understanding.
Section~\ref{Sect:demon-or-heatengine} addressed the question of whether an autonomous \cite{autonomous} Maxwell demon should simply be called a heat engine.  Section~\ref{sect:Ndemon} presented the N-demon, but said that it was different from a Maxwell demon. Now, we are a position to be more precise, and present a classification of different types of demon?

Historically, many things have been called Maxwell demons. Many papers use the term ``Maxwell demon'' for any system that appears to violate the second law of thermodynamics.  
However, as our understanding improves it is good to distinguish different types of demons, and their different physics.
In particular, the existence of N-demons has necessitated a more precise definition of a traditional Maxwell demon.
In this direction, Freitas and Esposito \cite{Freitas2021Mar} have recently
proposed a type of demon that they call {\it strict Maxwell demons}.   
This type of demon has a crucial property in common with the most famous Maxwell demon, described in Section~\ref{Sect:intro-Maxwell}, that is absent from all other systems described in this review thus far.  
This chapter gives the definition of such a strict Maxwell demon, and outlines a few examples.

\section{Classifying demons}
\label{Sect:Freitas}

Consider the most famous Maxwell demon, that described in Section~\ref{Sect:intro-Maxwell}, where the demon opens or closes the door based on what it observes about particles arriving at the door.  In the ideal case, the demon does not change the energy of any particle in the working fluid when it observes them, nor when it opens and closes the door.   Thus  
it operates without changing the energy of any particles in the working fluid.

This is clearly different from the N-demon in the previous chapter, which operated by redistributing energy between particles in the working fluid.   This is why we coined the term ``N-demon'' for it \cite{Whitney2016Jan}, rather than calling it a Maxwell demon.

Freitas and Esposito \cite{Freitas2021Mar} went further, and argued that ``strictly'' a system is a Maxwell demon, only if it does not change the energy of any particles in the working fluid.
To be more precise, they argued that a system is only a strict Maxwell demon, if the operating principle of a demon requires no change in the energy of any particle in the working substance. This more precise definition is for experimental applications; it is impossible to prove experimentally that a device has strictly zero effect on the energy of each particle, but it is possible to show that the operating principle does not rely on any such effect.

However, this definition is a very challenging condition to fulfil.
It clearly excludes the N-demons discussed above, however it also the trapdoor (or valve) that Maxwell proposed as a potential demon.
It also excludes a huge number of things called ``Maxwell demons'' in the literature.  
This should not discourage us, because 
I think it is very helpful to classify demons using this condition, while accepting that not everything we call a ``demon'' satisfies the condition.

In this context, one can imagine three categories of autonomous \cite{autonomous} demons;\footnote{A similar classification is possible for non-autonomous demons, but they are beyond the scope of this review.}
\begin{itemize}
\item {\bf Strict Maxwell demons:} As defined above, these are demons that make a system turn heat into work (reducing that system's entropy) without ever changing the energy of a particle in that system. Some examples are outlined in Section~\ref{Sect:strict-demon-examples} below, such as \cite{Sanchez2019Oct,Freitas2021Mar,Freitas2022Sep,Freitas2023Jan}.

\item {\bf N-demons:} Demons that  make a system turn heat into work (reducing that system's entropy) without an {\it average} change of energy of that system, but in which energy is given or taken at the level of individual particles. It does this by injecting a non-equilibrium distribution of particles into the system it acts, as described in the previous chapter, such as \cite{Whitney2016Jan,Sanchez2019Nov,Hajiloo2020Oct,Ciliberto2020Nov,Deghi2020Jul}.
\item {\bf Demonic heat engines:}  Demons that make a system turn heat into work (reducing that system's entropy) by extracting heat from that system.  Macroscopically, these look like normal heat engines exploiting a temperature difference, in which the demon is coupled to a reservoir that is colder than system it is acting upon. However, their microscopic mechanism looks demonic, 
involving a microscopic measurement followed by feedback. Then the cold reservoir is typically necessary to reset the demon's state after each measurement-feedback cycle. This heats the cold reservoir, as it absorbs entropy from the demon each time it resets the demon's state. Examples include the damped trapdoor in Section~\ref{Sect:damped-trapdoor} and the systems discussed in Chapter~\ref{Chap:experiments} such as \cite{Strasberg2013Jan,Koski2015Dec}.
\end{itemize}

The last category (demonic heat engines) should be compared to normal heat engines that use {\it non-demonic} mechanisms to make a system turn heat into work (reducing that system's entropy). Such non-demonic heat engines use a temperature difference to extract heat from that system, but cannot be interpreted in term of a measurement-feedback cycle.
In principle, there is a bit of subjectivity about what may or may not be interpreted as a measurement-feedback cycle. However, in practice, I have not seen a disagreement on this point, so this subjectivity is not a significant issue.  

A second subjective question is whether demons in the last category (demonic heat engines) can be thought of as strict Maxwell demons in  limit where the demon reservoir's temperature $T_{\rm demon} \to 0$. Take the damped trapdoor in Section~\ref{Sect:damped-trapdoor} (for which $T_{\rm ph}$ plays the role of $T_{\rm demon}$) or the systems discussed in Chapter~\ref{Chap:experiments} such as \cite{Strasberg2013Jan,Koski2015Dec}.
The mechanism in such systems relies on the energy taken from each particle in the system being of order $k_{\rm B} T_{\rm demon}$ or larger. In the limit, $T_{\rm demon} \to 0$, one can take this energy to zero as well. So one could say it is a strict Maxwell demon in this limit.
However, for any finite temperature $T_{\rm demon}$ no matter how small, the operating principle of this demon relies on the demon extracting a small but finite amount of heat from the system it acts on.  Thus one could equally say the system is never a strict Maxwell demon.
I slightly prefer to say it is never a strict Maxwell demon (as zero temperature is never achievable), but would not object if someone prefers to say it is a strict Maxwell demon in the limit of zero temperature.

Finally we note that some demons may be {\it hybrids} between the various categories.
This occurs when their operating principles fulfil more than one category at the same time.
A simple example is a demon that extracts particles from the system it acts on, replacing them with a non-equilibrium distribution, except that the non-equilibrium distribution has a slightly different average energy from the particles extracted \cite{Hajiloo2020Oct}. Then the demon is effectively operating as an N demon and a demonic heat engine {\it at the same time}.
Similarly, a system that looks like a hybrid of an N demon and the strict Maxwell demon is given in Ref.~\cite{Whitney2016Jan}.
If one forgets that the demon may be such a hybrid, running on multiple operating principles at the time, it may appear that the system is generating more useful work that allowed by the laws of thermodynamics.   However carefully taking into account all operating principles will show that there is no violation of the second law.

\section{\mbox{Experimental distinction} \mbox{between categories of demon}}

Freitas and Esposito \cite{Freitas2021Mar} suggest an elegant way to quantify the difference between these different categories of demon,
through experimental observables that do not require looking at the detailed workings of the demon.
Thus if one had a demon acting on a system, one could say if it was a strict Maxwell demon, an N-demon or a demonic heat engine by just studying the energy flows in the system (without measuring anything about the demon).

The demonic heat engine can be distinguished from the others by measuring average energy flows in the system; if this average heat flow is not conserved, then it is neither a 
strict Maxwell demon nor an N demon, so it is  a demonic heat engine (although it could also simply be a non-demonic heat engine).

The distinction  between a strict Maxwell demon and an N demon lies in the long-time fluctuations in the currents through the demon.
One starts by defining a time-window $t$, and defining the average energy current 
that flowing into reservoir $\alpha$ in that time-window in the usual manner as
$\bar{I}^{\rm E}_\alpha = (U_\alpha(t) - U\alpha(0))/t$, where $U_\alpha (t)$ 
is the internal energy of reservoir $\alpha$ at time $t$.
One then repeatedly measures this in reservoirs  1 and 2 many times, to get good statistics on it.
One then defines the Pearson correlation coefficient which measures correlations in the fluctuations about the average;
\begin{eqnarray}
R &=& {1 \over \sigma_1 \sigma_2} \Big\langle\big(\bar{I}^{\rm E}_1-\big\langle\bar{I}^{\rm E}_1\big\rangle\,\big)  \,\big(\bar{I}^{\rm E}_2-\big\langle\bar{I}^{\rm E}_2\big\rangle\,\big) \Big\rangle, \qquad 
\end{eqnarray}
where the average $\langle \cdots \rangle$ is over the repeated measurements, and the standard deviation $\sigma_\alpha =  \sqrt{\langle\big(\bar{I}^{\rm E}_\alpha-\big\langle\bar{I}^{\rm E}_\alpha\big\rangle }$. 
If the system is a strict Maxwell demon, then in the long time limit $\bar{I}^{\rm E}_1=- \bar{I}^{\rm E}_2$ in all realizations and $\lim_{t\to\infty}  R=-1$. \footnote{One should not expect that $\bar{I}^{\rm E}_1=- \bar{I}^{\rm E}_2$ at short times, because particles must spend some time (often a variable amount of time) in the region of interaction with the demon, so the energy current leaving reservoir 1 at time $t$ will correspond to an energy current into reservoir 2 at a slightly later time.}

We can guess the range of values of $\lim_{t\to\infty}  R$ for 
certain N-demons, using the following handwaving argument.  How this argument needs confirming with more serious calculations.
In the extreme case, where the N-demon injects huge fluctuations about equally into reservoirs 1 and 2, then the long-time limit could have $\bar{I}^{\rm E}_1-\big\langle\bar{I}^{\rm E}_1\big\rangle \sim \bar{I}^{\rm E}_2-\big\langle\bar{I}^{\rm E}_2\big\rangle$ in every repetition of the experiment, even though the system is tuned so the N-demon injects no energy on average, meaning $\big\langle\bar{I}^{\rm E}_1\big\rangle=-\big\langle\bar{I}^{\rm E}_2\big\rangle$. In this extreme case, the N-demon would have 
a long-time Pearson coefficient $\lim_{t\to\infty}  R\to 1$.
In the opposite extreme case, of a very weak N-demon that injects only very small fluctuations
(much smaller than the thermal fluctuations of reservoir 1 and 2), then we can expect 
$\bar{I}^{\rm E}_1 \sim -\bar{I}^{\rm E}_2$ in every repetition of the experiment, and so 
$\lim_{t\to\infty}  R \simeq -1$. However, this very weak N-demon would (at best) only reduce the entropy of reservoirs 1 and 2 by a very small amount.
Thus while a N-demon might have any value of $\lim_{t\to\infty}  R$ satisfying
$-1<\lim_{t\to\infty}  R < 1$, it seems likely that values close to $-1$ will only be possible at the expense of the N-demon having almost no effect.
It would be great to quantify this, and see if there is a strict bound on how negative
$\lim_{t\to\infty}  R$ can be for an N-demon with a given strength (as quantified by the amount by which it reduces the entropy of reservoirs 1 and 2).

Thus for now we can say that a demonic system that significantly reduces
the entropy of reservoir 1 and 2, while having the long-time Pearson coefficient, 
$\lim_{t\to\infty}  R$, within error-bars of $-1$ is  almost certainly a strict Maxwell demon rather than an N-demon.

Note that many of the demons in the literature that rely on measurement and feedback 
do not fulfil the conditions to be a ``strict'' Maxwell demon. Indeed, at the time of writing I  only know of three theoretical proposals that fulfil this conditions, all of them are recent and none have yet been demonstrated experimentally.

\begin{figure}
\centerline{\includegraphics[width=0.8\columnwidth]{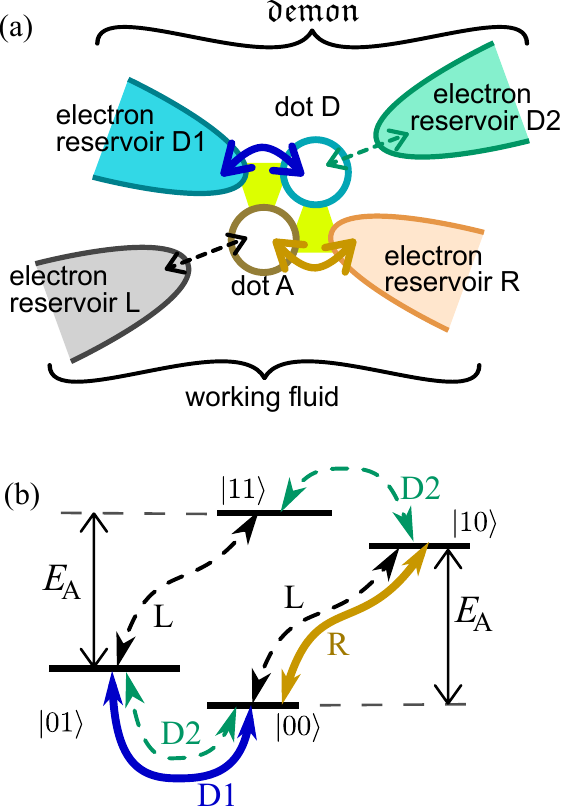}}
\caption{
\label{Fig:strict-demon}
A version of the strict Maxwell demon proposed by Freitas and Esposito \cite{Freitas2021Mar}. 
Dashed arrows indicate weak tunnelling, while thick arrows indicate strong tunnelling, however the strong tunnelling can only occur when the adjacent dot is empty. (b) The levels are labelled $|n_{\rm A}\ n_{\rm D}\rangle$, where $n_{\rm A}$ and $n_{\rm D}$ are the occupations of dots A and D.  The transitions are due to tunnelling events, 
and the colour of the arrow is the same as the equivalent tunnelling event in (a).}
\end{figure}

\subsection{A few strict Maxwell demons}
\label{Sect:strict-demon-examples}

The first proposal that I know of that fulfils the strict demon conditions is that 
of S\'anchez, Samuelsson and Potts~\cite{Sanchez2019Oct}, which slightly predates Freitas and Esposito's definition of a strict demon. It has some advantages over Freitas and Esposito's subsequent proposal \cite{Freitas2021Mar}, but I explain Freitas and Esposito's proposal first, because it is a bit simpler.

A version of the proposal of Freitas and Esposito \cite{Freitas2021Mar} is shown in Fig.~\ref{Fig:strict-demon}.  The demon consists of reservoirs D1 and 2 with dot D between them. It relies on a strong bias (or a strong temperature difference) between reservoir D1 and D2, so electrons want to flow from D1 to D2 through dot D.  
The demon acts on a working fluid consisting of dot A with reservoirs L and R. 
If dot A is empty then electrons tunnel more easily from reservoir D1 to dot D than 
from dot D to reservoir D2, so dot D will be full. However when dot A contains an electron, it blocks tunnelling between reservoir D1 and dot D, so dot D empties (into reservoir D2). 
Thus the state of dot D is a microscopic measurement of dot A's state; if dot D is empty there is a high chance dot A is occupied, and vice versa.
Then dot D has a feedback effect on the working fluid, making tunnelling from dot A to Reservoir R much more likely than tunnelling back into reservoir L.
Thus the demon makes it more likely that electrons flow from reservoir L to reservoir R than the reverse. 
This remains the case if reservoir R has a slightly higher bias than reservoir L, meaning that work is being generated in the working fluid.
This can be thought of as a kind of electron drag; using a large bias to force a strong electron flow from D1 to D2, drags electrons from L to R, even against a small bias.
This means the entropy generated in the demon is large than the entropy reduction in the working fluid, so the second law is satisfied.

The reason that this is a strict demon is that electron entering and leaving dot A always have energy $E_{\rm A}$, irrespective of the occupation of dot D, see Fig.~\ref{Fig:strict-demon}.  Thus the demon is not changing the energy of any electrons in the working substance, it is just moving them from left to right. So the work generated by the flow from L to R must be coming entirely from a reduction of heat in reservoirs L and R.

While this is a simple and elegant theoretical model, 
it will be extremely hard to achieve experimentally.
It is only a strict demon if electrons enter and leave dot A with the same energy, $E_{\rm A}$. This is only the case if electrons in the two dots do not repel each other the energy of state, for which the energy of $|11\rangle$ equals the sum of the energies of states $|10\rangle$ and $|01\rangle$. If this were not the case, then the transition marked by $E_{\rm A}$ in Fig.~\ref{Fig:strict-demon} will have different energies. The transition from $|00\rangle$ to $|10\rangle$ will continue to have energy $E_{\rm A}$, but the transition from $|01\rangle$ to $|11\rangle$ will have energy $E_{\rm A}+U$, where $U$ is the Coulomb energy due to repulsion between dots A and D.  Thus an electron will often enter dot A with energy $E_{\rm A}+U$ and leave with energy $E_{\rm A}$, meaning it is no longer a strict Maxwell demon.
Instead it becomes a hybrid between a strict Maxwell demon and a demonic heat engine.
To achieve the regime where it is purely a strict Maxwell demon, one must implement a situation
where electron repulsion can stop tunnelling into dot D when there is an electron in dot A, but that there is no repulsion between electrons when they are in dots A and D.  This seems almost impossible, so this does not seem a viable route to an experimental demonstration of a strict Maxwell demon.

Ref.~\cite{Sanchez2019Oct} avoids this problem by allowing the electron in one dot to affect the energy-level of the electron in another, but adds leads with band gaps that accept only the electrons whose energy has not changed, thereby enforcing the requirements of a strict  Maxwell demon.  They also had a slightly more complex device, involving a pair of dots in the demon.

\begin{figure}
\centerline{\includegraphics[width=0.9\columnwidth]{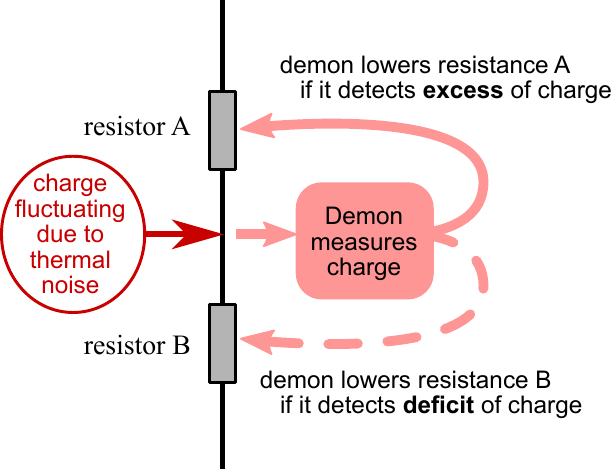}}
\caption{
\label{Fig:Esposito-resistances-demon}
A sketch of the operating principle of the macroscopic Maxwell demon proposed in Ref.~\cite{Freitas2023Jan}. 
As the charge fluctuates in the central region (between resistors A and B), the demon raises lowers resistances A or B, to ensure that charge flows from bottom to top on average.
In Ref.~\cite{Freitas2022Sep,Freitas2023Jan}, the resistors are actually MOS transistors, so the demon change their resistance by applying a voltage to them.  
It is then shown that the demon itself can be made  
from a pair of MOS transistors on another wire, making an autonomous four-terminal device that acts as a Maxwell demon.}
\end{figure}

Freitas and Esposito found another way to avoid this problem, in their  
very recent proposal for a {\it macroscopic} demon \cite{Freitas2022Sep,Freitas2023Jan}. It is similar in principle to that shown in Fig.~\ref{Fig:strict-demon}, but it is made of macroscopic wires and MOS transistors. In this case, there is a wire containing two resistors (A and B) in series, both with controllable resistances (each one actually being a MOS transistor), see Fig.~\ref{Fig:Esposito-resistances-demon}. Thermal noise cause the charge fluctuations in the region between the two resistors, which the demon detects. Then whenever the 
the fluctuation creates an excess of charge in this region, the demon lowers the resistance on resistor A. In contrast, whenever the fluctuation causes a deficit of charge in this region,the demon lowers the resistance on resistor B. One can see that this will turn thermal fluctuations into an average electrical current from bottom to top, which can be used to do work. 
They then show how the demon can be made  
from another pair of MOS transistors on another wire which has a voltage bias across it. They show that the demon can push current against a bias on the wire it is performing measurement and feedback on (the wire sketched in Fig.~\ref{Fig:Esposito-resistances-demon}), so long as that bias is less than the voltage bias on the demon's wire.
Thus while while the demon is reducing the entropy of the system it is acting one, it is generating more entropy, so the second law hold globally.  
Under reasonable experimental conditions, this demon's effect on the energy of the particles
that it detects in negligible, so it is a strict demon under the above definition.

\subsection{\mbox{A strict Maxwell demon needs} \mbox{two reservoirs}}

It is notable that the strict Maxwell demon (like the N demon) needs more than one reservoir operate.  That is why the example given here are four reservoir systems; two reservoirs for the demon and two reservoirs for the system it acts on.  The reason the demon needs more than one reservoir is simple, we can see it from the second law of thermodynamics as follows.
Since there is no energy flow into the demon, its reservoir cannot heat up, so its entropy cannot increase.  As a result, it cannot cause an entropy decrease in another system.
Hence the laws of thermodynamics forbid a strict Maxwell demon from operating with only has a single reservoir.

In contrast, if the demon has two (or more) reservoirs that are not in equilibrium with each other, they can cause degrees-of-freedom in the demon (such as dot D in Fig.~\ref{Fig:strict-demon}) to have non-equilibrium dynamics.  These non-equilibrium dynamics then allow the demon to operate as required.  Thermodynamically, the flow of particles between the demon's two reservoirs generates entropy, so it is allowed to reduce the entropy of another system.  The second law only forbids that it reduces another system's entropy by more than its own increase in entropy. The remarkable and counter-intuitive thing is that it can  reduce a system's entropy without 
changing the energy of any particles in that system.

\chapter{A fully quantum analysis of the two-level trapdoor}
\label{Chap:quantum-trapdoor}

This chapter briefly discusses a method called {\it real-time diagrammatics} or {\it real-time transport theory} \cite{Schoeller1994Dec,Konig1996Dec,Schoeller1997,Konig2000,Governale2008Apr,Saptsov2012Dec,Saptsov2014Jul,Schulenborg2016Feb,Whitney2018Aug,Lindner2019May}, for treating the two-level trapdoor in a fully quantum manner. 
One can use it to prove that
the results of the rate-equations in chapter~\ref{Chap:rate-eqns} can be derived from a weak-coupling approximation of this fully quantum model. I give a flavour of the proof here, but skip some technical details.
I indicates how to calculate quantum corrections to the rate equation results in
chapter~\ref{Chap:rate-eqns}, but do not actually so here.

\begin{figure}[b]
\centerline{\includegraphics[width=0.85\columnwidth]{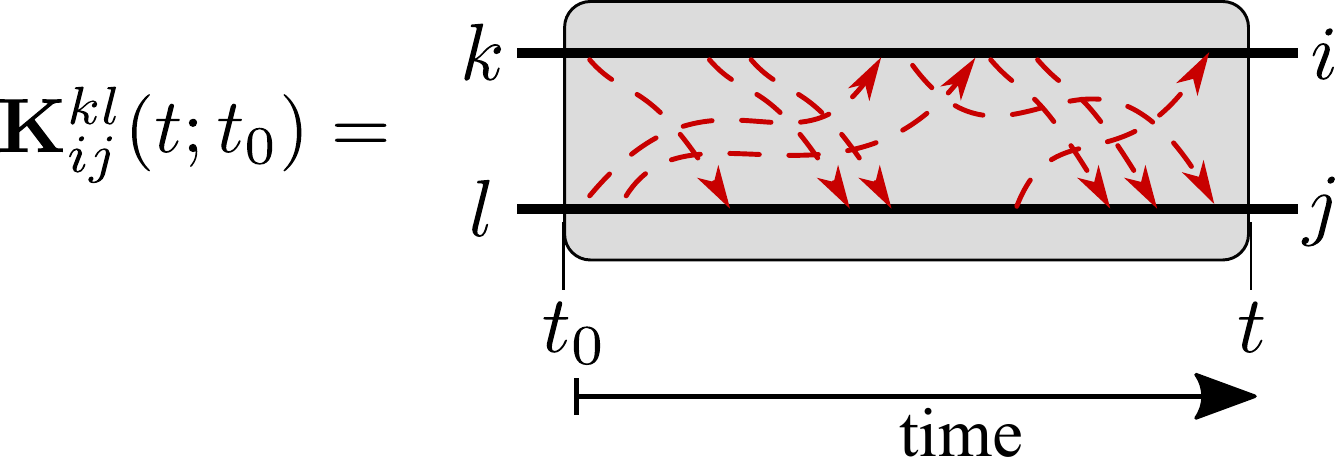}}
\caption{\label{fig:propagator} 
A sketch of the propagator of the density matrix, showing its double-line structure corresponding to the Keldysh contour. }
\end{figure}

\section{\mbox{Treating non-equilibrium} \mbox{quantum systems}}
\revisit{
The critical point in treating non-equilibrium quantum systems is that one needs to work with the density matrix of the quantum system, not its wavefunction.  More strictly one needs to work with the density matrix of the quantum system and the reservoirs, and then carefully eliminate (trace out) the reservoir degrees-of-freedom that are of little interest, but not their effect on the 
state of the small quantum system, and the flows in and out of that small quantum system.
}

The time-evolution of the density matrix of the quantum system and the reservoirs takes the form of \cite{Fano1957Jan}
\begin{eqnarray}
\rho^{\rm total}(t) = {\cal U}(t;t_0) \ \rho^{\rm total}(t_0)\ {\cal U}^{\dagger}(t;t_0)
\label{Eq:rho}
\end{eqnarray}
where the density matrix $\rho(t)$ is $n \times n$ and thus so is the  
evolution operator ${\cal U}(t;t_0)$ from time $t_0$ to time $t$.
If the full Hamiltonian for system and reservoirs is time-independent, as we will assume here then we simply have\footnote{For time-dependent Hamiltonians,  $ {\cal U}(t;t_0)$ is given by a product of infinitesimal short slices of evolution, with the time-evolution in any given slice being that under the Hamiltonian at that time.} 
\begin{eqnarray}
 {\cal U}(t;t_0) = \exp\big[-\rmi {\cal H}_{\rm total} \ (t-t_0)\big]
\end{eqnarray}
for the $n \times n$ Hamiltonian of system and reservoirs, ${\cal H}_{\rm total}$.

This means that, in general, 
the value of $ij$th element of the density matrix at time $t$ depends on the value of the $kl$th element of the density matrix at time $t_0$, via
\begin{eqnarray}
\rho^{\rm total}_{ij}(t) = \sum_{kl}{\cal U}_{ik}(t;t_0) \ \rho_{kl}^{\rm total}(t_0)\ {\cal U}_{lj}^{\dagger}(t;t_0).
\end{eqnarray}
Hence there is a double-index structure for evolution in time.
In the Keldysh way of seeing this, this is a double-line evolving in time (the Keldysh contour), in which the upper-line is the first index on the density matrix and lower line    
is the second. 
Eq.~(\ref{Eq:rho}) can be written as 
\begin{eqnarray}
\rho^{\rm total}_{ij}(t) = \big[\mathbf{K}^{\rm total}(t;t_0)\big]_{ij}^{kl}  \ \rho^{\rm total}_{kl}(t) 
\end{eqnarray}
where $\mathbf{K}^{\rm total}(t;t_0)$ is the propagator from the density matrix at time $t_0$ to the system density matrix at time $t$.
So $\big[\mathbf{K}^{\rm total}(t;t_0)\big]_{ij}^{kl} = {\cal U}_{ik}(t;t_0)\,{\cal U}^\dagger_{lj}(t;t_0)$.
If we trace out the reservoir degrees-of-freedom (and skip many details), this double-index structure remains, and
\begin{eqnarray}
\rho^{\rm sys}_{ij}(t) = \mathbf{K}_{ij}^{kl}(t;t_0)  \ \rho^{\rm sys}_{kl}(t) 
\end{eqnarray}
but the formula giving $\mathbf{K}(t;t_0)$ is much more complicated than that for  
$\mathbf{K}^{\rm total}(t;t_0)$, as we will see below.
Here, $ \mathbf{K}_{ij}^{kl}(t;t_0)$ is a weight relating a given matrix element at time $t_0$ to a different one at time $t$, and that this weight is the same for all initial density matrices.  In this context, it is natural to sketch  $\mathbf{K}_{ij}^{kl}(t;t_0)$
as the double-line in Fig.~\ref{fig:propagator}.
In other contexts this double-index structure is called a super-operator, 
however we will always think of it in terms of the Keldysh contour.

\section{Modelling the trapdoor system}

We take the Hamiltonian of the system described in Eqs.~(\ref{Eq:Hsys}-\ref{Eq:Hres}) 
and treat the system-reservoir couplings as a perturbation acting on the exact dynamics 
that the system and reservoirs would have if they were not coupled.
We do this in the interaction representation, which physically means finding the system's many body eigenstates and treating the system in the basis which rotates with those 
eigenstates. And then doing the same for the reservoir.  In this interaction picture, the only dynamics are those caused by the system-reservoir couplings, which greatly simplifies the analysis.  Of course, we will see that when calculating observables, we have to transform back to the laboratory basis. This transformation that will bring back the dynamics of the system under its Hamiltonian, and the dynamics of the reservoir under its Hamiltonian.

I start by writing the trapdoor system (dot A + double dot) in its many-body eigenbasis of four-states, just as in chapter~\ref{Chap:rate-eqns}.
We then write its 4-by-4 density matrix in this basis;
\begin{eqnarray}
\rho_{\rm sys}(t) = \left(\!\!\begin{array}{cccc} 
\rho_{0\widetilde{{\rm c}};0\widetilde{{\rm c}}}(t) & 
\rho_{0\widetilde{{\rm c}};0\widetilde{{\rm o}}}(t)\!\! & 0 & 0\\ 
\rho_{0\widetilde{{\rm o}};0\widetilde{{\rm c}}}(t) &
\rho_{0\widetilde{{\rm o}};0\widetilde{{\rm o}}}(t)\!\! & 0 & 0\\  
0 & 0 &
\rho_{1\widetilde{{\rm c}};1\widetilde{{\rm c}}}(t) & 
\rho_{1\widetilde{{\rm c}};1\widetilde{{\rm o}}}(t) \\
0 & 0 & 
\rho_{1\widetilde{{\rm o}};1\widetilde{{\rm c}}}(t) &
\rho_{1\widetilde{{\rm o}};1\widetilde{{\rm o}}}(t)
\end{array} \!\!\right) 
\nonumber \\
\end{eqnarray}
The zeros are here by anticipation; it turns out that the dynamics of the Hamiltonian that we consider
will make these matrix elements (which correspond to coherent superposition of 0 and 1 electron in dot A) decay to zero, so it is simpler to choose them to be zero from the beginning.
We go to the basis which rotates with this density matrix under the system Hamiltonian,
defining $\widetilde{{\rm c}} \to +$ and  $\widetilde{{\rm o}} \to -$ in this basis,
so for example 
\begin{eqnarray}
\rho_{0\widetilde{{\rm c}};0\widetilde{{\rm o}}}(t) 
= \exp\big[-\rmi (E_{0\widetilde{{\rm c}}} - E_{0\widetilde{{\rm o}}}) t\big]\rho_{0+;0-}(t).
\label{Eq:phasefactor-example} 
\end{eqnarray}
From now on we will only work in this rotating basis, we use $+$ and $-$ in place of 
$\widetilde{{\rm c}}$ and $\widetilde{{\rm o}}$ to remind us of this, and to remind us that the final state must be transformed back to the laboratory basis.  
It is extremely convenient 
for everything that follows to write the 8 non-zero density matrix elements as a column vector,
\begin{eqnarray}
\underline{\rho}(t) = \left(
\begin{array}{c} 
\rho_{\rm 0+;0+}(t) \\ 
\rho_{\rm 0+;0-}(t) \\ 
\rho_{\rm 0-;0+}(t) \\ 
\rho_{\rm 0-;0-}(t) \\  
\rho_{\rm 1+;1+}(t) \\ 
\rho_{\rm 1+;1-}(t) \\ 
\rho_{\rm 1-;1+}(t) \\ 
\rho_{\rm 1-;1-}(t)
\end{array} \right)
\end{eqnarray}
This is because its full dynamics (due to the reservoir coupling),
can then be written as 
\begin{eqnarray}
\frac{\rmd}{\rmd t} \,\underline{\rho}(t) &=& \underline{\underline{U}}  \ \underline{\rho}(t)
\label{Eq:rho-evolution-eqn}
\end{eqnarray}
where the matrix elements of the vector $\underline{\rho}(t)$ are now those of the reduced density-matrix for the system state after the reservoir has been traced out.
Our job is now to calculate $\underline{\underline{U}}$ using real-time diagrammatics.

\begin{widetext}
\section{Real-time diagrams}
We now write down the diagrams that we need to sum, without providing a detailed explanation.
The detailed explanation is technical, and is reviewed in Ref.~\cite{Schoeller1997}.
Recalling the definition of $L_\beta^\alpha$ in Eq.~(\ref{Eq:overlap_L}), 
we have the following matrix elements for system-reservoir interactions
\begin{eqnarray}
\raisebox{8pt}{$0\alpha$} \,\raisebox{8pt}{$\mbox{-}\hskip -1.3mm\oplus\hskip -1.2mm\mbox{-}$} 
\hskip -2.5mm\raisebox{-2pt}{\textbrokenbar} \hskip 2.5mm\raisebox{8pt}{$1\beta$}&=&  \gamma (E) \langle 1\beta| 0\alpha \rangle \ =\ \gamma(E) \ L^\alpha_\beta\ ,
\\
\nonumber
\\
\nonumber
\\
\raisebox{8pt}{$1\alpha$} \,\raisebox{8pt}{$\mbox{-}\hskip -1.3mm\ominus\hskip -1.2mm\mbox{-}$} 
\hskip -2.5mm\raisebox{-2pt}{\textbrokenbar} \hskip 2.5mm\raisebox{8pt}{$0\beta$}&=&  
\gamma (E) \langle 0\beta|1\alpha \rangle  \ =\ \gamma (E)\ L^\alpha_\beta \ ,
\\
\nonumber
\\
\nonumber
\\
\raisebox{-8pt}{$1\alpha'$} \,\raisebox{-8pt}{$\mbox{-}\hskip -1.3mm\oplus\hskip -1.2mm\mbox{-}$} 
\hskip -2.5mm\raisebox{2pt}{\textbrokenbar} \hskip 2.5mm\raisebox{-8pt}{$0\beta'$}&=&  \gamma (E) \langle 1\alpha' |0\beta'\rangle  \ =\ \gamma (E)\ L^{\alpha'}_{\beta'} \ ,
\\
\nonumber
\\
\nonumber
\\
\raisebox{-8pt}{$0\alpha'$} \,\raisebox{-8pt}{$\mbox{-}\hskip -1.3mm\ominus\hskip -1.2mm\mbox{-}$} 
\hskip -2.5mm\raisebox{2pt}{\textbrokenbar} \hskip 2.5mm\raisebox{-8pt}{$1\beta'$}&=&  \gamma (E) \langle 0\alpha' |1\beta'\rangle  \ =\ \gamma (E)  \ L^{\alpha'}_{\beta'}\ .
\end{eqnarray}

\revisit{ We can bolt these interaction vertices together with propagators
for the reservoir modes,  and write the integral over all reservoir modes as an integral over energy.  Skipping technical details, we then get diagrams like this:
}
\begin{eqnarray}
{\cal A}_{0\beta,0\alpha'}^{0\alpha,0\alpha'}  &=& \ 
\tikzmarkin{a}(0.3,-0.8)(-0.2,1.1) \ 
\raisebox{18pt}{$0\alpha$}\hskip -4mm \raisebox{-15pt}{$0\alpha'$}\hskip -2mm
\raisebox{2em}{
\xymatrix{
\raisebox{0.2pt}{$\oplus$} \hskip -1.4mm \ar@{-}[r]^{1+}  \ar@{--}@/_12pt/[r] & \hskip -1.4mm \raisebox{0.2pt}{$\ominus$}
 \\
\ar@{-}[r] &} \ } 
\hskip -2mm
\raisebox{18pt}{$0\beta$}\hskip -4mm \raisebox{-15pt}{$0\alpha'$}
\ \ + \ \ 
\raisebox{18pt}{$0\alpha$}\hskip -4mm \raisebox{-15pt}{$0\alpha'$}\hskip -2mm
\raisebox{2em}{
\xymatrix{
\raisebox{0.2pt}{$\oplus$} \hskip -1.4mm \ar@{-}[r]^{1-}  \ar@{--}@/_12pt/[r] & \hskip -1.4mm \raisebox{0.2pt}{$\ominus$}
 \\
\ar@{-}[r] &} \ } 
\hskip -2mm
\raisebox{18pt}{$0\beta$}\hskip -4mm \raisebox{-15pt}{$0\alpha'$}
\tikzmarkend{a} 
\qquad = 
 -L_+^\alpha  L_\beta^{+} \,g\left(E_{1+}-E_{0\alpha}\right) 
 -L_-^\alpha  L_\beta^{-} \,g\left(E_{1-}-E_{0\alpha}\right), \qquad \qquad 
\label{Eq:diagram_A_first}
\end{eqnarray}
\begin{eqnarray}
{\cal A}_{1\beta,1\alpha'}^{1\alpha,1\alpha'}  &=& \ 
\tikzmarkin{b}(0.3,-0.8)(-0.2,1.1) \ 
\raisebox{18pt}{$1\alpha$}\hskip -4mm \raisebox{-15pt}{$1\alpha'$}\hskip -2mm
\raisebox{2em}{
\xymatrix{
\raisebox{0.2pt}{$\ominus$} \hskip -1.4mm \ar@{-}[r]^{0+}  \ar@{--}@/_12pt/[r] & \hskip -1.4mm \raisebox{0.2pt}{$\oplus$}
 \\
\ar@{-}[r] &} 
} 
\hskip -1mm
\raisebox{18pt}{$1\beta$}\hskip -4mm \raisebox{-15pt}{$1\alpha'$}
\ \ +\  \ 
\raisebox{18pt}{$1\alpha$}\hskip -4mm \raisebox{-15pt}{$1\alpha'$}\hskip -2mm
\raisebox{2em}{
\xymatrix{
\raisebox{0.2pt}{$\ominus$} \hskip -1.4mm \ar@{-}[r]^{0-}  \ar@{--}@/_12pt/[r] & \hskip -1.4mm \raisebox{0.2pt}{$\oplus$}
 \\
\ar@{-}[r] &} 
}
\hskip -1mm
\raisebox{18pt}{$1\beta$}\hskip -4mm \raisebox{-15pt}{$1\alpha'$}
\tikzmarkend{b} 
\qquad = 
 -L_+^\alpha  L_\beta^{+} \,h^*\!\left(E_{1\alpha}-E_{0+}\right) 
 -L_-^\alpha  L_\beta^{-} \,h^*\!\left(E_{1\alpha}-E_{0-}\right), \qquad \ \ \ 
\end{eqnarray}
%
\begin{eqnarray}
{\cal B}_{1\beta,1\beta'}^{0\alpha,0\alpha'}  &=& \ 
\tikzmarkin{c}(0.3,-0.9)(-0.2,1) \ 
\raisebox{18pt}{$0\alpha$}\hskip -4mm \raisebox{-19pt}{$0\alpha'$}\hskip -2mm
\raisebox{2em}{
\xymatrix{
\raisebox{0.2pt}{$\oplus$} \hskip -1.4mm \ar@{-}[r]  \ar@{--}[dr] & 
 \\
\ar@{-}[r] & \hskip -1.4mm \raisebox{0.2pt}{$\ominus$}}} 
\hskip -1mm
\raisebox{18pt}{$1\beta$}\hskip -4mm \raisebox{-19pt}{$1\beta'$}
\ \ + \ \ 
\raisebox{18pt}{$0\alpha$}\hskip -4mm \raisebox{-19pt}{$0\alpha'$}\hskip -2mm
\raisebox{2em}{
\xymatrix{
\ar@{-}[r]   & \ar@{--}[dl] \hskip -1.4mm \raisebox{0.2pt}{$\oplus$} 
 \\
\raisebox{0.2pt}{$\ominus$}\hskip -1.4mm\ar@{-}[r] & }} 
\hskip -1mm
\raisebox{18pt}{$1\beta$}\hskip -4mm \raisebox{-19pt}{$1\beta'$}
\tikzmarkend{c} 
\qquad = \ L_\beta^\alpha  L_{\beta'}^{\alpha'} \Big( g\left(E_{1\beta}-E_{0\alpha}\right) + \,g^*\!\left(E_{1\beta'}-E_{0\alpha'}\right)\Big), \qquad \qquad \ \ \ \ 
\end{eqnarray}
\begin{eqnarray}
{\cal B}_{0\beta,0\beta'}^{1\alpha,1\alpha'}  &=& \ 
\tikzmarkin{d}(0.3,-0.9)(-0.2,1) \ 
\raisebox{18pt}{$1\alpha$}\hskip -4mm \raisebox{-19pt}{$1\alpha'$}\hskip -2mm
\raisebox{2em}{
\xymatrix{
\raisebox{0.2pt}{$\ominus$} \hskip -1.4mm \ar@{-}[r]  \ar@{--}[dr] & 
 \\
\ar@{-}[r] & \hskip -1.4mm \raisebox{0.2pt}{$\oplus$}}} 
\hskip -1mm
\raisebox{18pt}{$0\beta$}\hskip -4mm \raisebox{-19pt}{$0\beta'$}
\ \ +\  \  
\raisebox{18pt}{$1\alpha$}\hskip -4mm \raisebox{-19pt}{$1\alpha'$}\hskip -2mm
\raisebox{2em}{
\xymatrix{
\ar@{-}[r]   & \ar@{--}[dl] \hskip -1.4mm \raisebox{0.2pt}{$\ominus$} 
 \\
\raisebox{0.2pt}{$\oplus$}\hskip -1.4mm\ar@{-}[r] & }} 
\hskip -1mm
\raisebox{18pt}{$0\beta$}\hskip -4mm \raisebox{-19pt}{$0\beta'$}\ 
\tikzmarkend{d} 
\qquad = \ L_{\beta}^{\alpha} L_{\beta'}^{\alpha'} \Big( h^*\!\left(E_{1\alpha}-E_{0\beta}\right) + h\left(E_{1\alpha'}-E_{0\beta'}\right)\Big), \qquad \qquad \ \ \
\end{eqnarray}
\begin{eqnarray}
{\cal C}_{0\alpha,0\beta'}^{0\alpha,0\alpha'}  &=& \ 
\tikzmarkin{f}(0.3,-0.9)(-0.2,1) \ 
\raisebox{18pt}{$0\alpha$}\hskip -4mm \raisebox{-15pt}{$0\alpha'$}\hskip -2mm
\raisebox{2em}{
\xymatrix{
\ar@{-}[r] & 
\\
\raisebox{0.2pt}{$\ominus$} \hskip -1.4mm \ar@{-}[r]_{1+} \ar@{--}@/_-12pt/[r] &\hskip -1.4mm \raisebox{0.2pt}{$\oplus$}
}} 
\hskip -1mm
\raisebox{18pt}{$0\alpha$}\hskip -4mm \raisebox{-15pt}{$0\beta'$}
\ \ +\ \ 
\raisebox{18pt}{$0\alpha$}\hskip -4mm \raisebox{-15pt}{$0\alpha'$}\hskip -2mm
\raisebox{2em}{
\xymatrix{
\ar@{-}[r] & 
\\
\raisebox{0.2pt}{$\ominus$} \hskip -1.4mm \ar@{-}[r]_{1-} \ar@{--}@/_-12pt/[r] &\hskip -1.4mm \raisebox{0.2pt}{$\oplus$}
}} 
\hskip -1mm
\raisebox{18pt}{$0\alpha$}\hskip -4mm \raisebox{-15pt}{$0\beta'$}
\tikzmarkend{f} 
\qquad = \ \Big[ {\cal A}_{0\beta',0\alpha,}^{0\alpha',0\alpha,} \Big]^*,  \hskip 6.5cm \ 
\end{eqnarray}
\begin{eqnarray}
{\cal C}_{1\alpha,1\beta'}^{1\alpha,1\alpha'} &=& \ 
\tikzmarkin{e}(0.3,-0.9)(-0.2,1) \ 
\raisebox{18pt}{$1\alpha$}\hskip -4mm \raisebox{-15pt}{$1\alpha'$}\hskip -2mm
\raisebox{2em}{
\xymatrix{
\ar@{-}[r] & 
\\
\raisebox{0.2pt}{$\oplus$} \hskip -1.4mm \ar@{-}[r]_{0+} \ar@{--}@/_-12pt/[r] &\hskip -1.4mm \raisebox{0.2pt}{$\ominus$}
}} 
\hskip -1mm
\raisebox{18pt}{$1\alpha$}\hskip -4mm \raisebox{-15pt}{$1\beta'$}
\ \ + \ \ 
\raisebox{18pt}{$1\alpha$}\hskip -4mm \raisebox{-15pt}{$1\alpha'$}\hskip -2mm
\raisebox{2em}{
\xymatrix{
\ar@{-}[r] & 
\\
\raisebox{0.2pt}{$\oplus$} \hskip -1.4mm \ar@{-}[r]_{0-} \ar@{--}@/_-12pt/[r] &\hskip -1.4mm \raisebox{0.2pt}{$\ominus$}
}} 
\hskip -1mm
\raisebox{18pt}{$1\alpha$}\hskip -4mm \raisebox{-15pt}{$1\beta'$}
\tikzmarkend{e} 
\qquad  =\  
\Big[{\cal A}_{1\beta',1\alpha}^{1\alpha',1\alpha}\Big]^*, \hskip 6.5cm \ 
\label{Eq:diagram_C_last}
\end{eqnarray}
where I define for compactness that
\begin{eqnarray}
g(E) &=& \Gamma(E)f(E) \ +\  i \, P\hskip -3.5mm \int_{-\infty}^{\infty} \frac{\rmd E'}{\pi} \ \frac{\Gamma(E') \ f(E')}{E'-E}
\\
h(E) &=& \Gamma(E)[1-f(E)] \ +\  i \, P\hskip -3.5mm \int_{-\infty}^{\infty}  \frac{\rmd E'}{\pi} \ \frac{\Gamma(E') \ [1-f(E')]}{E'-E}
\end{eqnarray}
in which ${\scriptstyle P}\hskip-2.6mm \int$ indicates a principal value integral, and the reservoir coupling is given by the parameter 
\begin{eqnarray}
\Gamma(E) = \pi [\gamma(E)]^2\nu(E).
\end{eqnarray}
The integrals in $g(E)$ and $h(E)$ are real, so the first terms in both
$g(E)$ and $h(E)$ are their real parts, and the second terms are their imaginary parts.
\revisit{
In most situations the real part of these terms thus gives
decoherence and relaxation, while the imaginary parts give oscillations or Lamb shifts.
}

There is a great-deal to say about when we are allowed to drop all diagrams beyond the above Fermi golden rule diagrams, which are only the lowest order in system-reservoir coupling.  In short, we can do this when the amplitude of the above diagrams is small compared to the inverse reservoir memory time.  So that the reservoir memory time is fast on the scale of the dynamics induced by the system-reservoir.  This is similar to a Markov approximation, but it is not as strict, because the system dynamics can still be fast on the scale of the reservoir memory time.  This approximation is the diagrammatic equivalent of the Bloch-Redfield approximation \cite{Bloch1957Feb,Redfield1957Jan} (that some call simply the Redfield approximation).   
The difficulty of knowing when this approximation is applicable is to calculate the reservoir memory time, which depends on the reservoir spectrum and diverges as the fermionic reservoir's temperature goes to zero. Indeed, there is also the question of proving that the memory decays fast enough for 
a diagrammatic expansion to converge.  A well-known case where the above approximation definitely breaks down is the Kondo regime. The above approximation is reasonable at temperatures well-above the Kondo temperature (which goes like the system-reservoir coupling), but fails as one lower the temperature to approaches the transition to the Kondo state.

Here we simply assume that the coupling is small enough, and the temperature is high-enough that we are allowed to use this approximation.
Then the above diagrams are the only diagrams that we need to calculate, and the evolution is given by Eq.~\ref{Eq:rho-evolution-eqn} with   
\begin{eqnarray}
\underline{\underline{U}}
\!&=&\!\left(
\begin{array}
{cccccccc} 
\ \phantom{\bigg|}{\cal D}^{0+,0+}_{0+,0+} \ & {\cal C}^{0+,0-}_{0+,0+} & 
{\cal A}^{0-,0+}_{0+,0+} & 0 & 
{\cal B}^{1+,1+}_{0+,0+} & {\cal B}^{1+,1-}_{0+,0+}  & 
{\cal B}^{1-,1+}_{0+,0+}   & {\cal B}^{1-,1-}_{0+,0+} 
\\
{\cal C}^{0+,0+}_{0+,0-} & \ \phantom{\bigg|} {\cal D}^{0+,0-}_{0+,0-} \ & 
0 & {\cal A}^{0-,0-}_{0+0-} &  
{\cal B}^{1+,1+}_{0+,0-} & {\cal B}^{1+,1-}_{0+,0-}  & 
{\cal B}^{1-,1+}_{0+,0-}   & {\cal B}^{1-,1-}_{0+,0-} 
\\
 {\cal A}^{0+,0+}_{0-,0+} & 0 &  
 \ \phantom{\bigg|} {\cal D}^{0-,0+}_{0-,0+} & {\cal C}^{0-,0-}_{0-,0+}\ &
{\cal B}^{1+,1+}_{0-,0+} & {\cal B}^{1+,1-}_{0-,0+}  & 
{\cal B}^{1-,1+}_{0-,0+}   & {\cal B}^{1-,1-}_{0-,0+} 
\\
0 & {\cal A}^{0+,0-}_{0-,0-} & 
{\cal C}^{0-,0+}_{0-,0-} &  \ \phantom{\bigg|} {\cal D}^{0-,0-}_{0-,0-}\ &
{\cal B}^{1+,1+}_{0-,0-} & {\cal B}^{1+,1-}_{0-,0-}  & 
{\cal B}^{1-,1+}_{0-,0-}   & {\cal B}^{1-,1-}_{0-,0-} 
\\
{\cal B}^{0+,0+}_{1+,1+} & {\cal B}^{0+,0-}_{1+,1+}  & 
{\cal B}^{0-,0+}_{1+,1+}   & {\cal B}^{0-,0-}_{1+,1+} &
\ \phantom{\bigg|}{\cal D}^{1+,1+}_{1+,1+} \ & {\cal C}^{1+,1-}_{1+,1+} & 
{\cal A}^{1-,1+}_{1+,1+} & 0 
\\ 
{\cal B}^{0+,0+}_{1+,1-} & {\cal B}^{0+,0-}_{1+,1-}  & 
{\cal B}^{0-,0+}_{1+,1-}   & {\cal B}^{0-,0-}_{1+,1-} &
{\cal C}^{1+,1+}_{1+,1-} & \ \phantom{\bigg|} {\cal D}^{1+,1-}_{1+,1-} \ & 
0 & {\cal A}^{1-,1-}_{1+,1-} 
\\
{\cal B}^{0+,0+}_{1-,1+} & {\cal B}^{0+,0-}_{1-,1+}  & 
{\cal B}^{0-,0+}_{1-,1+}   & {\cal B}^{0-,0-}_{1-,1+} &
 {\cal A}^{1+,1+}_{1-,1+} & 0 &  
 \ \phantom{\bigg|} {\cal D}^{1-,1+}_{1-,1+} & {\cal C}^{1-,1-}_{1-,1+}\ 
\\ 
{\cal B}^{0+,0+}_{1-,1-} & {\cal B}^{0+,0-}_{1-,1-}  & 
{\cal B}^{0-,0+}_{1-,1-} & {\cal B}^{0-,0-}_{1-,1-}  &
0 & {\cal A}^{1+,1-}_{1-,1-} & 
{\cal C}^{1-,1+}_{1-,1-} &  \ \phantom{\bigg|} {\cal D}^{1-,1-}_{1-,1-}\
\end{array}\right)\qquad
\label{Eq:Umatrix}
\end{eqnarray}
where I define  for compactness
\begin{eqnarray}
{\cal D}^{0\alpha0\beta}_{0\alpha0\beta}
= {\cal A}^{0\alpha0\beta}_{0\alpha0\beta}+{\cal C}^{0\alpha0\beta}_{0\alpha0\beta},
\\
{\cal D}^{1\alpha1\beta}_{1\alpha1\beta}
= {\cal A}^{1\alpha1\beta}_{1\alpha1\beta}+{\cal C}^{1\alpha1\beta}_{1\alpha1\beta},
\end{eqnarray}
\end{widetext}

If we were to simply forget the off-diagonal terms in this matrix, we would recover exactly those rate-equations discussed in chapter~\ref{Chap:rate-eqns}.  Four of the diagonal elements (the first, fourth, fifth and eighth) would give the evolution of the population of the states (i.e.~occupation probability of each state), and the remaining four diagonal elements would tell us that coherences decay to zero.
However, we see that the off-diagonal elements of $\underline{\underline{U}}$ couple the coherences to the populations, stopping the coherences decaying to zero.  As a result, in general the steady-state of the trapdoor system will not have the same populations as those given by the rate equations, and it will not be a mixed state in the basis of its many-body eigenvalues because it will contain quantum coherences.

At the end of this chapter, I will explain that the off-diagonal terms can be systematically neglected in the weak-coupling limit, where the system-reservoir coupling is small enough that the matrix elements of  $\underline{\underline{U}}$ are much smaller than the level-spacing in the system.   In this limit, and only in this limit, do the rate-equation of chapter~\ref{Chap:rate-eqns} become a good approximation of the fully quantum dynamics.

\section{\mbox{Symmetries of this} \mbox{evolution matrix}}

\subsection{Symmetries and Positivity}

Dynamics are said to be {\it positive}, if it only ever generates probabilities that are real and in the range from 0 to 1. All legitimate dynamics must be positive.\footnote{There is a stronger condition, called {\it complete positivity} that all quantum dynamics must obey, but it is too technical to consider here.  In any case, dynamics cannot be completely positive without being positive.}  

To ensure this positivity, $\underline{\underline{U}}$ must have symmetries that ensure that the density matrix remains hermitian under evolution in time for any initially hermitian density matrix. If this were not the case, then some probabilities would become complex.
To ensure that it is the case, 
we require that
\begin{eqnarray}
\underline{\underline{U}}\,^{m,n}_{i,j} = 
\Big[\underline{\underline{U}}\,^{n,m}_{j,i} \big]^* 
\label{Eq:U-symmetries1}
\end{eqnarray}
where in our case each index ($i$, $j$, $m$ or $n$) has four possible states;
$0+$, $0-$, $1+$ and $1-$.
The proof that Eq.~(\ref{Eq:U-symmetries1}) ensures the Hermiticity of the density matrix can be done as follows.  We start with 
${\rmd\over \rmd t}\rho_{ij} = \sum_{m,n} U^{m,n}_{i,j} \rho_{mn}$.
which means that 
${\rmd\over \rmd t}\rho_{ji}^* = \sum_{mn} \Big[U^{m,n}_{j,i}\Big]^* \rho_{mn}^*
= \sum_{m,n} \Big[U^{n,m}_{j,i}\Big]^* \rho_{mn}$,
where we get the second equality by using $\rho_{mn}^*=\rho_{nm}$ and then exchange dummy indices $n$ and $m$.
Now the fact that $\rho$ is Hermitian at all times means that 
${\rmd\over \rmd t}\rho_{ij}= {\rmd\over \rmd t}\rho_{ji}^*$,
this can only be the case if $\underline{\underline{U}}$ obeys 
Eq.~(\ref{Eq:U-symmetries1}).

A second requirement for positivity is that the trace of the density matrix sum is one, to ensure probabilities sum to one, at all times. 
To ensure that this is the case we require that
\begin{eqnarray}
\sum_i 
\underline{\underline{U}}\,^{m,n}_{i,i} = 0 
\label{Eq:U-symmetries2}
\end{eqnarray}
for all $i,m,n$. 
The proof that this ensure density matrix has a trace equal to one at all times of the following. If the trace of the density matrix is constant in time, then $\sum_i{\rmd\over \rmd t}\rho_{ii}=0$ hence we require that $\sum_i \underline{\underline{U}}^{m,n}_{i,i} \rho_{mn}=0$,
and this is satisfied for any $\rho_{mn}$ by having Eq.~(\ref{Eq:U-symmetries2}) .

\begin{widetext}
In our case, Eq.~(\ref{Eq:U-symmetries1}) mean the matrix $\underline{\underline{U}}$ must have the following structure
\begin{eqnarray}
\underline{\underline{U}}
&=&\left(
\begin{array}
{cccccccc}
\myBoxed\mathtt{A}\phantom{^*}  & 
\myBoxed\mathtt{a}\phantom{^*} & 
\myBoxed\mathtt{a}^* &  
\myBoxed 0\phantom{^*} & 
\myBoxed\mathtt{B}\phantom{^*} & 
\myBoxed\mathtt{b}\phantom{^*} & 
\myBoxed\mathtt{b}^* &  
\myBoxed\mathtt{C}\phantom{^*} 
\\
\mathtt{c}\phantom{^*} & 
\mathtt{d}\phantom{^*} & 
0\phantom{^*} & 
\mathtt{e}\phantom{^*} &
\mathtt{f}\phantom{^*} & 
\mathtt{g}\phantom{^*} & 
\mathtt{h}\phantom{^*}  & 
\mathtt{i}\phantom{^*} 
\\
\mathtt{c}^* & 
0\phantom{^*} & 
\mathtt{d}^* & 
\mathtt{e}^* &
\mathtt{f}^* & 
\mathtt{h}^* & 
\mathtt{g}^* & 
\mathtt{i}^*
\\
\myBoxed 0\phantom{^*}&  
\myBoxed\mathtt{j}\phantom{^*} & 
\myBoxed\mathtt{j}^*&  
\myBoxed\mathtt{D}\phantom{^*}&  
\myBoxed\mathtt{E}\phantom{^*}& 
\myBoxed\mathtt{k}\phantom{^*} & 
\myBoxed\mathtt{k}^* & 
\myBoxed\mathtt{F}\phantom{^*}
\\
\myBoxed\mathtt{G}\phantom{^*} &  
\myBoxed\mathtt{l}\phantom{^*} & 
\myBoxed\mathtt{l}^* &  
\myBoxed\mathtt{H}\phantom{^*} &  
\myBoxed\mathtt{I}\phantom{^*} & 
\myBoxed\mathtt{m}\phantom{^*} & 
\myBoxed\mathtt{m}^* & 
\myBoxed 0\phantom{^*} 
\\
\mathtt{n}\phantom{^*} & 
\mathtt{p}\phantom{^*} & 
\mathtt{q}\phantom{^*} & 
\mathtt{r}\phantom{^*} &
\mathtt{s}\phantom{^*} & 
\mathtt{t}\phantom{^*} & 
0\phantom{^*} & 
\mathtt{u}\phantom{^*} 
\\
\mathtt{n}^* & 
\mathtt{q}^* & 
\mathtt{p}^* & 
\mathtt{r}^* &
\mathtt{s}^* & 
\mathtt{0}\phantom{^*} & 
\mathtt{t}^* & 
\mathtt{u}^* 
\\
\myBoxed\mathtt{J}\phantom{^*} & 
\myBoxed\mathtt{v}\phantom{^*} & 
\myBoxed\mathtt{v}^* &
\myBoxed\mathtt{K}\phantom{^*} &
\myBoxed 0\phantom{^*} & 
\myBoxed\mathtt{w}\phantom{^*} & 
\myBoxed\mathtt{w}^* &
\myBoxed\mathtt{L}\phantom{^*}
\\
\end{array}
\right)
\label{Eq:Umatrix-symmetries}
\end{eqnarray}
where capital letter ($\mathtt{A,B,C,}\cdots$) represent real numbers, 
and small letters ($\mathtt{a,b,c,}\cdots$) represent complex numbers.
Eq.~(\ref{Eq:U-symmetries2}) means that the sum of the elements in boxes in each column is zero; so the first column has $\mathtt{A + G +J =0}$, the second column has 
$\mathtt{a + j +l +v =0}$, and so on.
Note that there is no symmetry reason to have zeros on the two diagonals of the matrix,
this comes from the particularities of the model we study here, we simply copied the zeros across from Eq.~(\ref{Eq:Umatrix}).
Happily, by inspection of Eq.~(\ref{Eq:diagram_A_first}-\ref{Eq:diagram_C_last}), we see that they respect the symmetries in Eq.~(\ref{Eq:U-symmetries1}-\ref{Eq:U-symmetries2}).

\end{widetext}

These conditions are necessary for positivity, but they are not sufficient, the dynamics could still generate a Hermitian density matrix whose diagonal elements sum to one, but where some diagonal elements leave the range from 0 to 1, becoming unphysical.  The proof of positivity is only complete, if one also proves that no probability is ever negative \footnote{If one proves both that no probability is ever negative, and that the probabilities sum to one, then it means that all probabilities are in the allowed range from 0 to 1.}.
I do not currently believe that there is a simple requirement on $\underline{\underline{U}}$ to ensure that no probability becomes negative.  I studied it for a simple model for which  I proved positivity\cite{Whitney2008Apr}, however the proof that no probability became negative was rather complicated, even for that simple model.  I hope that the future will bring simpler proofs for more complicated systems.  

However, here we can apply an (arguably better) approach to the question of positivity and complete positivity, that is unavailable for phenomenological models of open quantum systems. 
Here we start from a Schr\"odinger equation for the system and its reservoirs, and we know that this equation respects positivity and complete positivity.  If we make no approximation when tracing out the reservoir degrees-of-freedom, then {\it it is proven} that the 
dynamics of the remaining degrees-of-freedom will respect positivity and complete positivity.
This means that any violations of positivity can only come from the approximations that we make.  These approximations need to be controlled, with estimated error bars, irrespectively of whether the approximate dynamics respect positivity or not. If the error bars on probabilities are of size $\delta$, we should not be surprised if the approximate dynamics occasionally generates a negative probability of order $-\delta$.  This simply means the probability is indistinguishable from zero at our level of approximation.  

Within this point of view, it is clear that it is of little interest to place ``positivity'' on a pedestal above the precision of the approximation\footnote{In contrast, positivity and complete positivity are considered gold-standard tests of phenomenological models of open quantum systems, because there are few other ways to evaluate their reliability.}. This makes me critical of works in which one makes additional approximations (such as course-graining the dynamics) to restore the positivity of the dynamics at the cost of reducing the precision of the dynamics.

\subsection{Symmetry leading to the second law}
I identified a symmetry at the level of diagrams in my 2018 work \cite{Whitney2018Aug}, 
which leads the system obeying the second law of thermodynamics.
The symmetry is as follows.

Firstly, for every diagram (such as those in Eqs.~(\ref{Eq:diagram_A_first}-\ref{Eq:diagram_C_last}) there is a diagram that corresponds 
to it rotated by 180$^\circ$ in the plane of the page.
For example, the first diagram in 
${\cal A}_{1\beta,1\alpha'}^{1\alpha,1\alpha'}$ is the 180$^\circ$ rotation of the first diagram in ${\cal C}_{0\alpha,0\beta'}^{0\alpha,0\alpha'}$, under 
the exchange of indices $\beta'\leftrightarrow\alpha $ 
and $\alpha'\to\alpha$, $\beta\to\alpha'$ and $\alpha\to\beta'$.
Similarly, the first diagram in 
${\cal B}_{1\beta,1\beta'}^{0\alpha,0\alpha'}$ is the 180$^\circ$ rotation of the first diagram in ${\cal B}_{0\beta,0\beta'}^{1\alpha,1\alpha'}$, under 
the exchange of indices $\beta'\leftrightarrow\alpha $ 
and $\beta\leftrightarrow\alpha'$.

Secondly, there is a simple relationship between the values of these diagrams which are twins under a 180$^\circ$ rotation.
If we give a twin pair of diagrams the values ``$\bm{{\rm V}}$'' and ``$\bm{\Lambda}$'', then the relationship reads
\begin{eqnarray}
\bm{\Lambda} = \bm{{\rm V}} \times \exp \big[- \Delta S_{\rm res}[\bm{{\rm V}}] \big] 
\label{Eq:my-180-degree-symmetry}
\end{eqnarray}
where $\Delta S_{\rm res}[\bm{{\rm V}}]$ is the change in the reservoir entropy associated with the diagram $\bm{{\rm V}}$.
This can be observed diagram by diagram for those shown in Eqs.~(\ref{Eq:diagram_A_first}-\ref{Eq:diagram_C_last}).
This entropy change is zero for the diagrams labelled ${\cal A}$ and ${\cal C}$,
however it is finite for diagrams labelled ${\cal B}$. 
These entropy changes have the Clausius form, so that
the diagrams in ${\cal B}_{1\beta,1\beta'}^{0\alpha,0\alpha'}$, which involve
the addition of a electron with energy $E$ to reservoir $j$, have an entropy change of 
\begin{eqnarray}
 \Delta S_{\rm res} =  (E-\mu_j)/T_j.
\end{eqnarray}
Similarly, the diagrams in ${\cal B}_{0\beta,0\beta'}^{1\alpha,1\alpha'}$ involve the removal of an electron with energy $E$ form reservoir $j$, and have an entropy change of 
\begin{eqnarray}
 \Delta S_{\rm res} =  -(E-\mu_j)/T_j.
\end{eqnarray}
To verify that all the diagrams shown in Eqs.~(\ref{Eq:diagram_A_first}-\ref{Eq:diagram_C_last}) do indeed satisfy Eq.~(\ref{Eq:my-180-degree-symmetry}), one needs to recall that the Fermi function obey 
\begin{eqnarray}
{f_j(E) \over 1-f_j(E)} = \exp [-(E-\mu_j)/T_j].
\end{eqnarray}
which is the relation the led directly to the local detail balance relation in Eq.~(\ref{Eq:local-detailed-balance}).
Hence we could call Eq.~(\ref{Eq:my-180-degree-symmetry}) a quantum version of the local-detailed balance relation.

Remarkably, these properties are enough to prove that the quantum system and its reservoirs together obey various fluctuation theorems 
(those briefly summarized in chapter~\ref{Chap:fluctuations}), which means that the system can never violate the second law of thermodynamics.
The proof of this it too long to summarize here, 
but the details are in Ref.~\cite{Whitney2018Aug}.

Interestingly, to the best of my knowledge, no equivalent proof currently exists for the other type of Keldysh theory, that known as Nonequilibrium Green's functions (NEGF) \cite{Haug-Jauho-book}; although some strides have been made in this direction
for non-interacting systems (quadratic Hamiltonians)\cite{Ludovico2014Apr,Esposito2015Feb,Bruch2016Mar} or
adiabatic driving \cite{Ludovico2016Jul}.
So, in general, those using NEGF methods have to look at their results on a case-by-case basis, to check that their modelling and approximations are not violating the second law.

\begin{figure*}[t]
\centerline{\includegraphics[width=0.7\textwidth]{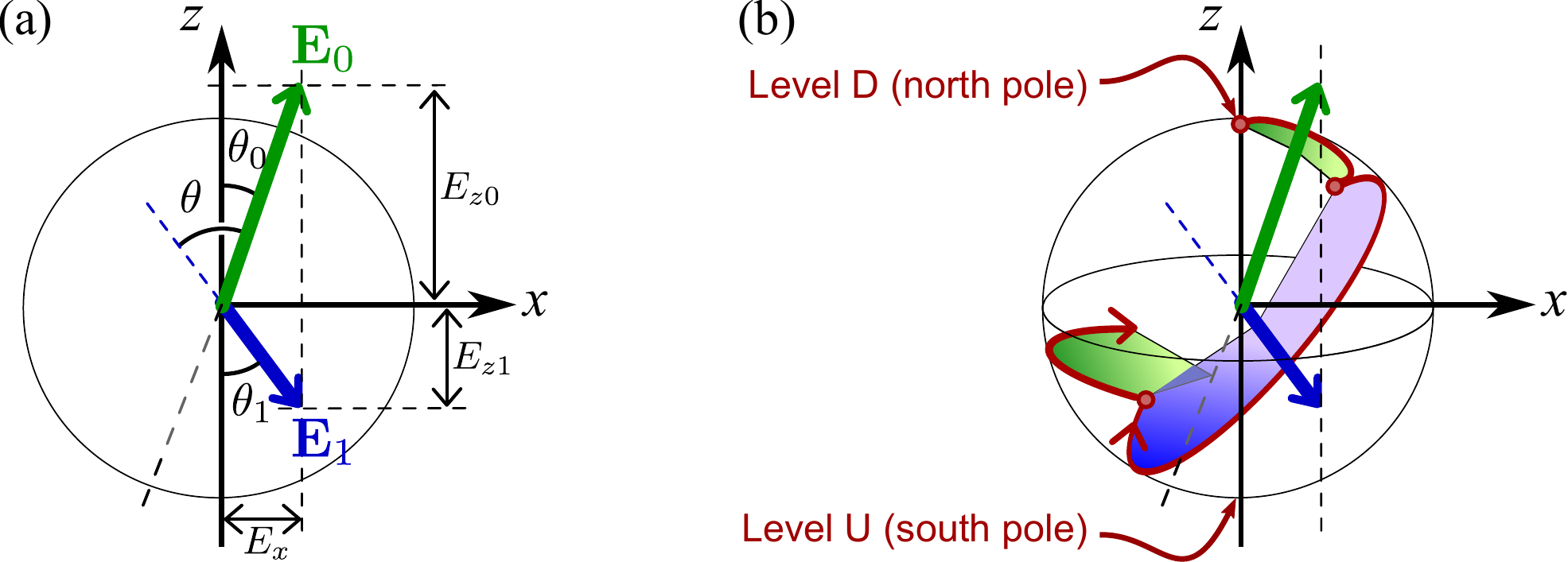}}
\caption{\label{fig:spin-space} 
(a) A Bloch sphere sketch of the two-level trapdoor's dynamics. When dot A is empty, the two-level trapdoor's state evolves around ${\bf E}_0$ (the green vector), as if the green vector is a magnetic field acting on a spin-half. 
When dot A is full, the two-level trapdoor's state evolves around $\bf{E}_1$ (the blue vector). Both  ${\bf E}_0$ and $\bf{E}_1$ are in the $x$-$z$ plane.
(b) A Cartoon of the dynamics, in which we imagine the dynamics of dot A as a classical random switching between state 0 and state 1 (rather than a quantum process of tunnelling). When the state of dot A changes, the two-level system switches between precessing around the green and blue vectors.  As these switches
occur randomly, the overall effect is to push the two-level system towards a
steady-state (not show) that is mixed, and correlated with the state of dot A.} 
\end{figure*}

\section{The steady state}

To find the steady-state, we must take the density matrix in the rotating 
basis, $\underline{\rho}(t)$, and transform back to the density matrix in the laboratory basis, $\underline{\rho}^{\rm lab}(t)$.
The two are related by
\begin{eqnarray}
\underline{\rho}(t) = \underline{\underline{R}}(t) \underline{\rho}^{\rm lab}(t) 
\end{eqnarray}
where $\underline{\underline{R}}$ is a diagonal matrix of phase factors like Eq.~(\ref{Eq:phasefactor-example}).
Substituting this into Eq.~(\ref{Eq:rho-evolution-eqn}) and using the chain rule for the time-derivative, we get
\begin{eqnarray}
\frac{\rmd}{\rmd t} \,\underline{\rho}^{\rm lab}(t) &=& \left[-\underline{\underline{R}}^\dagger(t) 
\left(\frac{\rmd}{\rmd t} \underline{\underline{R}}\right) \ +\  
\underline{\underline{R}}^\dagger(t) \underline{\underline{U}}\ 
\underline{\underline{R}}(t) \right] \underline{\rho}^{\rm lab}(t)
\nonumber \\
\label{Eq:rho-evolution-eqn2}
\end{eqnarray}
Thus the steady state density matrix is given by
\begin{eqnarray}
0 &=& \left[-\underline{\underline{R}}^\dagger(t) 
\left(\frac{\rmd}{\rmd t} \underline{\underline{R}}\right) \ +\  
\underline{\underline{R}}^\dagger(t) \underline{\underline{U}}\ 
\underline{\underline{R}}(t) \right] \underline{\rho}^{\rm lab}_\steady(t)
\nonumber \\
\label{Eq:rho-ss-eqn}
\end{eqnarray}
As $\underline{\underline{R}}$ is a diagonal matrix of phase-factors, it is easy to show that $\underline{\underline{R}}^\dagger(t) 
\left(\rmd \underline{\underline{R}} \big/\rmd t\right)$ is a diagonal matrix
whose diagonal elements are of the form $-\rmi \Delta E$, where $\Delta E$ is the energy
difference between the two eigenstates associated with that density matrix element;
meaning it is zero for diagonal density matrix elements.

\section{Weak-coupling limit}
If we now take the limit where the system-reservoir coupling is much smaller than the energy spacings in the trapdoor system, we can expect Eq.~(\ref{Eq:rho-ss-eqn}) to be dominated by the first term in the square-brackets.
If we throw away the second term completely, it would decouple into 8 independent equations, tell us that all density matrix elements decay to zero.
This works for the off-diagonal elements of density matrix,
but is wrong for the diagonal ones because 
$\underline{\underline{R}}^\dagger(t) 
\left(\rmd \underline{\underline{R}} \big/\rmd t\right)$ is zero for all diagonal elements of the density matrix, so they are principally determined by the 
second term in the square-brackets.
A careful inspection of the second term in the square-bracket indicates that
the elements of $\underline{\underline{U}}$ associated with the diagonal elements of the density matrix coincide with those in the rate equations used in chapter~\ref{Chap:rate-eqns}.

This approximation then gives us the justification for the rate-equations in 
chapter~\ref{Chap:rate-eqns}.  It tells us that it the system-reservoir coupling 
is much less than the level-spacing between many-body eigenbasis of the system Hamiltonian, the steady-state will exhibit no coherences in that basis, and will
simply be the solution of the classical rate equations.
Thus in this limit we need go no further, as the results were already calculated in 
chapter~\ref{Chap:rate-eqns}.

\subsection{Beyond this weak-coupling limit}
It is easy to see that when the system-reservoir coupling starts to be significant on the scale of the level spacing, then coherent superpositions of the system's many-body eigenstates  will start to enter the dynamics.
Then the density matrix of the steady state given by Eq.~{\ref{Eq:rho-ss-eqn}) will not be diagonal in the basis of many-body eigenstates of the system Hamiltonian.
As Eq.~(\ref{Eq:rho-ss-eqn}) is simply a set of 8 simultaneous equations, there must be an analytic algebraic solution of Eq.~(\ref{Eq:rho-ss-eqn}), but it would be crazy to find it by hand.  Rather than going further with finding it using computer algebra here, I prefer to give a simple picture of it.

Without doing any mathematics, we can get an intuitive feeling 
from the cartoon in Fig.~\ref{fig:spin-space}.  It treats the tunnelling of particles in and out of dot A as if it was a classical process, in which we remove the Hamiltonian of dot A and the reservoirs, and replace it is a classical charge on the dot which changes
at random intervals between 0 and 1 (like an experimenter randomly flicking a switch that changes the charge on the dot A). When the charge is absent the two-level trapdoor state evolves around the axis $\bm{E}_0$ in its Bloch sphere, but when the charge is present
it evolves around axis $\bm{E}_1$.  As the charge changes randomly the two-level system will reach a steady state in the long-time limit that is somewhere in the Bloch sphere that does not correspond to a mixed state of its eigenstates.  Even in the long-time limit the average state of the two-level trapdoor when the change is present will be different from when the charge is absent, so it will be correlated with the state of dot A.  This correlations are likely to also be present in the full quantum problem,
suggesting the steady state will be a correlated state of dot A and the two-level trapdoor.  What is not clear without detailed calculation is whether this 
will be an entangled state of the two quantum systems or just a correlated state.

Of course, after finding the steady-state, one has to calculate the currents of particles and energy in this steady state, and use that to calculate the entropy of the reservoirs. 
Luckily there is one thing we can be sure of, these diagrams have the symmetry identified in my 2018 paper \cite{Whitney2018Aug}, which makes it impossible for this system to violate the laws of thermodynamics.

\chapter{When do fluctuations matter in nanoelectronics?}
\label{Chap:when-fluctuations-are-important}

The previous chapters looked at average steady-state flows 
(particle flows, heat flows, etc) in various types of demons. 
However, we know that there are thermal fluctuations in these flows, even in the steady state. 
They are the origin of fluctuations of the entropy; fluctuations that mean that there are short periods of time during which the entropy is reducing rather than increasing.
They are what invalidated the formulations of the second law that existed prior to Smoluchowski's formulation, given in section~\ref{sec:2ndlaw}.

With this in mind, you should immediately ask the following; how do we know when the fluctuations are small enough that the flows are well-approximated by the average steady-state values calculated in previous chapters?
For this reason, this chapter provides estimates of magnitude of fluctuations in nanoelectronics, and identifies when they are of similar size as the average flows.  This provides a rough guide to whether knowing only the average flows allows you to predict what will be observed in any given experimental measurement. 
We will see that the fluctuations are small enough to neglect in most experimental situations.  They are only as big as the average on very short timescales, for which both the fluctuations and the average flows are very small.  
So an experiment will be dominated by the average flows, unless that experiment is  
very sensitive and very fast, where below I make simple estimates of how sensitive and how fast.  This means that the results given for average flows in previous chapters will capture most experiments on the various types of demons discussed there.

At the same time, these simple estimates will draw attention to how challenging it is to  
to reach the regime where fluctuations are as large as the average flows.  This is the regime where fluctuations that reduce the entropy will be common enough to observe with ease.  Thus 
it will not be easy to experimentally observe fluctuations that reduce entropy in nanoelectronic systems.

\section{Fluctuations in a classical gas}
\label{sect:air-fluctuations}
When we first learn about thermodynamics, we learn that
thermal motion is random, but that average properties dominate over fluctuations in macroscopic systems. 
{\it In principle} all air molecules in a room could all be in the same corner of the room, leaving no air in the rest of the room. However, the probability of this happening is negligible.
The argument is that average properties grow like the number of particles in the gas, while the fluctuations grow like the {\it square-root} of the number of particles.

If one applies this to the air pressure at a given position in the room, we
have an average air pressure $\langle P\rangle\propto N_{\rm air}$, for $N_{\rm air}$ air molecules in the room, while the magnitude of a typical fluctuation in air pressure,\footnote{Here the magnitude of a typical fluctuation of $P$ is the root-mean-squared average over the fluctuations; $\Delta P_{\rm rms} = \sqrt{\big\langle \big(P -\langle P\rangle \big)^2\big\rangle}$.} $\Delta P_{\rm rms} \propto \sqrt{N_{\rm air}}$. 
This means the perceived change in the pressure 
\begin{eqnarray}
\frac{\Delta P_{\rm rms}}{\langle P\rangle}\ \propto \ \frac{1}{ \sqrt{N_{\rm air}}} \ \xrightarrow{\ {\rm large}\ N_{\rm air}\ }  \ \mbox{negligible}.  
\end{eqnarray}

If the air in the room were in thermal equilibrium, the main effect of thermal fluctuations would be to cause Brownian motion of very small particles, like dust particles.
In real life, this effect is dwarfed by other non-thermal 
fluctuation effects which are due to the fact that air in a room is never in perfect thermal equilibrium; there are drafts, people moving in the room, convection due to temperature gradients, etc.
Yet, even these non-thermal fluctuations only have a small effect; 
they move dust particles in the air, but rarely move the papers on a desk.

\section{\mbox{Fluctuations in nanoscale} \mbox{flows}}

Following the logic of section~\ref{sect:air-fluctuations} above, one can guess that fluctuations could be significant in the flow through nanoscale systems, simply because  nanoscale systems are small.
However, to judge if this is the case, we have to be more quantitative.
The sorts of questions we need to answer are
\begin{itemize}
\item
When do we decide that fluctuations are {\it significant}?
\item
What parameter plays a role in nanoscale flows similar to $N_{\rm air}$ in  section~\ref{sect:air-fluctuations} above? In some cases it could be number of electrons in the nanostructure, however I will argue that it is often the number of electrons that flow {\it through} the nanostructure in a given observation time, $t_{\rm obs}$.
\item
What are other factors that can be used to increase or decrease the ratio of fluctuations 
to average flows?
\end{itemize}
To answer these questions in a semi-quantitative manner, we can take well-known results from scattering theory.

\subsection{When are fluctuations significant?}

There are two answers to this question.
One answer is that fluctuations are significant whenever they are large enough to be measurable.
This response is applicable whenever we are interested in what the fluctuations tell us about the fluctuating system.  The oldest example
is Brownian motion, which helped confirm the atomistic picture that gases and liquids are not continuous fluids, they are made up of discrete molecules.
Examples abound in mesoscopic physics, where studying the noise tells us much about a nanostructure that we do not learn from the average flow alone.
The most famous example being the identification of fractionally charged particles in the fractional quantum 
Hall state \cite{De-Picciotto1997Sep,Saminadayar1997Sep}.
This was summed up in Landauer's well-known 1998 statement that 
\textsf{the noise is the signal} \cite{Landauer1998Apr}.

However, here I give another answer to the question.
In the context of understanding the second law of thermodynamics, we are interested in
when the fluctuations cause a qualitative change in the physics, so that knowing the average flows ceases to be enough to capture the result of a given experimental measurement.
In this context,  we know the average flows always increase the entropy, but a fluctuation can reduce the entropy, but this can only happen if the fluctuation is as big as the average. Hence, a good rule of thumb is that that fluctuations become {\it significant} (meaning that they qualitatively change the physics) when they become of similar order as the average flows.

\subsection{Which fluctuations?}

The fluctuations should be those associated with an observable that we are measuring.  
It is possible to measure the state of the nanostructure in real time
\cite{Gustavsson2006Feb}, for which the fluctuating observable is the number of the electrons
in the nanostructure (or in part of the nanostructure).  However, this is  experimentally challenging, and easy to estimate theoretically. 

Instead, here I will take an observable which is much easier to measure experimentally; the electrical current, i.e.~the electron flow through the nanostructure.
In contrast, it is not straight-forward to guess when the fluctuations in this flow are similar in magnitude to the average flow, which makes it worthwhile making such an estimate here.  The estimate can be applied to other flows, such as heat flow.

\section{\mbox{Estimating fluctuations} \mbox{using scattering theory}}

The calculation of fluctuations within scattering theory is reviewed in 
Ref.~\cite{Blanter2000Sep} (see also chapter 6 of Ref.~\cite{Moskalets-scattering-book}).
It is important to recall that scattering theory does not include 
electron-electron interactions, so it does not apply to the trapdoor or experimental demon in previous chapters.
Despite this, it will give us a basic estimate of the magnitude of fluctuations in nanoscale systems.

Let ${\cal N}(t_{\rm obs})$ be the number of electrons that flow in the time $t_{\rm obs}$ during which the system is observed.
Then the average of ${\cal N}(t_{\rm obs})$ is given in terms of the electrical current, $I(t_{\rm obs})$, by
\begin{eqnarray}
\big\langle {\cal N}(t_{\rm obs})\big\rangle 
&=& \frac{1}{e}\int_0^{t_{\rm obs}} \rmd t_1 \big\langle {\cal I}\big\rangle \ = \ t_{\rm obs} \,\big\langle {\cal I}\big\rangle/e\, , \qquad
\label{Eq:averageN}
\end{eqnarray}
where we use the fact that $\big\langle {\cal I}\big\rangle$
is independent of $t_1$. The factor of electronic charge $e$ is 
here, because ${\cal N}(t_{\rm obs})$ is the number of electrons that have flowed in a time $t_{\rm obs}$, when $t \,\big\langle {\cal I}\big\rangle$ is the charge that flows in time $t_{\rm obs}$.
The typical magnitude of the fluctuations of ${\cal N}(t_{\rm obs})$, 
given by the root-mean-square average, is
\begin{eqnarray}
\Delta {\cal N}_{\rm rms}(t_{\rm obs})
&=& \frac{1}{e}\ \sqrt{\int_0^{t_{\rm obs}} \rmd t_1 \rmd t_2 \ S(t_1-t_2) }
\nonumber 
\\
& & \hskip 6mm \approx\ \frac{1}{e}\ \sqrt{\,t_{\rm obs}  \int_{-\infty}^\infty  \rmd \tau \ S(\tau)}\, , \qquad \quad
\label{Eq:DeltaNrms}
\end{eqnarray}
where $S(t_1-t_2)$ is defined in 
Blanter and B\"uttiker's review on noise in scattering theory \cite{Blanter2000Sep}, as 
\begin{eqnarray}
S(t_1-t_2) = \tfrac{1}{2} 
\Big\langle \Delta I(t_1)  \Delta I(t_2) + \Delta I(t_2)  \Delta I(t_1) \Big\rangle.
\end{eqnarray}
The approximation in Eq.~(\ref{Eq:DeltaNrms}) relies on the fact that
we assume the time $t_{\rm obs}$ is much bigger than the timescale on which 
$S(\tau)$ decays, which is the timescale on which electrons traverse the scatterer.

To estimate Eqs.~(\ref{Eq:averageN}-\ref{Eq:DeltaNrms}), we take a two-terminal scatterer in the linear response regime, where the bias is much smaller than temperature; $eV \ll k_{\rm B}T$.  
Then
\begin{eqnarray}
\big\langle {\cal N}(t_{\rm obs})\big\rangle &=& t_{\rm obs}\ G\ V/e
\end{eqnarray}
where $G$ is the two-terminal conductance.
The scattering theory formula for the equilibrium thermal noise, is given in Ref.~\cite{Blanter2000Sep}'s Eq.~(54)
as
\begin{eqnarray}
S(\omega=0) = 4 k_{\rm B} T \, G
\label{Eq:equilibrium-noise}
\end{eqnarray}
where $S(\omega)$ is the Fourier transform,
\begin{eqnarray}
S(\omega) = \int_{-\infty}^\infty \rmd\tau \e^{\rmi \omega \tau} \,S(\tau).
\end{eqnarray}
The noise for a small bias ($eV \ll k_{\rm B}T$) will be similar to the equilibrium noise.  Thus Eq.~(\ref{Eq:DeltaNrms}) gives
\begin{eqnarray}
\Delta {\cal N}_{\rm rms}(t_{\rm obs})
&\approx& \frac{1}{e} \sqrt{\,t_{\rm obs} \ S(\omega=0)}
\nonumber \\
&=& \sqrt{ 4 \,\big\langle {\cal N}(t_{\rm obs})\big\rangle \,\frac{k_{\rm B}T}{eV}} \ . \qquad
\label{Eq:DeltaNrms-final}
\end{eqnarray}
Thus the fluctuations are only significant, in the sense that 
$\Delta {\cal N}_{\rm rms}(t_{\rm obs}) \gtrsim \big\langle {\cal N}(t_{\rm obs})\big\rangle$,
when 
\begin{eqnarray}
\big\langle {\cal N}(t_{\rm obs})\big\rangle &\lesssim& 4 \,\frac{k_{\rm B}T}{eV}\, .
\label{Eq:condition-fluctuations-significant}
\end{eqnarray}
where we recall that $\big\langle {\cal N}(t_{\rm obs})\big\rangle$ grows linearly with the observation time $t_{\rm obs}$.
Once enough time passes that $\big\langle {\cal N}(t_{\rm obs})\big\rangle$ becomes much bigger than this,  the number of particles that flow through the nanostructure in time $t_{\rm obs}$ will have fluctuations much smaller than its average.

If we imagine that $eV/(k_{\rm B}T) \sim 0.1$, then the fluctuations
are only of order (or bigger than) the average when the observation time $t_{\rm obs}$ 
is short enough that less than about 40 electrons have passed through the nanostructure, $\big\langle {\cal N}(t_{\rm obs})\big\rangle\lesssim 40$.  
If many more electrons than this pass through the nanostructure in the observation time then
fluctuations will be insignificant compared to the average flow.

\subsection{Nanostructure size should be similar to an electron wavelength}

The condition for significant fluctuations in Eq.~(\ref{Eq:condition-fluctuations-significant})
requires that the conductance in units of $e^2/h$ 
to be 
\begin{eqnarray}
\frac{G}{e^2/h} &\lesssim&  \left(\frac{k_{\rm B}T}{eV}\right)^2  \frac{h}{k_{\rm B}T \, t_{\rm obs}} 
\label{Eq:condition-fluctuations-significant1}
\end{eqnarray}
It is worth noting that experiments rarely 
measure particle flows on timescale less than a nanosecond, so 
$h\big/(k_{\rm B}T\, t_{\rm obs}) < 0.01$ when $T\sim 1$\, K. 
Hence, if we take $eV/(k_{\rm B}T) \sim 0.1$, the fluctuation are only similar in magnitude to (or bigger than) the average if the conductance 
\begin{eqnarray}
\frac{G}{e^2/h} &\lesssim& 1
\label{Eq:condition-fluctuations-significant2}
\end{eqnarray}
However, it is should be much smaller, if $t_{\rm obs}$ is larger than 1\,ns.

Now, in scattering theory we have
\begin{eqnarray}
\frac{G}{e^2/h} &\sim& \hbox{(number of transverse modes)} 
\nonumber 
\\
& & \qquad \quad \times \hbox{(transmission per mode)}, \qquad
\end{eqnarray}
So we see immediately that the conditions in Eq.~(\ref{Eq:condition-fluctuations-significant1})  or (\ref{Eq:condition-fluctuations-significant2})  typically requires that the scatterer has few transverse modes,
meaning its width should be of order the electron's Fermi wavelength.
This is generally the case for nanostructures, and definitely the case when the nanoscale systems is a quantum dot (like the trapdoor or demon systems in previous chapters),
since the quantum dot has only one transverse mode and it has low transmission.

We recall that these estimates are based on a theory of non-interacting electrons. Electron-electron interactions of the type discussed in previous chapters will modify the details of these estimates, but are unlikely to change the orders of magnitude involved.

\subsection{\mbox{A counter-example: fluctuations} \mbox{in macroscopic wires}}

The condition in Eq.~(\ref{Eq:condition-fluctuations-significant1}) is very hard to fulfil in a macroscopic system with a huge number of transverse modes (a millimetre diameter metal wire has about $10^{14}$ transverse modes). However, it is not impossible to fulfil.
The oldest and best-known example is Johnson-Nyquist noise in very long telegraph cables (typically up to hundreds of kilometres long).  There, the condition is fulfilled by the fact that the transmission per transverse mode is absolutely tiny in such a long wire. The probability of transmission in a diffusive wire of length $L$ with mean-free path of $l$ is $l/L$.
In typical wires $l\sim$ 10\,nm, so the transmission
probability of order $10^{-13}$ for a wire of length $L=$100\,km. 
This means a millimetre diameter wire with $L=$100\,km has a conductance of order $10e^2/h$,
placing it in a regime where fluctuations can be significant.\footnote{This conductance estimate using scattering theory, would only be rigorously correct if the electron's diffusive motion were due to static disorder in the wire.  In reality, it is predominantly due to electron-phonon scattering in the wire; something scattering theory does not capture.  None the less,  for the order of magnitude estimates made here, it is sufficient to make the brutal assumption of treating electron diffusion induced by phonon scattering as if it were diffusion induced by static disorder, and then make estimates based on scattering theory.}
If the observation time $t_{\rm obs}$ is a millisecond, then thermal fluctuations will be bigger than the average signal whenever that signal is less than about a micro-volt (i.e.~$eV/(k_{\rm B} T) \lesssim 10^{-4}$).

\chapter{How stochastic thermodynamics changes our vision of entropy }
\label{Chap:fluctuations}


Most of this review discussed the average flows in autonomous demons made of quantum systems coupled to reservoirs.  However, as we recalled in section~\ref{sec:2ndlaw}, it has long been understood that thermal fluctuations can be significant in small systems,
and certain such fluctuation can reduce the entropy for a short time, 
in an unpredictable and stochastic manner.  
In recent years the theory of {\it stochastic thermodynamics} has been developed
that quantifies such fluctuations, and allows us to derive very elegant fluctuation theorems for such systems. The second law follows as a direct consequence of these fluctuation theorems, so a demonic system that appears to violate the second law will also appear to violate fluctuation theorems.
At the same time the fluctuations contain much more information than the second law. They are equalities, when the second law is only an inequality. They tell use what fluctuations can or cannot do, when the second law only tells us what the average behaviour can or cannot do.

Here, I briefly review these fluctuation theorems (sometimes called fluctuation relations or fluctuation equalities), and then cite the proofs that they apply to all the demonic systems considered in the review.  These proofs use stochastic thermodynamics (or generalizations of it). They are enough to prove that demons considered in this review (such as the trapdoor or the N-demon) all obey the fluctuation theorems, even if they appear superficially to violate them.

I position my brief review of stochastic thermodynamics at a more basic level than more technical reviews, such as Refs.~\cite{Seifert2012Nov,VandenBroeck2015Jan,Benenti2017Jun}.   The reason is that I will use this basic level to argue that the stochastic thermodynamics forces us to change our view of what entropy actually is, for the first time in about 100 years. 
{\it Entropy is no longer a number. It is a distribution!}

This is an uncomfortable change, because it makes entropy a more complicated concept to grasp (when it was already a more complicated concept to grasp than energy, etc).
So we need to justify why the change is worthwhile.
For this I start by listing the important physical consequences of fluctuation theorems, derived by taking the stochastic thermodynamics view of entropy.  At the same time it is worth noting that stochastic thermodynamics is regularly used to calculate detailed properties of individual systems.  Thus it is worth taking seriously any conceptual issues that
it raises about the nature of entropy.

The version of stochastic thermodynamics that we use applies for systems with discrete states, it was introduced in Ref.~\cite{Schmiedl2007Jan} (shortly after the more technical version for continuous states \cite{Seifert2005Jul}). It is well-explained in reviews like\cite{VandenBroeck2015Jan,Benenti2017Jun}.  It has since been fully extended into the quantum domain, including for cases in which there is quantum measurement \cite{Elouard2017Mar}, 
or fully quantum dynamics that could include strong memory effects, including quantum correlations between system and reservoir \cite{Whitney2018Aug}.

\section{\mbox{Steady-state fluctuation} \mbox{theorem}}
\label{Sect:Evans-Searles}

There are multiple {\it fluctuation theorems}  with different names
and different regimes of applicability.  We use the term ``theorem'' for them, because that is 
the most often used in the literature, however their status as ``theorems'' in the mathematical sense is very variable.\footnote{For example my derivations of them in non-Markovian quantum systems was performed with a rigour typical in theoretical physics \cite{Whitney2018Aug}. They relies on a diagrammatic perturbation theory which has never been placed on a sufficiently rigorous mathematical footing to claim that they are theorems in the mathematical sense.}
Each fluctuation theorem tells us about not just about the average entropy change, but also the fluctuations in entropy change.
Each one is also an equality so it clearly tells us more about the system's dynamics
than an inequality like the second law of thermodynamics.

I believe that the easiest fluctuation theorem to understand is the 
one that will be of most interest to us here; {\it the steady-state fluctuation theorem} of Evans and Searles \cite{Evans1994Aug}.
This steady-state fluctuation theorem is a special case of the more recent 
Crook's fluctuation theorem \cite{Crooks1999Sep}.
Taking $P(\DS)$ as the probability of the entropy changing by $\DS$,
the steady-state fluctuation theorem reads
\begin{eqnarray}
P(-\DS) = P(\DS) \e^{-\DS}.
\label{Eq:Evans-Searles}
\end{eqnarray}
In other words, there is a finite probability of the total entropy decreasing as well as increasing, however the entropy is always more likely to  
increase than decrease, so the second law of thermodynamics (as formulated in section~\ref{sec:2ndlaw} above) is satisfied.

All systems described by the theory of rate equations
used in chapter \ref{Chap:rate-eqns}, obey this fluctuation theorem  
when they are in their steady-state. 
Ref.~\cite{Whitney2018Aug} shows that it also hold for a broad class of quantum systems, including those with non-Markovian dynamics (although in non-Markovian cases $P(\DS)$ in Eq.~\ref{Eq:Evans-Searles} must be replaced by a suitably defined conditional probability).

Eq.~(\ref{Eq:Evans-Searles}) has numerous consequences, any system obeying it will have the following properties.
\begin{itemize}
\item[{(i)}] 
{\bf It obeys the second law of thermodynamics} given in section~\ref{sec:2ndlaw}, as explained above.

\item[{(ii)}] {\bf Fluctuations in equilibrium or reversible dynamics do not change the entropy.} In other words, if $\langle\Delta S\rangle = 0$ then all fluctuations have zero entropy change. This is  because the only $P(\DS)$ that satisfies Eq.~(\ref{Eq:Evans-Searles}) and has $\langle\Delta S\rangle = 0$ is a Dirac delta-function centred at zero; $P(\DS)=\delta(\DS)$. 

\item[{(iii)}] {\bf Whenever average entropy is growing, $\langle\DS\rangle > 0$, there is always a finite probability of an entropy reduction.} Thus even a system producing 
a huge amount of entropy on average will occasionally have a fluctuation that reduces the entropy.

\item[{(iv)}] {\bf Significant entropy reductions are exponentially rare.} 
Since $P(\DS)$ cannot exceed one, we can see from  Eq.~(\ref{Eq:Evans-Searles})
that $P(-\DS) < \e^{-\DS}$ so a reduction of the entropy by $\DS$ is exponentially rare. 

\item[{(v)}] {\bf The entropy distribution for negative $\DS$ is completely determined by the entropy distribution for 
positive $\DS$.}
\end{itemize} 

\section{Integral fluctuation theorem}
\label{Sect:integral-fluct-theorem}

Eq.~(\ref{Eq:Evans-Searles}) does not apply to systems that are not in their steady-state, so it does not apply to a system decaying to its steady-state from an arbitrary initial state.
However, there is another fluctuation theorem called the {\it integral fluctuation theorem} \cite{Seifert2005Jul,Schmiedl2007Jan,Seifert2012Nov,VandenBroeck2015Jan,Benenti2017Jun}  that is obeyed by any system described by stochastic thermodynamics (and generalizations like Ref.~\cite{Whitney2018Aug}) irrespective of their initial or final states.\footnote{A similar relation derived under different assumptions previously appeared under the name {\it non-equilibrium partition
identity} \cite{Yamada1967Nov,Morriss1985Feb,Carberry2004Nov}).}
The integral fluctuation theorem reads
\begin{eqnarray}
\Big\langle \exp\big[-\Delta S\big]\Big\rangle \ =\ 1
\label{Eq:Int-Fluct-theorem}
\end{eqnarray}
where $\DS$ is the total entropy change, and the average 
\begin{eqnarray}
\langle \cdots \rangle\equiv \int \rmd (\DS) (\cdots) P(\DS).
\end{eqnarray}
This result is surprising, naively one would expect the exponential to decay to zero as the entropy grows. 
This equality says that this does not happen.

The integral fluctuation theorem is an equality, so it clearly contains much more information that the second law of thermodynamics, which is an inequality.
However it is much harder to interpret than the steady-state theorem 
in Eq.~(\ref{Eq:Evans-Searles}).  
So what are the main consequences of the integral fluctuation theorem?
The answer is not so obvious, or indeed well-known, so we list four consequences that correspond to the first four of the five consequences of the steady-state theorem in the previous section, although they are harder to prove here than they were there.
Any system obeying Eq.~(\ref{Eq:Int-Fluct-theorem}) has the four following properties:
\begin{itemize}
\item[{(i)}] 
{\bf It obeys the second law of thermodynamics} given in section~\ref{sec:2ndlaw}. This has long been know, and one can use the fact that $x \geq 1 - \e^{-x}$ for all
$x$ to show that $\langle\DS\rangle \geq 1- \left\langle \exp[-\DS]\right\rangle =0 $, which is the second law. Formally, this proof is an example of using the Jensen's inequality.

\item[{(ii)}] {\bf Fluctuations in equilibrium or reversible dynamics do not change the entropy.} In other words, if $\langle\Delta S\rangle = 0$ then all fluctuations have zero entropy change, meaning $P(\DS)$ is a Dirac delta-function centred at zero; $P(\DS)=\delta(\DS)$.  
To prove this let us assume the reverse, and see the contradiction.
If we were to assume $P(\DS)$ corresponded to a spread of $\DS$ then we could use the fact that $\e^{-x} >1 - x$ for all $x\neq 0$ to show that,
$\left\langle \exp[-\DS]\right\rangle >1 - \langle\Delta S\rangle$.
Now if $\langle\Delta S\rangle = 0$, this would imply that $\left\langle \exp[-\DS]\right\rangle >1$ in contradiction to Eq.~(\ref{Eq:Int-Fluct-theorem}). 
Thus Eq.~(\ref{Eq:Int-Fluct-theorem}) and $\langle\Delta S\rangle = 0$ 
can only both be satisfied if  $P(\DS)=\delta(\DS)$. Then $\DS$ is not just zero on average, it is also zero in any and all fluctuations.  

\item[{(iii)}] {\bf Whenever the average entropy is growing, $\langle\DS\rangle > 0$, there is always a finite probability of an entropy reduction.} 
In other words, if $P(\DS)$ is the probability of an entropy change  $\DS$, 
one {\it cannot} have $P(\DS)=0$ for all negative $\DS$.
The easiest way to see this is to assume the opposite, that $P(\DS)=0$ for all negative $\DS$. If this were the case, one would have $\int_0^\infty \rmd (\DS) P(\DS)=1$, since the integral over all probabilities must be one. Then the right hand side of Eq.~(\ref{Eq:Int-Fluct-theorem}) would read 
$\int_0^\infty \rmd (\DS) \e^{-\DS} P(\DS)$. This would be less than $\int_0^\infty \rmd (\DS) P(\DS)$, since $\e^{-\DS} < 1$ for positive $\DS$, which would mean that  $\left\langle \exp[-\DS]\right\rangle <1$.
Thus Eq.~(\ref{Eq:Int-Fluct-theorem}) cannot be satisfied if there is zero probability of negative $\DS$.

\item[{(iv)}] {\bf Significant entropy reductions are exponentially rare.} The probability that the entropy change $\DS \leq -X$ is less than $\e^{-X}$ for any $X$. In other words 
\begin{eqnarray}
\int_{-\infty}^{-X} \rmd (\DS) \,P(\DS) \,\leq\, \e^{-X} \ \mbox{ for all } X.
\end{eqnarray}
The proof is straight-forward (see Eq.~(20) of \cite{Jarzynski2011Feb}); 
it relies on the fact that 
$\e^{X}\Theta[\DS+X] \leq \e^{-\DS}$ for all $X$, where $\Theta[x]$ is a Heaviside theta-function (this inequality is easily seen graphically).
Writing the right hand side of Eq.~(\ref{Eq:Int-Fluct-theorem}) as $\int_{-\infty}^{\infty} \rmd (\DS) \e^{-\DS} P(\DS)$ and using the inequality to replace the exponential by the theta function, we see that  $\e^{X} \int_{-\infty}^{-X} \rmd (\DS) P(\DS) \leq 1$, giving the desired result. Formally this proof is an example of using the exponential Chebyshev inequality (also called the exponential moment of the  Chebyshev inequality) \cite{wiki-Chebyshev,Rassoul-Agha}. 

 \end{itemize}  

\section{\mbox{Questioning the foundations} \mbox{of stochastic thermodynamics}}

The fluctuation theorems given above can be derived
using the theory of {\it stochastic thermodynamics}.
This theory was initially developed for small systems with classical dynamics (without memory effects).  It has since been extended to a broad class of quantum systems, including those in which there is quantum measurement \cite{Elouard2017Mar},
or in which quantum mechanics induces strong memory effects \cite{Whitney2018Aug}.
However, the assumptions about entropy that form the foundation of stochastic thermodynamics were not easy for me to accept initially. 
They appear in the definition of $\DS$ in the fluctuation theorems in Eqs.~(\ref{Eq:Evans-Searles}\ref{Eq:Int-Fluct-theorem}), and by extension give meaning to the distribution of entropy changes, $P(\DS)$. 
So what they are and why we should accept them?

Firstly, stochastic thermodynamics implies a system does not have a unique well-defined entropy.  We can no longer say a system has an entropy equal to some numerical value.  Instead, the entropy of a system is quantified by a probability distribution, as we will explain in section~\ref{sect:entropy-dist}. Hence, the system in a given state can have many values for its entropy, each one occurring with a certain probability.  What we consider to be the standard entropy is simply the average of the distribution, but the system has a finite probability to have an entropy very different from this average. 
This is explained in detail in section~\ref{sect:entropy-dist}. 

Secondly, stochastic thermodynamics takes the standard Clausius definition of entropy change in a reservoir. This is different from the definition taken for the small system that is coupled to the reservoir, and seems to be a potential for a contradiction.

These two points are the ones that I had the most difficulty accepting at the beginning, but other people may have different points of doubt.
A pragmatic response to such doubts is to make the analogy with conserved quantities in dynamics. 
Conserved quantities in the dynamics of a system are consequences of conservation laws (conservation of energy, conservation of momentum, etc). They tell us much about a system before we have even solved the dynamics; they give a direct view of what a system will {\it never do}. For example, we know without any calculation that an isolated classical system with low initial energy will never subsequently be
found in a high energy state, because energy is a conserved quantity. 
Some of these conserved quantities (such as energy and momentum)
are sufficiently intuitive to be learnt in school, however others are highly non-intuitive and can only be understood with years of study. Such non-intuitive examples regularly appear in N-body problems in one-dimension, such as the Lieb-Liniger Model, Hubbard model, Kondo model, etc, and are crucial to our understanding of the dynamics of such systems.\footnote{A critical point is to identify and count the number of conserved quantities, since this will determine if the system's dynamics are integrable or not.  If the dynamics is integrable, then it can be found using the Bethe ansatz.} 
Thus, we should be delighted whenever we see a conserved quantity in physics,
even if that quantity is not very intuitive. Eq.~(\ref{Eq:Int-Fluct-theorem}) tells us that 
$\langle \e^{-\DS}\rangle$ is a conserved quantity for a suitable definition of 
$\DS$. Even if this definition of $\DS$ is a bit counter-intuitive, we should accept it because Eq.~(\ref{Eq:Int-Fluct-theorem}) tells us much about the quantum system.
We already know that entropy is rarely an intuitive concept for systems
that are not both (i) macroscopic and (ii) in internal thermal equilibrium.
So why not take the  stochastic thermodynamics definition of entropy in place of the habitual definition, if it allows us to understand things better.

If this point of view does not convince you. Then I suggest that you keep the habitual definition of entropy, and then accept that the $\DS$ in stochastic thermodynamics is not an entropy change, but it is an {\it entropy-like quantity} (which we will explain in the next section), whose average is the habitual entropy change.

\section{\mbox{Entropy is not a number;} \mbox{it is a distribution!}}
\label{sect:entropy-dist}

Here I introduce the main difference between stochastic thermodynamics and traditional thermodynamics.
Traditionally, one take the entropy of an $N$-level system (using the Shannon or von Neumann entropy\footnote{For systems without quantum coherences Eq.~(\ref{Eq:Shannon-entropy}) is Shannon entropy.  For systems with quantum coherences, it is the von Neumann entropy, after rotation to the basis where the quantum system's density matrix is diagonal.}) to be
\begin{eqnarray}
S = \sum_{n} p_n \ln [1/p_n]
\label{Eq:Shannon-entropy}
\end{eqnarray}
where $p_n$ is the probability the system is in level $n$. 
{\it This is NOT what is done in stochastic thermodynamics.}
Instead, an entropy is assigned to each level of the system via the following definition.
\begin{itemize}
\item[] \textbf{\textsf{Definition:}} The entropy of the  $n$th level of a system is defined as $\ln[1/p_n]$, where $p_n$ is the probability that that level is occupied.
\end{itemize}

However, as the system is probabilistic, with a certain probability of being in each level, it means the entropy is also probabilistic, taking a different value depending which level the system is in.
This means that instead of saying the system has entropy $S = \sum_{n} p_n \ln [1/p_n]$, one says it has a 
probability $p_1$ of having entropy $\ln[1/p_1]$, a probability $p_2$ of having entropy $\ln[1/p_2]$, etc.
Thus its entropy is not a number, {\it it is a distribution}. This distribution consisting of $N$ Dirac delta-functions,
\begin{eqnarray}
P(S)&=& p_1 \ \delta \big(S-\ln[1/p_1]\big) 
\nonumber\\
& & + p_2\ \delta \big(S-\ln[1/p_2]\big)
\ +\ \cdots \, .
\end{eqnarray}
Then the usual entropy (Shannon or von Neumann entropy)  
formula in Eq.~(\ref{Eq:Shannon-entropy}) is given by $\langle \DS\rangle$, where the average 
\begin{eqnarray}
\langle \cdots\rangle \ \equiv\ \int dS P(S) (\cdots) \ \ =\\sum_n p_n (\cdots) .
\end{eqnarray}
More generally we see that  the entropy distribution has moments given by
\begin{eqnarray}
\langle S^\alpha \rangle &=&  \sum_n p_n \big(\ln[1/p_n]\big)^\alpha.
\end{eqnarray}
This is an object most people have little experience with, and it 
fools my intuition whenever I forget that I must treat the entropy as a distribution rather than a number. For example, it is easy to find cases 
where $\langle S \rangle <1$ but $\langle S^2 \rangle > 1$, since 
$\langle S^2 \rangle \neq \langle S \rangle^2$.

\subsection{\mbox{Entropy change distribution with} \mbox{a single reservoir}}
Stochastic thermodynamics usually deals with entropy {\it changes} during a system's dynamics,  rather than that system's entropy itself. These dynamics can be quantified as follows; the 
system initially has a probability $p_m^{(0)}$ to be in level $m$, 
goes to a level $n$ with probability $p_{m\to n}$, so the probability to be in the final level $n$ is 
\begin{eqnarray}
p_n =\sum_m p_m^{(0)}p_{m\to n}.
\end{eqnarray}
Assigning entropy to each state as defined in the previous section, 
the initial state $m$ has entropy $\ln\big[1/p_m^{(0)}\big]$ and the final state $n$ has entropy $\ln[1/p_n]$, so the transition from state $m$ to $n$, has an entropy change of $\ln[1/p_n]-\ln[1/p_m^{(0)}]$ in the system.
At the same time, this transition causes changes in the entropy of the reservoirs.
If there is a single reservoir then the transition $m\to n$ injects an energy $E_m-E_n$ into that reservoir, this causes an entropy change in the reservoir
$\big(E_n-E_m - \mu\big)\big/(\kB T)$,
where we use the Clausius definition of entropy here, which is applicable because
we take the reservoir to be in thermal equilibrium at temperature $T$ and electro-chemical potential $\mu$.
So the total entropy change associated with the transition $m\to n$ is
\begin{eqnarray}
\DS(m\to n) = \ln\left[\frac{p_m^{(0)}}{p_n}\right]+ {E_n-E_m-\mu \over \kB T}.
\end{eqnarray}
Then, for a system interacting with a single reservoir, the distribution of total entropy change reads
\begin{eqnarray}
P(\DS) \,=\, \sum_{mn} p_m^{(0)}\,p_{m\to n} 
\ \delta \Big(\DS -\DS(m\to n)\Big).
\label{Eq:P_of_DS-1reservoir}
\end{eqnarray}
Hence, it consists of $N^2$ Dirac delta-functions for an $N$-level system coupled to reservoirs.
In this case, $\alpha$th moment of the entropy production,
\begin{eqnarray}
\langle \DS^\alpha \rangle \!\!\! &=& \!\!\! \sum_{mn} p_m^{(0)}\,p_{m\to n}  
\big(\DS(m\to n) \big)^\alpha
\label{Eq:moments_of_DS-1reservoir}
\end{eqnarray}

\begin{figure*}
\centerline{\includegraphics[width=\textwidth]{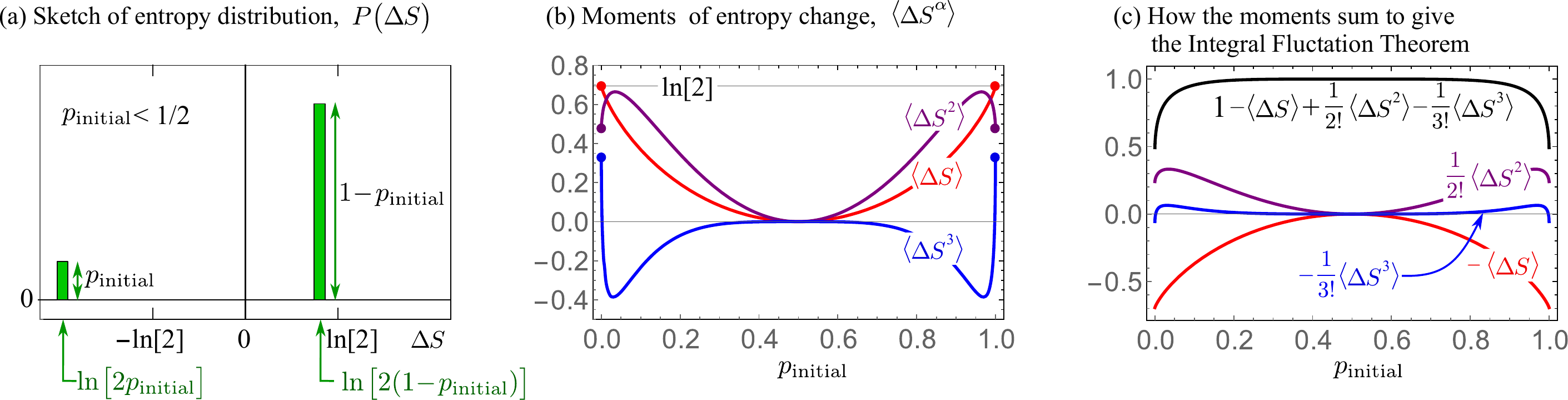}}
\caption{\label{fig:constructing-IFT} 
Entropy change in a two-level system thermalizing to a maximally mixed state.
The system initially has probability $p_{\rm initial}$ to be in one level, 
and $1-p_{\rm initial}$ to be in the other level.  Through its coupling to a very hot reservoir, it thermalizes to a state with probability of $1/2$ in each level.
(a) A sketch of the distribution of entropy change, $\Delta S$, for a 
given $p_{\rm initial}$ (shown for $p_{\rm initial} <1/2$).  
Each rectangle represents a Dirac $\delta$ functions,
so $P(\DS)$ is given by Eq.~(\ref{Eq:P_of_DS-two-level}).  (b) The first few moments 
of the entropy change. (c) The colored curves are the first few moments weighted as in the Taylor expansion of $\langle \exp[-\DS]\rangle = 1- \langle \DS \rangle + \langle \DS^2 \rangle/2! - \langle \DS^3 \rangle/3! + \cdots$. The black curve is this expansion up to 3rd order; the expansion converges rapidly to $1$ as terms 
are added to the Taylor expansion.
} 
\end{figure*}

\subsection{Simple two-level example}

To get a feel for such entropy distributions, 
let us consider a two-level system thermalizing with a reservoir in the high temperature limit ($\kB T$ very much greater than the energy gap between the two-levels).
Then the reservoir's entropy does not change, because $T$ is so large that
$\big(E_n-E_m - \mu\big)\big/(\kB T)\to 0$, but the entropy of the system increases as it thermalizes.
Irrespective of the initial state of the system, its final state 
will be 50\% in one level and 50\% in the other level, which means
$p_{m\to n}=1/2$ for all $m,n$. 
Thus
if the system is initially in one level with probability $p_{\rm initial}$, and in the 
other level with probability $(1-p_{\rm initial})$, then
the distribution of total entropy, given by Eq.~(\ref{Eq:P_of_DS}), is  
\begin{eqnarray}
P(\DS)
&=& (1-p_{\rm initial}) \ \delta \big(\DS-\ln[2(1-p_{\rm initial})]\big) 
\nonumber \\
& &
+ P_{\rm initial} \ \delta \big(\DS-\ln[2p_{\rm initial}]\big) \qquad
\label{Eq:P_of_DS-two-level}
\end{eqnarray}
as sketched in Fig.~\ref{fig:constructing-IFT}a.  
In the limit  $p_{\rm initial}\to 1/2$, the system starts in equilibrium with the bath, then $P(\DS) \to \delta(\DS)$ and there is never any entropy change.
In the limit $p_{\rm initial}\to 0$, one delta function goes to $\ln[2]$ and the other delta-function goes to $-\infty$, so most of the time the entropy increase by a small amount, but very occasionally it reduces by a huge amount.
 
Eq.~(\ref{Eq:P_of_DS-two-level}) gives the moments of $\DS$ as 
\begin{eqnarray}
\langle \DS^\alpha \rangle &=& (1-p_{\rm initial})\left(\ln\big[2(1-p_{\rm initial})\big] \right)^\alpha
\nonumber
\\
& & + 
p_{\rm initial}\left(\ln\big[2p_{\rm initial}\big] \right)^\alpha
\end{eqnarray}
the first few of which are shown in Fig.~\ref{fig:constructing-IFT}b.
All moments are zero for $p_{\rm initial}=1/2$, since there is never any entropy change ($\DS=0$) when the system starts in equilibrium with the reservoir.
Higher even moments are strongly peaked near $p_{\rm initial}=0,1$,
while higher odd moments are strongly dipped (going negative) near $p_{\rm initial}=0,1$. Fig.~\ref{fig:constructing-IFT}c shows what happens if the moments are weighted as in the Taylor expansion of $\exp[-\DS]$; their sum then reproduces the 
Integral fluctuation theorem in Eq.~(\ref{Eq:Int-Fluct-theorem}).  We see that the sum rapidly converges to one as we add orders to the expansion, except near $p_{\rm initial}=0,1$ where it is the high moments that dominate.

\subsection{\mbox{Entropy change distribution with} \mbox{multiple reservoirs}}
If there are multiple reservoirs, the situation is qualitatively similar
to the single reservoir case,
but the details are significantly more complicated.
The entropy change in the reservoirs is not simply a function of the initial and final state of the system.  For example, the system could begin and end the evolution in the same level, 
but during its evolution it could have absorbed energy from a hot reservoir and emitted it to a cold one; thereby increasing total entropy.  To deal with this, one must add a sum over all system trajectories from level $m$ to level $n$ to  Eqs.~(\ref{Eq:P_of_DS-1reservoir}) and (\ref{Eq:moments_of_DS-1reservoir}). 
Suppose a given trajectory $\dtraj$ from level $m$ to level $n$ has a probability $p(\dtraj)$ and 
involves injecting energy $E_i(\dtraj)$ into reservoir $i$. Then 
the total entropy change of that trajectory 
(given by the entropy change of the system plus the sum of the entropy changes in each reservoirs) is
\begin{eqnarray}
\DS(\dtraj) = \ln\left[\frac{p_m^{(0)}}{p_n}\right]+ \sum_i{E_i(\dtraj)-\mu \over \kB T_i}.
\end{eqnarray}
where the sum is over all reservoirs,
and the distribution of total entropy change  
reads
\begin{eqnarray}
P(\DS) \,=\, \sum_{mn} \sum_\dtraj p_m^{(0)}\,p (\dtraj)
\ \delta \Big(\DS -\DS(\dtraj)\Big).
\label{Eq:P_of_DS}
\end{eqnarray}
where the $\dtraj$ sum is over all trajectories of the system going from level $m$ to level $n$.  Thus it is a set of Dirac delta-functions, with one delta function for each trajectory; that delta function is positioned at the entropy change of the trajectory, and has a prefactor given by the probability of that trajectory occurring.  
As the time of evolution increases, the number of trajectories 
typically increases exponentially.  
This can make it difficult to calculate the entropy distribution in a system with multiple reservoirs.

\section{\mbox{Stochastic thermodynamics} \mbox{for demons}}

Any demonic system described by the rate equations which obey local-detailed balance (reviewed here in Chapter~\ref{Chap:rate-eqns} for the two-level trapdoor) is guaranteed by to obey the fluctuation theorems presented here, by the theory of stochastic thermodynamics\cite{Schmiedl2007Jan,VandenBroeck2015Jan,Benenti2017Jun}. 
Thus we know without any calculation that the rate equation models of the two-level trapdoor (with or without phonon bath) in Chapter~\ref{Chap:rate-eqns}, and the other demonic systems in Chapter~\ref{Chap:experiments}, will obey these fluctuation theorems.

The fully quantum models in chapter~\ref{Chap:quantum-trapdoor} go beyond these rate equations, by including the physics of coherences and entanglement, however they have also been shown to obey these fluctuation theorems using the quantum diagrammatic generalization of stochastic thermodynamics in Ref.~\cite{Whitney2018Aug}. Thus the fully quantum version of the two-level trapdoor (and other autonomous demonic systems) will also obey these fluctuation theorems.

There is a technical difficulty with the N-demons described by scattering theory.
Formally one can always also describe the scatterer as a tight-binding network coupled to reservoirs, which is nothing but a many-level quantum system coupled to reservoirs.
Thus it is formally described by the quantum diagrammatic generalization of stochastic thermodynamics in Ref.~\cite{Whitney2018Aug}, which means it must also obey the above fluctuation theorems.   However, this diagrammatic method is unwieldy for specific calculations, as soon as the scattering region consists of a network of more than a few sites.   Instead it is often better to use full-counting statistics to derive various fluctuation theorems, for example see Ref.~\cite{Gaspard2015Oct,Barbier2020Aug}, although I am not aware of the status of the specific fluctuation theorems presented in this chapter.

While, both rate equations and fully quantum methods tell us that the autonomous demons obey fluctuation theorems, this is only when all entropy changes are taken into account (including those in reservoirs associated with 
the demon). The demon induces an apparent violation of these fluctuation theorems in the system it acts on.  
To see this, we note that the demon causes an apparent violation of the second law in the system that it acts on---as seen throughout this review---if one neglects the entropy change in the demon. And one cannot violate the second law while respecting the fluctuation theorems presented above\footnote{Sections~\ref{Sect:Evans-Searles} and \ref{Sect:integral-fluct-theorem} presented simple proofs that the steady-state fluctuation theorem and the integral fluctuation theorem both lead directly to the second law. It is easy to reverse those arguments, and show that a violation of the second law is impossible without a violation of these fluctuation theorems.}.  Hence, if one neglects entropy changes in the demon, all the demons in this review will also induce an apparent violation of the above fluctuation theorems
in the systems that they act upon.

\chapter{Concluding remarks}
\label{Chap:conclusion}

My main aim here was to show why autonomous quantum machines obey the second law of thermodynamics, even when there are superficial arguments to say that they could act in a similar way to a Maxwell demon.  Thus I argue that cracks in the second law, sometimes glimpsed in such systems,  have so-far turned out to be illusory.  I think nearly all experts agree that this will continue to be true as we go deeper into the quantum regime.

An alternative point of view is that this review is really about how we calculate the irreversible dynamics of quantum systems (as it applies to nanoelectronic systems).
It is about what those irreversible dynamics can do, and what they cannot do.
Part of that is about identifying a quantity called entropy that  never shrinking with increasing time, and thereby unambiguously indicates the direction of time's arrow. 
Demons are systems that superficially appear to go against the direction of time's arrow, 
so it is important to show why this is not the case.

In the context of identifying the quantity called entropy,
I take the entropy change to be given by the change in Clausius entropy of the macroscopic reservoirs, plus the change in von Neumann entropy of any small quantum systems.  The main justification for this choice is that it naturally leads to the second law. 
However, a question arises of why we take a different form of the entropy for the reservoirs and the quantum system,  when the reservoirs themselves are made of quantum systems. 
Does this choice only make sense when the reservoirs are infinite?
This has been partially addressed 
in the literature (as reviewed in Ref.~\cite{Landi2021Sep}), but I think there are still significant open questions there.  At the same time, the easiest way to prove that this entropy obeys the second law is to change perspective, abandoning the idea that entropy is a number, and accepting that it is a distribution (as outlined in the last chapter). This change of perspective also allows us to derive fluctuation theorems, which tell us much more than the second law about what irreversible dynamics can and cannot do.

Given my domain of expertise, I naturally address this question through theories and experiments in
quantum nanoelectronics, but of course the question is much broader.
In this context, it  has been enriching to discuss these questions
with people who think of them in terms of theories and experiments from other domains, such as quantum optics.
It is one of the treasures of the quantum thermodynamics community that we 
address the same questions, but with a broad variety of view points.

\section{Acknowledgements}

\small
I thank the authors of all the papers that I cite\footnote{My thanks and apologies also go to the authors of any papers that I will later discover have been forgotten in this review.} for informing and inspiring my work.
I am particularly grateful to my collaborators with whom I learnt the techniques discussed in this review. This includes those with whom I learnt scattering theory (Igor Lerner, Philippe Jacquod and Markus B\"uttiker);  those with whom I learnt the rate equations method (Rafael S\'anchez and Janine Splettstoesser); and those with whom I learnt real-time transport theory 
(Yuval Gefen, Yuriy Makhlin, and Alexander Shnirman, with whom I realised our spin-boson calculations were a simpler version of real-time transport theory, via J\'urgen K\"onig's PhD work).  Other collaborators have also greatly contributed to my understanding of demonic systems (include Alexia Auff\`eves, Fatemeh Hajiloo, Federica Haupt, and Ludovico Tesser),
and quantum thermodynamics more generally (including Giuliano Benenti, Giulio Casati, Masahiro Hasegawa, \'Etienne Jussiau, and Keiji Saito).
I am extremely grateful to the members of the jury of my Habilitation (Massimiliano Esposito, Serge Florens, Benjamin Huard, Rosa L\'opez, and Maarten Wegewijs) for extremely useful feedback on my Habilitation thesis, which became this review. 
Finally, I want to thank the wonderful community that formed within the European COST action {\it MP1209 - Thermodynamics in the quantum regime} (2013-2017), as represented by Ref.~\cite{Binder2018}.}

{\small
I acknowledges the support of the French Agence Nationale de la Recherche programme ``Investissement d'avenir'' (ANR-15-IDEX-02) via the Universit\'e Grenoble Alpes QuEnG project.  
My ongoing research is financed by an international project co-funded by the Singapore National Research Foundation and the French Agence Nationale de la Recherche, 
{\it Quantum computers in the presence of resource constraints (QuRes)} [ANR-21-CE47-0019], along with two  French Agence Nationale de la Recherche projects, 
{\it Testing Quantum Thermodynamics with local Probes (TQT)}
[ANR-20-CE30-0028], and 
{\it Spintronic harvesting of ambient thermal fluctuations (SpinElec)} 
[ANR-21-CE50-0039].}


%
%
%
 
\clearpage
\appendix

\fancyhead[RE]{\textsf{\thechapter : \leftmark}}


\chapter{A bit of useful mathematics}

This appendix gives some mathematical results that are useful in 
the calculations in the review.  These mathematical results are not in any way original, and are of no interest in themselves. 

\section{Solution of a useful matrix equation}
\label{Sect:matrix-equation}

In different situations in chapters \ref{Chap:rate-eqns} and 
\ref{Sect:3term-model}, we need the solution of an equation of the form in   
Eqs.~(\ref{Eq:basic-matrix-eqn}) below. So here we derive the solution.
In short, we want to find the occupation probability of the level $n$ 
in the steady-state, $P_n$.  In two different examples in chapters \ref{Chap:rate-eqns} and \ref{Chap:experiments}, this is given by
\begin{subequations}
\label{Eq:basic-matrix-eqn}
\begin{eqnarray}
\left(
\begin{array}{cccc}
\mathtt{A}   & 0  & \mathtt{B}  & \mathtt{C}  
\\
0 &  \mathtt{D} &  \mathtt{E} & \mathtt{F} 
\\
\mathtt{G}  &  \mathtt{H}  &  \mathtt{I}  &  0  
\\
\mathtt{J}  & \mathtt{K}  & 0  & \mathtt{L} 
\end{array}
\right)
\left(
\begin{array}{c}
P_{1}  \\ P_{2} \\ P_{3} \\ P_{4}
\end{array}
\right) &= &0\,.
\end{eqnarray}
The meaning of $P_1,\cdots P_4$ and the value of the matrix elements $(\mathtt{A},\mathtt{B}, \cdots)$ are different in the different examples.
However, in both examples, the sum of elements in each column must be zero,\footnote{This comes from the fact that Eq.~(\ref{Eq:basic-matrix-eqn}) is the steady-state solution of an evolution equation for probabilities.  Probabilities are conserved, so an increased probability in one level must be compensated by a reduced probability in other levels.} 
so one can substitute in the following relations,
\begin{eqnarray}
\mathtt{A} &=& -\mathtt{G}-\mathtt{J}\, ,
\\
\mathtt{D} &=& -\mathtt{H}-\mathtt{K}\, ,
\\
\mathtt{I} &=& -\mathtt{B}-\mathtt{E}\, ,
\\
\mathtt{L} &=& -\mathtt{C}-\mathtt{F}\, .
\end{eqnarray}
In addition, the probabilities must sum to one, so 
\begin{eqnarray}
P_1+P_2+P_3+P_4=1.
\label{Eq:Ps-sum-to-1}
\end{eqnarray}
\end{subequations}

To solve this, one can write it as four simultaneous equations. 
One can eliminate $P_4$ using Eq.~(\ref{Eq:Ps-sum-to-1}).
This leaves three simultaneous equations 
\begin{eqnarray}
 -(\mathtt{G}+\mathtt{J} + \mathtt{C}) P_1  - \mathtt{C}P_2 + (\mathtt{B}-\mathtt{C}) P_3 &=&-\mathtt{C} \, ,
 \nonumber \\
 - \mathtt{F}P_1 -(\mathtt{H}+\mathtt{K} + \mathtt{F}) P_2  + (\mathtt{E}-\mathtt{F}) P_3 &=&-\mathtt{F} \, ,\qquad \qquad 
 \\
\mathtt{G}P_1 +\mathtt{H} P_2  - (\mathtt{B}+\mathtt{E}) P_3 &=& 0 \ .
\nonumber 
\end{eqnarray}
Solving these for $P_1,P_2,P_3$ by hand is straightforward, but very tedious. We preferred to do it using computer algebra software (such as Wolfram's Mathematica).
Once one has found $P_1,P_2,P_3$, one can get $P_4$ using Eq.~(\ref{Eq:Ps-sum-to-1}).
The resulting expressions are
\begin{subequations}
\label{Eq:basic-matrix-solution}
\begin{eqnarray}
P_1 &=& \frac{\mathtt{W}}{\mathtt{W} + \mathtt{X} + \mathtt{Y} + \mathtt{Z}  },
\\
P_2 &=& \frac{\mathtt{X}}{\mathtt{W} + \mathtt{X} + \mathtt{Y} + \mathtt{Z}  },
\\
P_3 &=& \frac{\mathtt{Y}}{\mathtt{W} + \mathtt{X} + \mathtt{Y} + \mathtt{Z}  },
\\
P_4 &=& \frac{\mathtt{Z}}{\mathtt{W} + \mathtt{X} + \mathtt{Y} + \mathtt{Z}  },
\end{eqnarray}
where, for compactness, we define 
\begin{eqnarray}
\mathtt{W} &=&  
\mathtt{B}  \mathtt{C}  \mathtt{H}
+ \mathtt{B}  \mathtt{C}  \mathtt{K}
+ \mathtt{B}  \mathtt{F}  \mathtt{H}
+ \mathtt{C}  \mathtt{E}  \mathtt{K},
\\
\mathtt{X} &=&
\mathtt{B}  \mathtt{F}  \mathtt{J}
+ \mathtt{C}  \mathtt{E}  \mathtt{G}
+ \mathtt{E}  \mathtt{F}  \mathtt{G}
+ \mathtt{E}  \mathtt{F}  \mathtt{J},
\\
\mathtt{Y} &=&
 \mathtt{C}  \mathtt{G}  \mathtt{H}
 + \mathtt{C}  \mathtt{G}  \mathtt{K}
 + \mathtt{F}  \mathtt{G}  \mathtt{H}
 + \mathtt{F}  \mathtt{H}  \mathtt{J},
\\
\mathtt{Z} &=&
\mathtt{B}  \mathtt{H}  \mathtt{J}
+ \mathtt{B}  \mathtt{J}  \mathtt{K}
+ \mathtt{E}  \mathtt{G}  \mathtt{K}
+ \mathtt{E}  \mathtt{J}  \mathtt{K}.
\end{eqnarray}
\end{subequations}
I use these expressions a number of times in this review.
Unfortunately they are not very intuitive; at least I have 
have no intuition about why $\mathtt{W}$, $\mathtt{X}$, $\mathtt{Y}$ and $\mathtt{Z}$ have the form that they do.


\fancyhead[RE]{\textsf{\leftmark}}

\bibliographystyle{unsrturl}  
\bibliography{cite}

\end{document}